\documentclass{article}

\usepackage{authblk}
\usepackage[breaklinks]{hyperref}
\usepackage{graphicx}
\graphicspath{{figs/}}
\usepackage{bbm}
\usepackage{placeins}
\AtEndOfClass{\RequirePackage{microtype}}
\RequirePackage[utf8]{inputenc}
\RequirePackage[english]{babel}
\RequirePackage{amsmath,amsfonts,amssymb}
\RequirePackage{lmodern}
\RequirePackage[scaled]{helvet}
\RequirePackage{afterpage}
\RequirePackage{ifpdf,ifxetex}
\RequirePackage{xcolor}
\RequirePackage{tikz}
\RequirePackage[framemethod=tikz]{mdframed}
\RequirePackage{graphicx,xcolor}
\RequirePackage{colortbl}
\RequirePackage{booktabs}
\RequirePackage{algorithm}
\RequirePackage[noend]{algpseudocode}
\RequirePackage{changepage}

\RequirePackage[T1]{fontenc}
\RequirePackage{lettrine} 

\RequirePackage{afterpage}
\RequirePackage{ifpdf,ifxetex}
\RequirePackage{xcolor}
\RequirePackage{tikz}
\RequirePackage[framemethod=tikz]{mdframed}
\RequirePackage{graphicx,xcolor}
\RequirePackage{colortbl}
\RequirePackage{booktabs}
\RequirePackage{algorithm}
\RequirePackage[noend]{algpseudocode}
\RequirePackage{changepage}
\RequirePackage[explicit]{titlesec}
\RequirePackage{etoolbox}
\AtBeginEnvironment{tabular}{
\fontsize{7.5}{10}\selectfont
} 
\newcommand{\addtabletext}[1]{{\setlength{\leftskip}{9pt}\fontsize{7}{9}\selectfont#1}}

\RequirePackage{lineno}
\makeatletter
\providecommand{\@LN}[2]{}
\makeatother

\titleformat{\subsection}[runin]
  {\sffamily\bfseries}
  {\thesubsection.}
  {0.5em}
  {#1. }
  []

\RequirePackage[twoside,%
                letterpaper,includeheadfoot,%
                layoutsize={8.125in,10.875in},%
                layouthoffset=0.1875in,%
                layoutvoffset=0.0625in,%
                left=38.5pt,%
                right=43pt,%
                top=43pt,
                bottom=32pt,%
                headheight=0pt,
                headsep=10pt,%
                footskip=25pt,
                marginparwidth=38pt]{geometry}
\RequirePackage[labelfont={bf,sf},%
                labelsep=period,%
                figurename=Fig.]{caption}
\RequirePackage[labelfont={bf,sf},%
                labelsep=period,%
                figurename=Fig.]{caption}
\RequirePackage[numbers,sort&compress,merge,round]{natbib}
\RequirePackage{jabbrv}
\RequirePackage{fancyhdr}  
\RequirePackage{lastpage}  
\RequirePackage[explicit]{titlesec}

\usepackage{siunitx}
\usepackage{booktabs}
\def\sym#1{\ifmmode^{#1}\else\(^{#1}\)\fi}

\title{Networks and Identity Drive Geographic Properties of the Diffusion of Linguistic Innovation}
\author[a,*]{Aparna Ananthasubramaniam}
\author[a,b]{David Jurgens} 
\author[a,b,c]{Daniel M. Romero}

\affil[a]{School of Information, University of Michigan, 105 S State St, Ann Arbor, Michigan 48109}
\affil[b]{Computer Science and Engineering Division, Electrical Engineering and Computer Science Department, University of Michigan, 2260 Hayward St, Ann Arbor, MI 48109}
\affil[c]{Center for the Study of Complex Systems, University of Michigan, 500 Church St, Ann Arbor, MI 48109}
\affil[*]{To whom correspondence should be addressed}

\begin{document}

\maketitle

\begin{abstract}
Adoption of cultural innovation (e.g., music, beliefs, language) is often geographically correlated, with adopters largely residing within the boundaries of relatively few well-studied, socially significant areas. These cultural regions are often hypothesized to be the result of either (i) identity performance driving the adoption of cultural innovation, or (ii) homophily in the networks underlying diffusion. In this study, we show that demographic identity and network topology are both required to model the diffusion of innovation, as they play complementary roles in producing its spatial properties. We develop an agent-based model of cultural adoption, and validate geographic patterns of transmission in our model against a novel dataset of innovative words that we identify from a 10\% sample of Twitter. Using our model, we are able to directly compare a combined network + identity model of diffusion to simulated network-only and identity-only counterfactuals---allowing us to test the separate and combined roles of network and identity. While social scientists often treat either network or identity as the core social structure in modeling culture change, we show that key geographic properties of diffusion actually depend on both factors as each one influences different mechanisms of diffusion. Specifically, the network principally drives spread among urban counties via weak-tie diffusion, while identity plays a disproportionate role in transmission among rural counties via strong-tie diffusion. Diffusion between urban and rural areas, a key component in innovation diffusing nationally, requires both network and identity. Our work suggests that models must integrate both factors in order to understand and reproduce the adoption of innovation. 
\end{abstract}

\section{Introduction}
From religious beliefs \citep{kong1990geography,land1991religious} to popular music \citep{kellogg1987spatial,nash1996seven} to memes on social media \citep{kamath2013spatio,dang2019simulating}, the adoption of social and cultural innovations exhibit notable geographic variation, with adopters often residing in a consistent set of familiar regions (e.g., within U.S.A., the Deep South or the Mid-Atlantic) \citep{jackson2012maps}. For instance, when new words are coined, they are often adopted by speakers in geographic areas that reflect their social, cultural, and historical significance \citep{chambers1998dialectology,grieve2016regional}. In fact, many social science disciplines (e.g., cultural and social geography, sociolinguistics) use linguistic variables as a proxy for other phenomena like migration and political shifts \citep{labov1963social,labov2012dialect}. Language is a particularly useful proxy for cultural change as shifts in culture often drive language change, and conversely, adoption of linguistic innovation can hasten cultural change \citep{kramsch1998language,Beckner2009}. The principal objective of most geographic analysis of lexical innovation has been to understand what drives social, cultural, and the resulting linguistic shifts. Specifically, researchers often use geographic distributions of linguistic adoption to test putative mechanisms of diffusion \citep{trudgill1974linguistic,trudgill2002sociolinguistic,chambers2018handbook}: If a mechanism cannot explain why speakers adopt a new coinage in certain areas but not others, we can falsify our hypotheses about the significance of that linguistic shift. 

Existing studies often attribute the regionalization of language to effects of either \textit{identity} or \textit{network} \citep{jackson2012maps}. On one hand, speakers' desires to perform their social identity (e.g., race, socioeconomic status) may drive selection among variants \citep{sturtevant1947introduction,labov1963social,eckert2008variation}. This explanation is congruent with how variationist sociolinguistic theory connects social, cultural, and historical factors to language change, often explaining geographic variation as the byproduct of spatial assortativity in personal characteristics \citep{chambers1998dialectology,labov2007transmission,schwartz2015spatial}. On the other hand, spatially concentrated adoption may be a consequence of the social network \citep{bloomfield1933language,milroy1987language}. Since densely connected communities in the network are, themselves, geographically and demographically homophilous \citep{mcpherson2001birds,adamic2003friends,lizardo2006cultural,fagyal2010centers,takhteyev2012geography}, some branches of network and social theory treat language regions as simply the product of diffusion within this particular network topology. 

Our central hypothesis is that key spatial properties of linguistic diffusion depend on both social identity and network topology. Specifically, we contend that the complementary roles of and interactions between network and identity are core mechanisms underlying the adoption of innovation---and, as such, omitting either one leads to incomplete inferences about the latent process. Our hypothesis constitutes a major departure from existing analytic approaches, which tend to focus on either network or identity as the primary mechanism of diffusion. For instance, cultural geographers rarely explore the role of networks in mediating the spread of cultural artifacts \citep{jackson2000rematerializing}, and network simulations of diffusion often do not explicitly incorporate demographics \citep{newman2006structure}. Even within fields that acknowledge both network and identity as drivers of diffusion, like variationist sociolinguistics, identity-centered and network-centered explanations of language change are often described competing hypotheses \citep{labov2007transmission, blythe2012scurve}.

Testing our hypothesis requires comparing a combined network/identity model of diffusion to network-only and identity-only counterfactuals---and since network and identity are woven so deeply into the social fabric, we cannot empirically observe these baselines. Instead, we construct a novel agent-based model, inspired by cognitive and social theory, to model the spread of new words through a network of speakers. Using agent-based models affords us a unique view into how network and identity interact, because it allows us to simulate network- and identity-only counterfactuals \citep{marshall2015formalizing}. We combine our simulations with large-scale empirical analysis in order to validate our model. Specifically, we create a spatial time series dataset of lexical innovations on the microblog site Twitter, allowing us to study the spatial diffusion of language change online. Key simulation parameters, including network topology and demographic identity, are drawn from the microblog Twitter.

Bridging network and cultural theory, we identify a mechanism for the diffusion of cultural innovation that suggests that identity performance and network topology play fundamentally different roles in the diffusion of innovation. Specifically, network mediates the spread along pathways between urban counties through weak-tie diffusion, while identity promotes strong-tie diffusion along rural-rural pathways. Furthermore, network and identity jointly drive transmission from urban centers to rural locales. Interestingly, the urban/rural heterogeneity is an emergent property of the distributions of network ties and demographics (i.e., differences between urban and rural counties are present even though we do not explicitly include them in our model formulation). Taken together, we conclude that models omitting either network or identity are missing a crucial dynamic in the adoption of innovation. As a result, key properties of linguistic diffusion---both the geographic regions that innovation spreads to and the spatiotemporal pathways through which they diffuse---are better approximated by network and identity together than either one individually. 

\section{\label{model} Model}
We develop an agent-based model to evaluate the roles of social identity and network topology in the spatial patterns of cultural diffusion. To realistically model the adoption of innovation \citep{valente1996social}, our formulation draws heavily from social and cognitive theory, and underlying assumptions are empirically derived. Our model simulates the diffusion of a new word $w$. The model begins with a set of initial adopters introducing the word to the lexicon (Section~\ref{words}), and spreads across a directed network of $n$ agents $\{j\}_{j=1}^{n}$ (Sections~\ref{network} and~\ref{agent-identity}). The new word connotes a particular social identity $\Upsilon_w$ that is assigned based on the identities of its early users (Section~\ref{word-identity}). In our simulations, the word continues to spread through the network over several subsequent iterations (Section~\ref{diffusion-eqs}). Agents are exposed to the word when a network neighbor uses it. Agents are more likely to use the word if it signals an identity congruent with their own and if they were recently exposed by network neighbors with similar identities. See Section~\ref{model-si} for the full set of model equations and Section~\ref{param-est} for information about parameters and how they are inferred. Our model's limitations, along with our attempts to address them, are in Section~\ref{limitations-si}. Although we test our model against diffusion of linguistic innovation, its formulation is sufficiently general to describe the adoption of many cultural innovations beyond new words.

\subsection{\label{words}New Words and Initial Adopters}
We simulate the diffusion of widely-used lexical innovations originating on Twitter between 2013 and 2020. Starting from all 1.2 million non-standard slang entries in the crowdsourced catalog UrbanDictionary.com, we identify 76 new words that were tweeted rarely before 2013 and frequently after (see Section~\ref{si:identifying-new-words} for details of the filtration process). Consistent with prior studies of online innovation \citep{crystal2011internet,eisenstein2012mapping,Grieve2018}, the 76 new words in our study include terms describing popular culture phenomena (e.g., fanmix, sweaties), phonologically-motivated orthographical shifts (e.g., bawmb, whatchoo), part-of-speech changes (e.g., ubering, lebroning), abbreviations (e.g., ihml, profesh), concatenations (e.g., amaxing, sadboi), and even new coinages (e.g., gwuap, fleeky) ( Table~\ref{words-table} has more examples). These words diffuse in well-defined geographic areas that match prior studies of online and offline innovation \citep{eisenstein2012mapping,Grieve2018} (see Figure~\ref{top_pcs} for a detailed comparison).

Each run of our model simulates the diffusion of one of these 76 words, and we run five differently-seeded trials per word. Each simulation's initial adopters are the corresponding word's first ten users in our tweet sample (see Section~\ref{init-adopters-si}). Model results are not sensitive to small changes in the selection of initial adopters (Section~\ref{si-sensitivity-init}).

\subsection{\label{network}Network}
Patterns in the diffusion of contagion are often well-explained by the topology of speakers' social networks \citep{bloomfield1933language,milroy1987language,dimaggio2011cultural,breiger2015culture,dimaggio2021information}. Therefore, the word in our model diffuses through a network of agents. Nodes (agents) and edges (ties) in this network come from the Twitter Decahose, which includes a 10\% random sample of tweets between 2012 and 2020. Agents in our model correspond to Twitter users in this sample who are located in U.S.A. We draw an edge between two agents $i$ and $j$ if they mentioned each other at least once, and the strength of the tie from $i$ to $j$ $w_{ij}$ is proportional to the number of times $j$ mentioned $i$ from 2012 to 2019 \citep{Granovetter1973}. This directed network has nearly 4 million nodes (agents) and 30 million edges (dyads); the edge drawn from agent $i$ to agent $j$ parametrizes $i$'s influence over $j$'s language style (e.g., if $j$ weakly weighs input from $i$, $w_{ij}$ is small).

Importantly, model results are not specific to the Twitter network topology, generalizing to the Facebook Social Connectedness Index network as well (Section~\ref{si-sensitivity-fb}) \citep{bailey2018social}.

\subsection{\label{agent-identity}Agent Identity}

An individual often adopts a cultural innovation that signals their membership to the desired demographic or other social identity \citep{friedman1989culture,cote1996sociological,jones1997virtual,eckert2008variation}. In our model, demographics are proxies for identity, and agents are characterized by five categories shown to be important to language style: (i) location within the U.S.A. \citep{labov1963social,eisenstein2010latent}, (ii) race/ethnicity \citep{rickford1999african,fought2002chicano,stewart2014now,Jones2015}, (iii) socioeconomic status measured via income level, educational attainment, and workforce participation \citep{labov1990intersection,milroy1992social,abitbol2018socioeconomic}, (iv) languages spoken \citep{haugen1950analysis,lo1999codeswitching,stewart2018si}, and (v) political affiliation \citep{labov2012dialect,sylwester2015twitter}. Each category is parametrized by several related registers (e.g., for political affiliation, ``registers'' are Democrat, Republican, and Third Party). An agent may identify with each register to a different degree \citep{agha2005voice,eckert2008variation}, so each identity register can take on a value in $[0,1]$. 

We infer each agent's location from their GPS-tagged tweets, using Compton et al. (2014)'s high-precision algorithm \citep{compton2014geotagging} (see Section~\ref{location-si} for details). Since Twitter does not supply demographic information for each user, we infer their identities based on location \citep{xiong2020mobile,wrigley2020us}. Similar to prior work studying sociolinguistic variation on Twitter \citep{grieve2016regional,grieve2019mapping}, we estimate each agent's race/ethnicity, SES, and languages spoken from the composition of their inferred Census Tract in the 2018 American Community Survey. We also represent each agent's political affiliation using their Congressional District's results in the 2018 U.S.A. House of Representatives election. Since Census tracts and Congressional districts are small, often fairly homogenous, units of geography, we expect the corresponding demographic estimates to be fairly granular and accurate \citep{us1994census,powell2017assessing}.

\subsection{\label{word-identity}Word Identity}

Just as each agent has a demographic identity, cultural innovations can be used to signal different registers of this identity \citep{goffman1978presentation,oring1994arts,wagner2014gettin}. Each word may provide information about one or more of the categories from Section~\ref{agent-identity} like location, race, etc. \citep{Eckert2012}; for each word, we denote the relative importance of each category with weight vector $v_w \in [0,1]^D$. Unlike agent identity, words tend to connote affiliation with a specific register of identity (e.g., in Eckert 2000, high schoolers may associate with multiple social groups, but each linguistic variable signals membership to a particular group \citep{eckert2000language}). Therefore, word identities in our model are binary (i.e., a word either signals a given register of identity or it doesn't), and each we model word identities distributed in $\Upsilon_w \in \{0,1\}^d$ unlike agents' identities in $\Upsilon_j \in [0,1]^d$. 

A word's identity is often \textit{enregistered} based on the demographics of a small number of its early adopters \citep{agha2005voice}, signaling registers of identity that these speakers are far more likely than other agents to identify with. For instance, if the initial adopters tend to come from disproportionately Republican, African American, French-speaking areas like Louisiana, the word signals this sociodemographic identity: specifically, $v_w = \frac{1}{3}$ for the dimensions corresponding to the political affiliation, race, and language categories; $\Upsilon_w = 1$ for the dimensions corresponding to the Republican political affiliation, African American race, and French language registers; and other entries of $v_w$ and $\Upsilon_w$ are $0$ (see Sections~\ref{ideq-si}-\ref{wordid-si} for a more formal description). Agent identities remain unaltered by a word's enregisterment. Since speakers often quickly converge to a universal understanding of the identity signaled by the word \citep{Agha2003}, our model assigns a word's identity based on the word's first ten adopters.

\subsection{\label{diffusion-eqs}Diffusion}
After the initial adopters introduce the innovation and its identity is enregistered, the new word spreads through the network as speakers hear and decide to adopt it over time. Although we present a model for lexical adoption, the cognitive and social processes underlying the adoption of linguistic innovation generalize well to many other cultural innovations \citep{dimaggio1997culture,dimaggio2010culture}. In our model, agents do not use the word until they have been exposed to it by a network neighbor at least once. At each discrete time iteration $t$, agent $j$ decides whether they will use the word $w$ with dynamic likelihood $p_{jwt} \in [0,1]$, reflecting whether the word is salient to them \citep{Ellis2019}. This probability is contextually and temporally fluid, aggregating information from agents' exposures to the novel item \citep{Agha2003,kirby2014iterated}. 
On one hand, if agent $j$ was previously exposed to the word but is not exposed at iteration $t$, their attention to the new word, and their likelihood of adoption, fades \citep{Ellis2019}.\footnote{Since sites like Twitter have large amounts of innovative content \citep{weng2012competition,shalom2019fading}, we would assume that agents' attention to a new word would fade without exposure because of all of the other new content they are exposed to. This may not be as true in settings where there is less competing innovation. In these cases, we can eliminate Equation~\ref{decayeq} from the model.} Since attention tends to fade exponentially \citep{shalom2019fading}, we assume that agents retain fraction $r \in [0,1]$ of their attention to the word as modeled in Equation~\ref{decayeq}:
\begin{equation}\label{decayeq}
p_{j,w,t+1} = r \cdot p_{jwt}
\end{equation}

On the other hand, an agent $j$ updates their likelihood of using the word $p_{j,w,t+1}$ if $j$'s network neighbor $i \in N(j)$ uses the word at iteration $t$, $i \in adopt(t)$. At this point, agent $j$'s mental representations are dominated by characteristics of this most recent exposure \citep{Beckner2009} and the salience of the innovation is determined by five main characteristics: (i) \textbf{Novelty}: With greater exposure, a word's novelty wears off and its salience declines \citep{ellis2017salience}.\footnote{This assumption may not apply to more persistent innovations, whose adoption grows via an S-curve \citep{blythe2012scurve} Since new words that appear in social media tend to be fads whose adoption peaks and fades away with time (Figure~\ref{word-ts}), we model the decay of attention theorized to underly this temporal behavior \citep{weng2012competition,shalom2019fading}. We can model more persistent change by removing the $\eta$ parameter from the equation.} (ii) \textbf{Stickiness}: Some words are inherently ``stickier'' than others, tending to experience higher coinage and adoption because they are related to topics of growing importance, used across a variety of semantic contexts, or have notable linguistic properties \citep{schmid2016toward,stewart2017making,ryskina2020new}. (iii) \textbf{Relevance}: since speakers often use language to perform their own identity, agents may preferentially use words whose demographics more closely match their own \citep{labov1963social,eckert2008variation}; (iv) \textbf{Variety}: In addition to common identity, diverse exposure, from multiple people across multiple contexts, improves a word's salience \citep{tomasello2000first,smith2011cross}; and (v) \textbf{Relatability}: Since self-expression and social engagement are key motivators for use of social networking sites, input from agents with similar identity may weigh more heavily \citep{Granovetter1973,valente1996social,watts2002simple,mcleish2011social,alhabash2017tale}. 

Per equation~\ref{modeleq}, we model these characteristics by making $p_{j,w,t+1}$ proportionate to: (i) \textbf{Novelty}: a cosine decaying function of the number of exposures $j$ has had to the word $\eta_{jwt}$; (ii) \textbf{Stickiness:} the ``stickiness'' of the word $S_w$; (iii) \textbf{Relevance:} the similarity between $j$'s identity and their understanding of the word's identity, $\delta_{jw}$; (iv) \textbf{Variety}: the fraction of their network neighbors to have adopted the word at iteration $t$; and (v) \textbf{Relatability}: this fraction is weighted by the similarity in their identity $\delta_{ij}$ and tie strength $w_{ij}$. 
\begin{equation}\label{modeleq}
p_{j,w,t+1} = \delta_{jw} S_w \eta_{jwt} \frac{\sum\limits_{i \in N(j) \cap adopt(t)} w_{ij} \delta_{ij}}{\sum\limits_{k \in N(j)} w_{kj} \delta_{kj}}
\end{equation}

Identity comparisons ($\delta_{jw}$, $\delta_{ij}$) are done component-wise, and then averaged using the weight vector $v_w$ (Section~\ref{word-identity}). See Section~\ref{diffusion-si} for the full set of model equations.

We stop the model once the growth in adoption slows to under 1\% increase over 10 iterations. Since early iterations have low adoption, uptake may fall below this threshold as the word is taking off; we eliminate such false-ends by running at least 100 iterations after initialization.

\subsection{\label{experiments}Simulated Counterfactuals}
We directly assess the roles of network homophily and performance of identity in linguistic diffusion, by evaluating the impact of omitting each of these sets of variables from the model. Accordingly, we simulate four counterfactual conditions: 
\begin{enumerate}
\setlength{\itemsep}{0pt}
\setlength{\parskip}{0pt}
    \item \underline{Network+Identity}: the full model described above.
    \item \underline{Network-only}: where we eliminate agents performing identity, by simulating the spread through just the weighted networks ($\delta_{ij},\delta_{jw}=1$). 
    \item \underline{Identity-only}: where we shuffle the edges of the network. This configuration model-like procedure \citep{bollobas1980probabilistic} preserves each agent's degree, allowing us to isolate the impact of eliminating the single network characteristic most often hypothesized to drive regionalization, homophily, while holding constant other network-geographic confounds like population and degree distributions.
    \item \underline{Null (Shuffled Network+No Identity)}: where we use the shuffled network without identity variables. This holds constant several variables (e.g., population size, degree distribution, model formulation), allowing us to isolate the impact of homophily and socially-motivated selection by comparison to the network- and identity-only models.
\end{enumerate}

We evaluate each model using 380 simulations: 5 random trials on each of the 76 lexical innovations described in Section~\ref{words}.

\section{Network and Identity Predict Spatial Properties Jointly but not Individually}

\begin{figure}
    \centering
    \includegraphics[width=\textwidth]{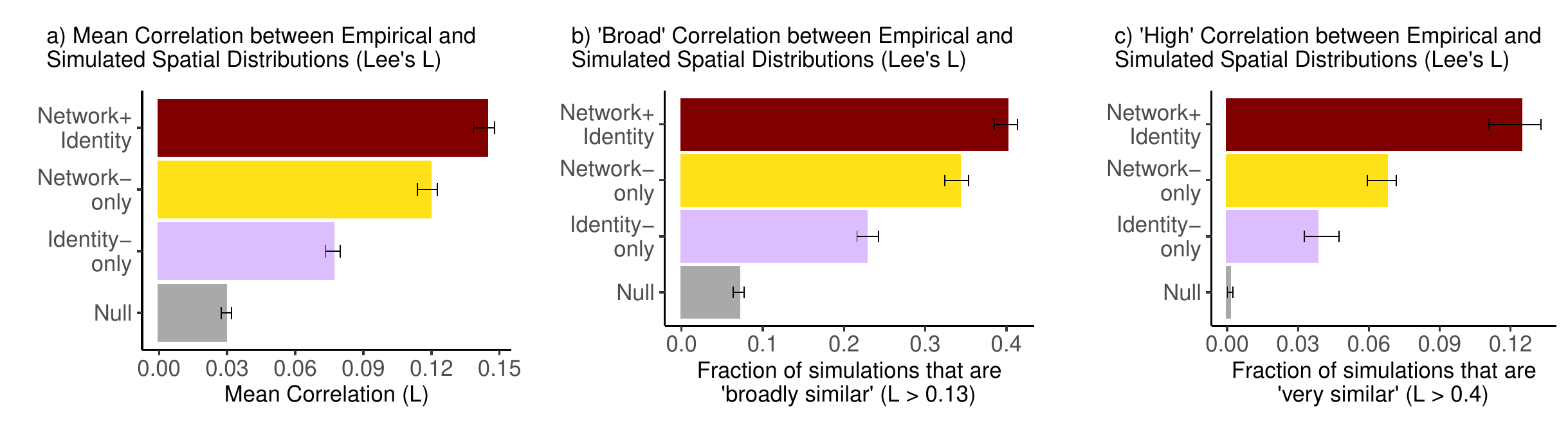} 
    \caption{The Network+Identity model trials most strongly correlates to empirical distributions, measured using: a) average Lee's L correlation over all trials, b) fraction of trials that are ``broadly similar'' ($L>0.13$); and c) fraction of trials that are ``very similar'' ($L>0.4$). Error bars are 95\% bootstrap confidence intervals.}
    \label{eval-lee}
\end{figure}

Cultural artifacts like language often diffuse in well-known geographic regions. Our model formalizes two interacting mechanisms thought to generate this spatial heterogeneity: 1) \textit{network}: edges tend to concentrate between demographically similar locales, meaning words may diffuse in areas well-connected by this network; and 2) \textit{identity}: linguistic variants are selectively adopted in (and subsequently transmitted from) areas where speakers identify with their social signal (e.g., a word like `democrap' will likely get more use in a Republican-leaning area). In this section, we show that these geographic regions are not only well-explained by network and identity, but the consequence of their joint contributions.

\subsection{Hypotheses} 
In studying culture change, cultural geographers often measure the frequency of adoption of innovation in different parts of the U.S.A. \citep{denevan1983adaptation, labov2008atlas, eisenstein2012mapping}, as well as a new word's propensity to travel from one geographic area (e.g., counties) to another \citep{denevan1983adaptation,labov2008atlas,eisenstein2012mapping}. In both the physical and online worlds, the spatial distribution of variants frequently carries signals about their cultural significance \citep{rose2016cultural, trudgill1974linguistic}, while spread between pairs of counties acts like ``pathways'' along which, over time, variants diffuse into particular geographic regions \citep{denevan1983adaptation,labov2008atlas,eisenstein2012mapping}. 
Using our model, we test the separate and joint effects of network and identity on a word's spatial and spatiotemporal diffusion:

\begin{enumerate}
    \item [H1] We hypothesize that our model most accurately predicts (i) the spatial distribution of each word's usage and (ii) spatiotemporal pathways between pairs of counties when it is given information about both homophily in the network topology and performance of identity. Specifically, we would expect the Network+Identity model to outperform all other models, and the Null (Shuffled Network+No Identity) model to perform the worst. 
\end{enumerate}

\subsection{Experimental Setup}
To test H1, we run identically-seeded trials across all four simulated counterfactuals (Section~\ref{experiments}) and track where in the country the adopters of a new word are located. We assess whether the Network+Identity model better matches empirical geographic properties than the Network- and Identity-only models, and whether those further boost the Null (No Network or Identity) model's performance.\footnote{Our model predicts the spatial diffusion and pathways of a new word from first principles, unlike machine learning models that often learn these macroscopic patterns from the data. Therefore, we do not penalize the full model for added complexity.} 

First, we assess whether each model trial diffuses in a similar region as the word on Twitter. We compare the simulated and empirical spatial distributions of adoption using Lee's $L$, an extension of Pearson's $R$ correlation that controls for the effects of spatial autocorrelation \citep{lee2001developing}. Based on Grieve et al. (2019)'s evaluation of this metric \citep{grieve2019mapping}, the simulated and empirical regions are ``very similar'' if the correlation between the two spatial distributions is $L \geq 0.4$, ``broadly similar'' if $L \geq 0.13$, and ``not similar'' otherwise (see Section~\ref{lees-l-si} for details). All reported differences are statistically significant at the level $\alpha=0.05$, using a bootstrap hypothesis test. 

Second, we compare the strength of empirical pathways against simulated pathways from the four counterfactuals. The strength of the pathway between counties $i$ and $j$ is $j$'s propensity to adopt the word after $i$ does---measured via the zero-inflated correlation $\tau$ \citep{pimentel2009kendall} between $i$'s level of adoption at iteration $t$ and $j$'s adoption at $t+1$. We compare empirical to simulated pathways by calculating the Bayesian likelihood of the empirical pathway strengths $\tau_{E}$ given the corresponding model pathway strengths $\hat{\tau_{N+I}}$, $\hat{\tau_{N}}$, or $\hat{\tau_{I}}$. (See Section~\ref{pathway-strength-si} for more details.)

\begin{figure*}[t]
    \centering
    \includegraphics[width=\textwidth]{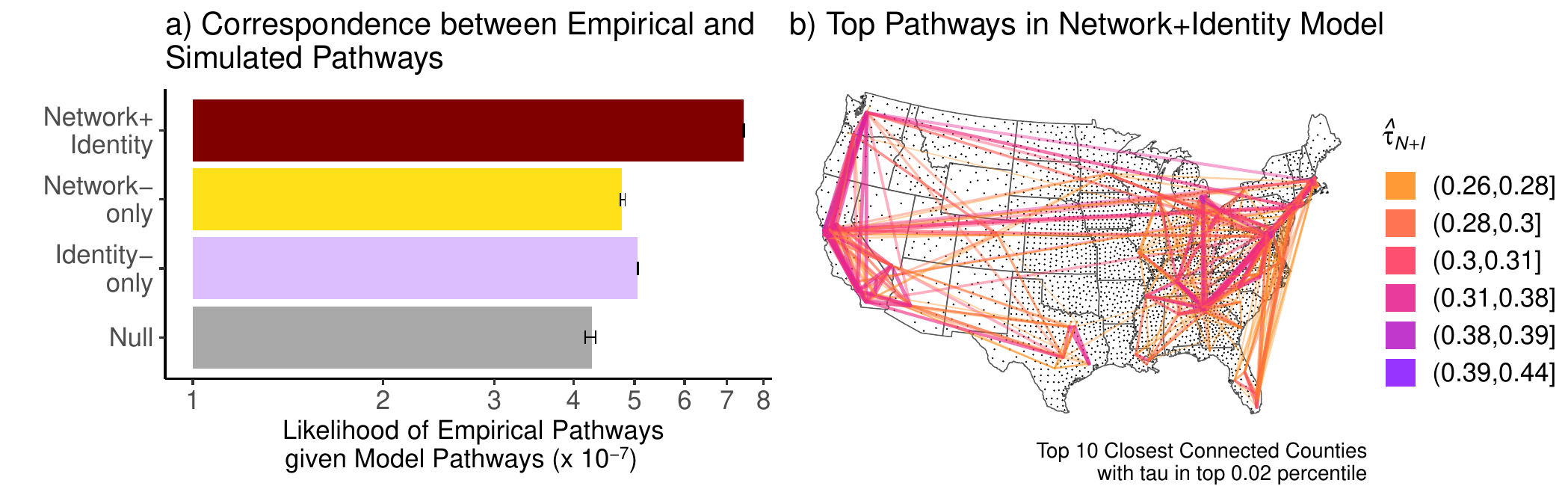} 
    \caption{The Network+Identity model best approximates the relative strengths of empirical pathways: a) The likelihood of the pathways observed on Twitter given each of the simulations is highest in the full model. b) The strongest pathways between pairs of counties in the Network+Identity model correspond to culturally significant regions (cf., Figure~\ref{pathways-cultural}). Pathways are shaded by their strength (purple is more strong, orange is less strong); if one county has more than 10 pathways in this set, just the 10 strongest pathways out of that county are pictured.}
    \label{pathway-map}
\end{figure*}

\subsection{Results}

The full Network+Identity model, using both realistic network topology and social identity, can often predict properties of a word's spatial diffusion. The Network+Identity model's simulated spatial distributions approximate empirical spatial distributions in the majority of trials. Nearly 40\% of simulations are at least ``broadly similar'' (Figure~\ref{eval-lee}b) and its adoption regions are, on average, ``broadly similar'' to those on Twitter (mean($L$)$\approx 0.15$) (Figure~\ref{eval-lee}a). Visually, the full model's strongest pathways correspond to well-known cultural regions (Figure~\ref{pathway-map}b). Some pathways extend from the mid-Atlantic into the South, where African American Language is most spoken \citep{Jones2015}; from Atlanta to other urban hubs, along pathways defined by the Great Migrations \citep{Jones2015}; along and between both coasts, which are politically, linguistically, and racially distinctive from the middle of the country \citep{labov2012dialect,sylwester2015twitter}; and within the economically significant Dallas-Austin-Houston ``Texas triangle'' \citep{cisneros2021texas}. The strong performance of the full model suggests that, together, network and identity can not only reproduce the spread of words on Twitter, but does so via socially significant pathways of diffusion. Our model appears to reproduce the mechanisms that give rise to several well-studied cultural regions.

Network and identity are both necessary to maintain the full model's efficacy. Specifically, the Network- and Identity-only models are outperformed by the Network+Identity simulations. Compared to the Network- and Identity-only models, the Network+Identity model trials were about 1.14 - 1.73 times as likely to be ``broadly similar'' to the corresponding empirical distribution (Figure~\ref{eval-lee}b), and 1.9 - 5x times as likely to be ``very similar'' (Figure~\ref{eval-lee}c). Moreover, the likelihood of the pathways observed on Twitter is more than 50\% higher given the Network+Identity model's pathways than the other models' pathways (Figure~\ref{pathway-map}a)---suggesting that the Network- and Identity-only models have diminished capacity to predict geographic distributions of lexical innovation may be attributable to the failure to effectively reproduce the spatiotemporal mechanisms underlying cultural diffusion. 

Finally, we note that the Null (Shuffled Network+No Identity) model fails to reproduce spatial distributions and pathways. The Network- and Identity-only models far overperform the Null simulations: e.g., geographic distributions produced by the Network- and Identity-only models were 2.5-3 times as likely to be ``broadly similar'' to the corresponding empirical distribution (Figure~\ref{eval-lee}b), while none of the Null model's trials were ``very similar'' to the corresponding empirical distribution (Figure~\ref{eval-lee}c). Moreover, using the Null model as a prior yields a lower likelihood of the empirical pathways than the Network- or Identity-only models (Figure~\ref{pathway-map}a). 

We conclude that spatial patterns of linguistic diffusion are the product of network and identity acting together, and that both network homophily and social identity account for some key diffusion mechanism that is not explained by structural factors alone (e.g., population density, degree distributions, cognitive mechanisms and other aspects of the model formulation). 

\section{\label{complementary-roles}Network and Identity Play Complementary, Interacting Roles}

Next, we show that Network- and Identity-only pathways play fundamentally unique roles in the spatiotemporal transmission of innovation.\footnote{Notably, the pathway strengths in the Network- and Identity-only models are strongly correlated (Pearson's $R = 0.78$, Spearman's $\rho = 0.81$). A strong correlation is to be expected, since network homophily often correlates with demographics \citep{mcpherson2001birds}. See Section~\ref{empirical-pathways-si} for more details about this correlation. Nonetheless, the Network- and Identity-only pathways exhibit important differences, which we discuss in this section.} Specifically, these two social structures take on complementary, interacting functions: identity pathways drive transmission among rural counties via strong-tie diffusion, while network pathways dominate urban-urban spread via weak-tie diffusion. Our framework unites two well-studied dichotomies in the diffusion of cultural innovation:

\paragraph*{Urban vs. Rural Adopters:} From music \citep{brunstad2010hip}, to opinion \citep{fischer1978urban}, to linguistic variation \citep{labov2003pursuing}, spatial patterns of cultural diffusion are often mediated by differences in adoption in urban and rural areas. Urban centers are larger, more diverse, and therefore often first to adopt new cultural artifacts \citep{stewart1958urban,agergaard2015rural}; innovation subsequently diffuses to rural areas and starts to signal a local identity \citep{trudgill18}. Evidence from social networking sites suggests that urban vs. rural heterogeneity persist online \citep{meyerhoff2003globalisation}. Consistent with prior studies of urban vs. rural areas \citep{gilbert2008network,trudgill18}, speakers in our Twitter sample exhibit differences in: (i) \textit{Tie Strength:} urban-urban pathways tend to have a higher fraction of weak ties running along them, while rural-rural pathways tend to be dominated by strong ties (Figure~\ref{fig-size}); and (ii) \textit{Diversity:} ties between rural counties tend to exhibit far more demographic similarities than ties between urban counties (Figure~\ref{fig-sharedid}). 

\paragraph*{Weak-Tie vs. Strong-Tie Mechanisms:} Social theory outlines two mechanisms for the adoption of innovation that relate to the strength and diversity of ties: (i) \textit{Weak-tie diffusion} suggests that new coinages tend to diffuse to diverse parts of the network via weak ties, or edges between individuals with more distant relationships \citep{milroy1992social,milroy1987language,zhu2021structure}; and (ii) \textit{Strong-tie diffusion} purports that variants tend to diffuse among demographically similar speakers (often connected by strong ties) who share a pertinent identity \citep{eckert1989woman,labov2006social}. In our Twitter sample, ties with lower edge-weight tend to share fewer demographic similarities than edges with higher weight (Table~\ref{reg-empnetwork}). Thus, weak and strong ties not only play different roles in the diffusion of information, but also spread novelty to different types of communities \citep{Granovetter1973,bakshy2012role}.

\begin{figure*}[t]
    \centering
    \includegraphics[width=\textwidth]{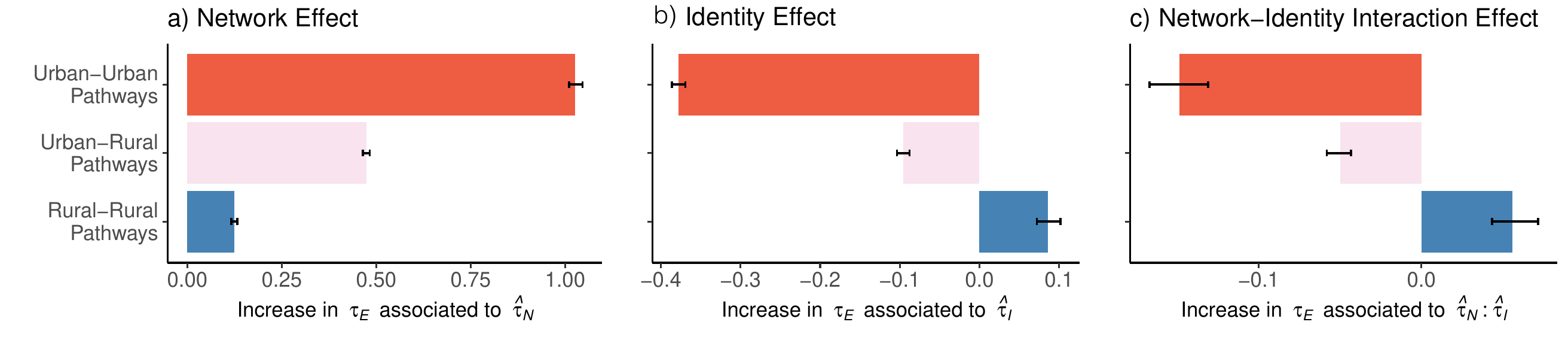} 
    \caption{Based on standardized coefficients from a linear regression predicting empirical pathway strength ($\tau_E$) from interactions between the strength of the pathways in the Network- and Identity-only models ($\hat{\tau_{N}}$,$\hat{\tau_{I}}$) and the type of pathway (urban vs. rural county): a) the strength of the Network-only model's pathways have the largest effect on the strength of the urban-urban empirical pathways and are positively associated with all pathways; b) conversely, identity strength has the largest effect on the strength of rural-rural pathways and is negatively associated with urban pathways; and c) urban strong network pathways are weakened by strong identity pathways---and conversely, rural-rural strong identity pathways are strengthened by strong network pathways. Error bars are 95\% bootstrap confidence intervals.}
    \label{fig-emerge}
\end{figure*}

\subsection{Hypotheses} 
Based on social science theory and our empirical observations about strong/weak-tie diffusion and urban/rural dynamics, we propose the following mechanism for spatial diffusion: The adoption of innovation among urban counties tends to happen via weak-tie diffusion; because of a preponderance of weak ties, demographically dissimilar speakers are exposed to words that have not yet entered their social circle. Among rural counties, new cultural artifacts spread via strong-tie diffusion; speakers are largely connected to demographically-like individuals via strong ties, and dyads tend to transmit variants that signal an identity that both parties share.

We test whether empirical diffusion pathways form in a way that is consistent with our proposed mechanism. Since we cannot empirically disentangle network from identity, we use our Network-only model to assess whether pairs of counties are connected via a \textit{strong network pathway} (i.e., when the Network-only model pathway strength is high, suggesting diffusion occurs on the basis of network ties) and the Identity-only model to determine whether they are connected via a \textit{strong identity pathway} (i.e., when the Identity-only model pathway strength is high, suggesting diffusion occurs on the basis of shared identity).\footnote{Note that our proposed mechanism is consistent with a purely empirical evaluation: Network characteristics explain a higher fraction of the variation in Twitter's urban-urban pathway strength, while similarity in identity explains more in rural-rural empirical pathways (Figure~\ref{fig-netid}). However, given the complexity of our model, these empirical characteristics likely have a nonlinear relationship with the strength of network- and identity-only pathways, which is what our hypotheses are truly regarding. Since a purely empirical analysis cannot account for the strength of separate network and identity pathways, we test our hypotheses using the results from our simulations.} Our proposed mechanism suggests that innovation may be adopted less selectively in urban areas, where populations are more diverse and more likely connected by weak ties, and that variants that signal common identities may diffuse along strong ties in the more homogenous rural areas. Specifically:
\begin{enumerate}
    \item [H2.1] \textbf{Urban-Urban Pathways}: We hypothesize that transmission between two urban counties\footnote{We use the U.S. Office of Management and Budget's definition of \textit{urban counties}. See Section~\ref{urban-rural-si} for details.} tends to occur via weak tie diffusion---i.e., new words tend to spread between dissimilar network neighbors connected by weak ties \citep{Granovetter1973}. As a result, we would expect an urban-urban pathway to be weaker if the counties were connected by a strong identity pathway  (i.e., diffusion on the basis of shared identity).
    Since weak-tie diffusion describes transmission along the network, we would also expect urban-urban pathways to be stronger when the counties are connected by a strong network pathway. Urban centers are more likely to be connected by edges with lower edge-weight, so a strong network pathway often suggests good diffusion along the network's weak ties. 
    \item [H2.2] \textbf{Rural-Rural Pathways}: We hypothesize that transmission between two rural (i.e., non-urban) counties tends to occur via strong-tie diffusion between network neighbors who share a pertinent identity. Therefore, we would expect rural-rural pathways to be stronger when the counties are connected by a strong identity pathway. Although the network is less important to strong-tie than weak-tie diffusion, strong identity rural-rural pathways should get stronger if counties are connected by a strong network pathway---since diffusion along, likely strong, network ties would make it easier for words to travel between the counties. 
    \item [H2.3] \textbf{Urban-Rural Pathways}: We would expect pathways between an urban and a rural county to fall in between urban-urban and rural-rural pathways---i.e., to rely more on the network than pathways connecting two rural counties and more on identity than urban counties. 
\end{enumerate}

\subsection{Experimental Setup}

In order to test whether network and identity play the hypothesized roles in spatiotemporal diffusion, we run a linear regression to predict the strength of each empirical pathway ($\tau_E$). The independent variables in this regression model are the interactions between the strength of pathways in the Network- and Identity-only models ($\hat(\tau_N)$, $\hat(\tau_I)$) and the type of pathway (urban-urban, rural-rural, or urban-rural); see Table ~\ref{tab-confluence} for details. 

\subsection{Results}
We find that network and identity play complementary roles in urban and rural diffusion. Figure~\ref{fig-emerge} shows the associations between the empirical pathway strength and the Network- and Identity-only strengths ($\hat{\tau_N}$, $\hat{\tau_I}$). As hypothesized: H2.1) empirical urban-urban pathways tend to be stronger when the Network-only pathway is strong (Figure~\ref{fig-emerge}a), and tend to be weaker when the Identity-only pathway is strong (Figure~\ref{fig-emerge}b); moreover, pathways among urban counties become weaker when strong network pathways interact with strong identity pathways (Figure~\ref{fig-emerge}c). H2.2) Empirical rural-rural pathways tend to be stronger when the Identity-only pathway is strong (Figure~\ref{fig-emerge}b), and, to a lesser extent than urban-urban pathways, also tend to be stronger when the Network-only pathway is strong (Figure~\ref{fig-emerge}a); moreover, pathways among rural counties become weaker when strong network pathways interact with strong identity pathways (Figure~\ref{fig-emerge}c). Finally, H2.3) the urban-rural pathways tend to rely more on identity than urban-urban pathways and more on the network than the rural-rural pathways (Figure~\ref{fig-emerge}a-c). 

Although differences in cultural diffusion between urban vs. rural areas have been well-documented \citep{stewart1958urban,fischer1978urban,labov2003pursuing,brunstad2010hip,agergaard2015rural,trudgill18,lengyel2020role}, few prior studies could explain how these differences came to be. We offer a well-reasoned proposal as to how network and identity produce these patterns. The interaction of network, identity, and type of pathway explains a high fraction (almost 70\%) of the variance in empirical pathway strength. Empirical pathways, then, are well-explained by our proposed mechanism, since most of the variance in the strength of pathways can be explained by urban/rural differences in weak- and strong-tie diffusion. 
Furthermore, as shown in Section~\ref{urban-rural-emerge}, urban/rural dynamics are likely explained by distributions of network and identity. The Network+Identity model was able to replicate most associations between network, identity, and pathway strength across the different subsets of counties (Figure~\ref{fig-emerge-si}), so empirical distributions of demographics and network ties likely drive many urban/rural dynamics. However, unlike empirical pathways, the full model's urban-urban pathways tend to be \textit{stronger} in the presence of strong identity pathways since agents select variants on the basis of shared identity. Taken together, we posit that urban-urban weak-tie diffusion may occur because i) urban speakers are less attentive to identity than rural speakers when selecting variants \citep{wirth1938urbanism,glenn1977rural}; ii) there are more low-strength ties among dissimilar individuals than in rural areas (Figure~\ref{fig-size}-\ref{fig-sharedid}); and iii) these weak ties tend to carry words along strong network pathways and weak identity pathways (Figure~\ref{fig-weight-mechanism-si}). Similarly, rural-rural strong-tie diffusion may occur because i) rural speakers are attentive to identity when selecting variants; ii) rural speakers are often connected to demographically similar people; and iii) strong network pathways often correspond to strong identity pathways.

\section{From Mechanisms to Performance}

Finally, we show that the mechanism described in Section~\ref{complementary-roles} influences the model's ability to replicate empirical pathways. Since the network is largely responsible for the dynamics underlying just urban-urban diffusion, and identity for rural-rural diffusion, each variable may individually model different subsets of the U.S.A. Consequently, in order to model diffusion at a national level, we must use both network and identity. 

\subsection{Hypotheses} Since network and identity play complementary roles in spatial diffusion, we expect that urban and rural pathways would be better modeled individually with just one set of variables:

\begin{enumerate}
    \item  [H3.1] \textbf{Urban-Urban Pathways}: We hypothesize that the Network-only model will outperform the other models in urban-urban diffusion. Using the network topology, the Network-only model would likely reproduce the hypothesized diffusion among weak ties in urban-urban pathways; conversely, the Identity-only model will likely perform poorly in urban-urban pathways, amplifying transmission among demographically similar ties, rather than promoting diverse diffusion. 
    \item [H3.2] \textbf{Rural-Rural Pathways}: We hypothesize that the Identity-only model will outperform the other models in rural-rural diffusion. The Identity-only model should correctly reproduce strong-tie diffusion among rural-rural pathways, increasing spread among only counties with relevant shared identities; conversely, the Network-only model will likely underperform, by inflating levels of diffusion among strongly connected individuals who lack a relevant shared identity (e.g., if two strongly-tied speakers share political but not linguistic identity, the identity-only model would differentiate between words signaling politics and language, but the network-only model would not). 
    \item [H3.3] \textbf{Urban-Rural Pathways}: We hypothesize that the Network+Identity model will best predict diffusion between urban and rural areas. Since urban-rural pathways depend less on identity than rural-rural pathways and less on network than urban-urban pathways, a model should need to include both factors in order to predict these pathway strengths.
\end{enumerate}

\subsection{Experimental Setup and Results}

\begin{figure}[t]
    \centering
    
    \includegraphics[width=\textwidth]{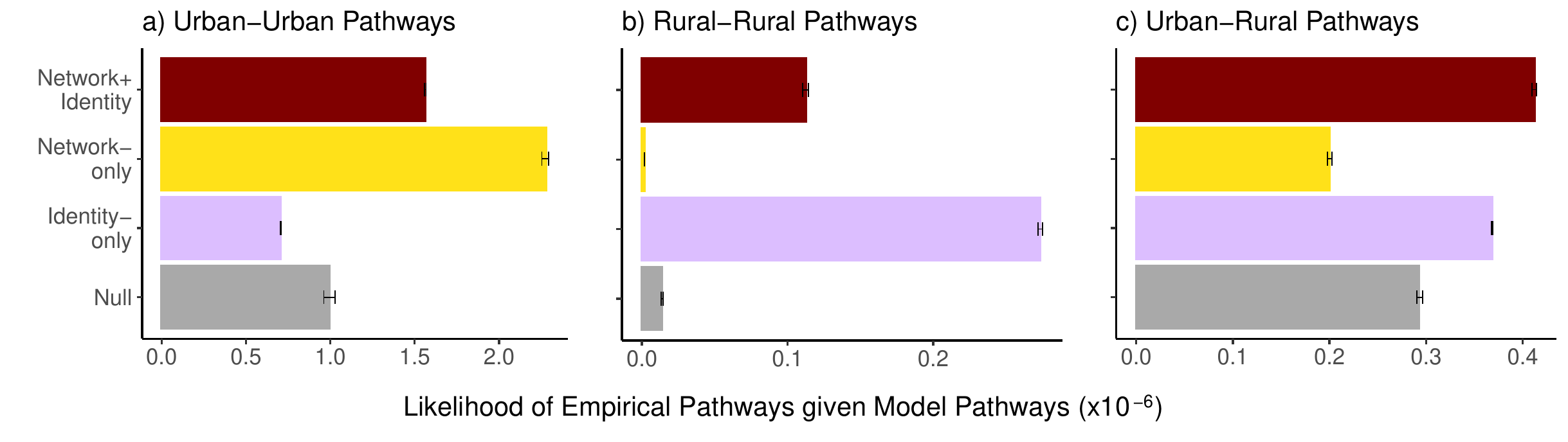} 
    \caption{Based on the likelihood of the pathways observed on Twitter given each of the simulations: a) The Network-only model best matches pathways containing an urban county; b) The Identity-only model best matches pathways among rural counties; and c) the Network+Identity model best matches pathways connecting an urban county to a rural county. Error bars are 95\% bootstrap confidence intervals.}
    \label{fig-pathways}
\end{figure}

We test our hypotheses by evaluating each model's ability to reproduce just urban-urban pathways, just rural-rural pathways, and just urban-rural pathways. As expected, we find that H3.1) the Network-only model best explains the strength of urban-urban pathways (Figure~\ref{fig-pathways}a); H3.2) the Identity-only model most closely approximates empirical rural-rural pathways (Figure~\ref{fig-pathways}b); and H3.3) the strength of urban-rural pathways is best captured by the joint Network+Identity model (Figure~\ref{fig-pathways}c).

Urban-rural diffusion is often described as a key mechanism underlying spatial diffusion at a national scale \citep{stewart1958urban,fischer1978urban,labov2003pursuing,brunstad2010hip,agergaard2015rural,trudgill18,lengyel2020role}, but has proven challenging to reproduce using methodologies that center either network or identity \citep{trudgill18,lengyel2020role}. Cultural scholars often unsuccessfully attempt to extend models that explain diffusion in urban areas or rural areas to the urban-rural case \citep{trudgill18}; although network- or identity-only models may show promising results in one type of geography, we have shown that these same models will not work in all subsets of the U.S.A. Moreover, our findings suggest that transmission between different types of counties may involve both network and identity---for instance, a network- or identity-only model of diffusion may not explain urban-rural diffusion well, because words may travel from urban center to a more sparsely populated area via both weak ties (diverse connections, bridging different geographic regions) \textit{and} strong ties (geographically distal but socially proximal connections, perhaps remnants of migrations or other forms of direct contact \citep{stewart1958urban}). 

Contrary to prior theories \citep{fischer1978urban,labov2003pursuing,gimpel2020urban}, properties like population size and the number of incoming and outgoing ties were insufficient to reproduce urban/rural differences. The Null model, which has the same population and degree distribution, underperformed the Network+Identity model in all types of pathways. However, notably, the Null model predicts urban-urban pathway strengths better than identity and rural-rural pathway strengths better than the network model, underscoring the fact that network and identity facilitate distinct mechanisms of diffusion that are each necessary in different parts of the U.S.A. Importantly, both network and identity are required to explain the adoption of innovation: omitting either one entails not only poorer prediction of spatial properties, but also losing a key determinant of diffusion. 

\section{Discussion}

We demonstrate that many existing models of cultural diffusion are missing a key dynamic in the adoption of innovation: models that consider identity alone ignore weak-tie diffusion between an urban resident and their diverse contacts; while models that use network alone are unable to consider shared identity and, as a result, likely dilute the diffusion of local variants to and from rural areas. One direct consequence, as demonstrated by the simulated counterfactuals in our model, is a loss of accuracy in reproducing spatial distributions and spatiotemporal pathways of diffusion. Moreover, the absence of either network or identity also hamstrings a model's ability to reproduce key macroscopic dynamics like urban-rural diffusion that are likely the product of both strong-tie and weak-tie spread. 

We also propose and test a mechanism through which words diffuse between and among urban or rural areas. Through this framework, we see that the adoption of cultural innovation is the product of complementary, interacting roles of network and identity. These ideas have powerful theoretic implications across disciplines. In the subfield of variationist sociolinguistics, our proposed mechanism for diffusion draws a link between identity- and network-based explanations of language change: showing how strong- and weak-tie theory require information about network and identity to work together. In network theory, this idea suggests how strong ties may influence diffusion when reinforced by node characteristics like identity \citep{milroy1992social}, and integrate Granovetter's theories on tie strength \citep{Granovetter1973} with cultural theory about the role of urban centers and rural peripheries in diffusion \citep{stewart1958urban,labov2003pursuing}. Moreover, in cultural geography, our analysis provides a key contribution to theory: since urban vs. rural differences are emergent properties of our model's minimal assumptions, urban/rural variation may not be the result of the factors to which it is commonly attributed (e.g., population size, edge distribution). Instead, people perform their spatially-correlated identities by choosing among variants that diffuse through homophilous networks; the differences in network topology and demographic distributions in urban and rural populations may also create the observed differences in adoption.

Although our hypotheses were tested on lexical diffusion, the results may apply to the spread of many other cultural innovations as well (e.g., music, beliefs). Linguistic variants often serve as proxies for cultural variables, since their adoption tends to reflect broader societal shifts \citep{labov1963social,kramsch1998language,chambers1998dialectology,labov2012dialect,jackson2012maps,grieve2016regional}. Moreover, the assumptions of our model are sufficiently general to apply to the adoption of most social or cultural artifacts. Importantly, our conclusions about the importance of network and identity, and the mechanisms we have identified for their interaction, may have applicability across a range of social science disciplines---and future work can use the agent-based model, developed in this paper, to test whether these findings generalize to other cultural domains.

In order to make more accurate predictions about how innovation diffuses, we call on researchers across disciplines to incorporate both network and identity in their (conceptual or computational) models of diffusion. Scholars develop and test theory about the ways in which other place-based characteristics (e.g., diffusion into specific cultural regions) emerge from network and identity. Our work offers one methodology, combining agent-based simulations with large-scale social datasets, through which researchers may create a joint network/identity model and use it to test hypotheses about mechanisms underlying cultural diffusion.

\section{Acknowledgements}
We thank Abigail Z. Jacobs, Nick Ellis, and members of University of Michigan's ``Sociolinguistics, Language Contact, and Discourse Analysis'' Seminar for their detailed feedback and useful discussion on an early version of this work, leading to changes in experimental design, analysis, and framing.

\bibliography{main}

\section*{\centering Supporting Information Appendix (SI)}

\appendix
\renewcommand{\thesection}{S\arabic{section}}
\renewcommand{\thesubsection}{S\arabic{section}.\arabic{subsection}}
\renewcommand{\thefigure}{S\arabic{figure}}
\renewcommand{\thetable}{S\arabic{table}}
\renewcommand{\theequation}{S\arabic{equation}}
\setcounter{section}{0}    
\setcounter{figure}{0}    
\setcounter{table}{0}    
\setcounter{equation}{0}    

\section{\label{twitter-si}Empirical Data}

We detail the methodology used to infer agent and network characteristics from Twitter data.

\subsection{Dataset}
We use data from our university's Twitter archives, which includes a sample of tweets between June 2012 and May 2020, to infer characteristics of the agents and network; we also validate our model by comparing its output to spatiotemporal patterns of lexical diffusion in this sample. For most of the timeframe, our archives were sampled via the Twitter Decahose, a 10\% (likely random \citep{pfeffer2018tampering}) sample of all tweets supplied by Twitter. From January 2017 through February 2018, we experienced a service interruption in the Decahose. Instead, from January through September 2017, our archives were sampled from the Twitter Gardenhose, a smaller 1\% sample of all tweets supplied by Twitter; and from October 2017 through January 2018, we did not receive any data from Twitter. 

\subsection{\label{location-si}Agent Location}

Agents in our model correspond to the nearly 4 million Twitter users who used the social networking site from within the United States. We select only Twitter users with five or more GPS-tagged tweets within a 15km radius, so that we have high certainty about their location. We estimate the user's geolocation to be the geometric median of the disclosed coordinates. This procedure uses conservative thresholds for frequency and dispersion, and has been shown to produce highly precise estimates of geolocation \citep{jurgens2013s,compton2014geotagging}. 

Since comparing urban and rural diffusion is a key research question, having accurate location estimates in both types of geographies is particularly important. Using ground truth (i.e., disclosed GPS coordinates) rather than inferred locations ensures that our estimates for location are comparably precise across urban and rural areas. Because of the sparseness of data and structural biases in models, location inference models often assign less precise and less accurate estimates to rural users than urban users \citep{johnson2017effect}. However, a user's disclosed GPS coordinates are unlikely to suffer from these biases. In our case, the inferred location needs only to be in the user's true Census tract. The quality of cellular service in rural areas is high enough to infer the user's Census tract; even in 2012, GPS data has been used in far more precise applications in the rural U.S.A. (cf., \citep{bora2012energy}). Moreover, although data from mobility studies suggest that rural Americans travel nearly 40\% more miles per day on average \citep{pucher2004urban}, rural Census tracts can be up to 10 times larger than urban Census tracts, suggesting that differences in mobility patterns are unlikely to result in higher rates of Census tract misclassification.

\subsection{\label{network-si}Network Topology}

We draw an edge between two agents if they mentioned each other at least once between 2012-2019, resulting in a graph of nearly 4 million nodes (agents) and 30 million edges (dyads). Although Twitter users are exposed to content from more users than they reciprocally mention (e.g., their follower network, public tweets), prior research has shown that the mention network captures edges likely influential in information diffusion \citep{Huberman2008}; furthermore, sets of similar lexical items diffuse along reciprocal ties \citep{romero2013interplay}, allowing us to better examine mechanisms across different types of new words. In this directed graph, the edge drawn from agent $i$ to agent $j$ parametrizes $i$'s influence over $j$'s language style. The tie strength $w_{ij}$ is estimated based on the number of times $j$ mentioned $i$, relative to the maximum number of times $j$ mentioned any of its neighbors (Equation~\ref{eq:weight}). Hence, $w_{ij}$ near 0 means that $j$ weakly weighs input from $i$. 

\subsection{Agent Identity}

We model the identity of each agent $\Upsilon_j$ using sociodemographic characteristics. In light of the debates around the methodological soundness and ethical acceptability of automated demographic recognition tools \citep{buolamwini2018gender,keyes2018misgendering}, we infer identity based on the agent's location. Therefore, we exclude attributes without meaningful spatial autocorrelation, such as gender \citep{eckert1989woman} and age \citep{nguyen2013old,eckert2017age}. We also necessarily exclude variables where such large-scale spatial distributions are unavailable, such as sexuality \citep{bucholtz2004theorizing,queen1997locating,queen2014sexual} and religion \citep{joseph2004language,darquennes2011language}. We select five spatially varied components of sociodemographic identity shown to be important to language style, each encoded as $d_k$-dimensional vectors of probabilities, including:

\textbf{Location}, $d_1=2$: We include the (lat,lon) GPS coordinates each agent tweets from, as linguistic innovations may signal place-based and regional identities \citep{labov1963social,eisenstein2010latent}.

\textbf{Race/Ethnicity}, $d_2=6$: Communities of practice in the U.S.A. often delineate along racial lines, and language style, in turn, often signals racial or ethnic identity \citep{rickford1999african,fought2002chicano,stewart2014now,Jones2015}. Accordingly, we include the percentage of residents in the Census Tract who self-identified their race or ethnicity as: White alone or in combination; Black or African American; Hispanic, Latino, or Spanish origin; American Indian or Alaska Native; Asian; and Native Hawaiian or Other Pacific Islander. We infer the agent's race based on the fraction of the agent's Census Tract identifying with each option in the 2018 American Community Survey; see Table~\ref{acs-vars} for details.

\textbf{Socioeconomic Status}, $d_3=9$: Speakers often signal their socioeconomic status through selection of phonetic and lexical variants \citep{labov1990intersection,milroy1992social,abitbol2018socioeconomic}. Following the Census Bureau's best practices on operationalizing SES \citep{green1970manual}, the percentage of residents in the Census Tract with: income below the Federal Poverty Line; highest educational attainment below high school, high school or GED, Associate's degree, or Bachelor's degree or higher; and employment in the civilian workforce, armed forces, or unemployed. We infer the agent's SES based on the fraction of the agent's Census Tract identifying with each option in the 2018 American Community Survey; see Table~\ref{acs-vars} for details.

\textbf{Languages Spoken}, $d_5=5$: Multilingual speakers often adopt novel lexical items that are borrowed from their non-English native languages and belong to linguistically distinctive speech communities \citep{haugen1950analysis,lo1999codeswitching,stewart2018si}. Thus, we include the percentage of residents in the Census Tract who self-identified as speakers of the five most common U.S.A. languages, apart from English: Spanish, Chinese languages, Tagalog, Vietnamese, and French. We infer the agent's race based on the fraction of the agent's Census Tract identifying with each option in the 2018 American Community Survey; see Table~\ref{acs-vars} for details.

\textbf{Political Affiliation}, $d_4=3$: The desire to signal political identity is hypothesized to have driven several major regional shifts in North American English the past century \citep{labov2012dialect,sylwester2015twitter}. Since the Census Bureau does not track this, we estimate political affiliation using the percentage of voters in the Congressional district who voted for the Democratic,\footnote{The Democratic party goes under three names in the data: ``democrat,'' ``democratic-farmer-labor,'' ``democratic-npl''.} Republican, or third party candidates in the 2018 U.S. House of Representatives Election.\footnote{Data can be found at: \url{https://dataverse.harvard.edu/dataset.xhtml?persistentId=doi:10.7910/DVN/IG0UN2}. We validate our results against news media coverage of the elections.}

Table~\ref{acs-vars} lists the variables we used from the American Community Survey codebook.

\begin{table}
\centering
\caption{ACS variables used for Race/Ethicicty, SES, and Languages Spoken identity categories}
\begin{tabular}{lrrrr}
\textbf{Identity Category} & \textbf{Identity Register} & \textbf{ACS Codebook Name} & \textbf{ACS Population Size} & \\
\midrule
Race/Ethnicity & White* & B02008\_001E & B02001\_001E &  \\
 & Black or African American* & B02009\_001E & B02001\_001E &  \\
 & American Indian or Alaska Native* & B02010\_001E & B02001\_001E &  \\
 & Asian* & B02011\_001E & B02001\_001E &  \\
 & Native Hawaiian or Other Pacific Islander* & B02012\_001E & B02001\_001E &  \\
 & Hispanic or Latino* & B03001\_001E & B02001\_001E &  \\
SES & Below Poverty Line & B17001\_002E & B08122\_001E &  \\
 & On SNAP & B22003\_002E & B22003\_001E &  \\
 & Employed & B23025\_004E & B23025\_001E &  \\
 & Unemployed & B23025\_005E & B23025\_001E &  \\
 & Armed Forces & B23025\_006E & B23025\_001E &  \\
 & Less than High School Education** & B15003\_003E & B15003\_001E & nursery \\
 &  & B15003\_004E & B15003\_001E & kindergarten \\
 &  & B15003\_005E & B15003\_001E & grade 1 \\
 &  & B15003\_006E & B15003\_001E & grade 2 \\
 &  & B15003\_007E & B15003\_001E & grade 3 \\
 &  & B15003\_008E & B15003\_001E & grade 4 \\
 &  & B15003\_009E & B15003\_001E & grade 5 \\
 &  & B15003\_010E & B15003\_001E & grade 6 \\
 &  & B15003\_011E & B15003\_001E & grade 7 \\
 &  & B15003\_012E & B15003\_001E & grade 8 \\
 &  & B15003\_013E & B15003\_001E & grade 9 \\
 &  & B15003\_014E & B15003\_001E & grade 10 \\
 &  & B15003\_015E & B15003\_001E & grade 11 \\
 &  & B15003\_016E & B15003\_001E & grade 12 \\
 & High School Education** & B15003\_017E & B15003\_001E & high school \\
 &  & B15003\_018E & B15003\_001E & GED \\
 &  & B15003\_019E & B15003\_001E & college under 1y \\
 &  & B15003\_020E & B15003\_001E & college over 1y \\
 & Associate's Degree & B15003\_021E & B15003\_001E & \\
 & Bachelor's Degree or Higher** & B15003\_022E & B15003\_001E & bachelors \\
 &  & B15003\_023E & B15003\_001E & masters \\
 &  & B15003\_024E & B15003\_001E & professional \\
 &  & B15003\_025E & B15003\_001E & doctoral \\
Languages Spoken & English & C16001\_002E & C16001\_001E &  \\
 & Spanish & C16001\_003E & C16001\_001E &  \\
 & French & C16001\_006E & C16001\_001E &  \\
 & Chinese & C16001\_021E & C16001\_001E &  \\
 & Vietnamese & C16001\_024E & C16001\_001E &  \\
 & Tagalog & C16001\_027E & C16001\_001E &  \\
\bottomrule
\label{acs-vars}
\end{tabular}
\FloatBarrier

\addtabletext{*Includes anyone who identifies with that race/ethnicity alone or in combination with one or more other races\\}
\addtabletext{**We combined education levels into fewer groupings, in order to keep the model tractable\\}
\vspace{-5.2581pt}
\end{table}

\section{\label{model-si}Model Equations}

We present the full equations and details of the agent-based model presented in the main paper. Table~\ref{parameters-table} summarizes model parameters and how we estimate them.

\subsection{Notation}

Our model simulates the diffusion of a new word $w$ through a network $G$. Agents $j \in V(G)$ either do or do not adopt the word at each iteration $t$ of the model, based on their network neighbors' adoption patterns. We denote the set of agents to adopt the word at iteration $t$ as $adopt(t) \subset V(G)$ and the set of $j$'s network neighbors as $N(j) = \{i | (i,j) \in E(G)\}$. The model begins with a set of initial adopters $A_w = adopt(0)$ introducing the word to the lexicon. 

\subsection{\label{ideq-si}Identity}

An important component of our model is social identity. We model each agent's social identity as consisting of $D$ categories (e.g., race/ethnicity, languages spoken), with $d_k$ registers of the $k^{th}$ category (e.g., if the ``category'' is race, ``registers'' are Black, Latino, white, etc.). Registers of agent identity vary continuously between 0 and 1, where agents closer to 0 affiliate weakly with that identity and agents closer to 1 strongly identify with the register. Thus, each agent has a $d$-dimensional social identity $\Upsilon_j \in [0,1]^d$, where $d = \sum_k d_k$. 

\subsection{\label{wordid-si}Word Identity Equations}

The identities of a word's early users inform which identities the new word connotes $\Upsilon_w \in \{0,1\}^d$. The word signals a particular component if early adopters tend to fall in the demographic's extreme. Specifically, per Equation~\ref{eq:wordid}, for each component of identity, the word signals that component ($\Upsilon_w = 1$) if the median identity among initial adopter falls in a sufficiently high percentile relative to other agents in $G$ (above some threshold $Q \in [0,1]$).

\begin{equation}\label{eq:wordid}
\Upsilon_w = \mathbbm{1}[\textrm{quantile}_{j \in V(G)}({\textrm{ median}_{i \in A_w}({\Upsilon_i}})) > Q]
\end{equation}

In two words $\Upsilon_w=0$ (i.e., all percentiles are below the threshold $Q$), so we find the largest threshold $Q_w$ such that one dimension of $\Upsilon_w$ is non-zero and use that instead of $Q$.

In thinking about the identity signaled by a word, we also consider the relative importance of each category. We model this quantity using the weight vector $v_w \in [0,1]^D$. Per Equation~\ref{eq:omega}, the $k^{th}$ category is weighted as 0 unless one of its registers is equal to 1 in $\Upsilon_w$ (i.e., the race category is weighted 0 unless $w$ signals one of Black, Hispanic, Native American, Asian, or white racial identity). In order to simplify notation in our formula, we define $\Tilde{d}_k = \sum_{l=1}^k d_l$. 

\begin{equation}\label{eq:omega}
\Tilde{v}_{wk} = \mathbbm{1}[]sum(\Upsilon_{w,\Tilde{d}_k:\Tilde{d}_{k+1}}) > 0]
\end{equation}

Since $v_w$ is a weight vector, we normalize it using $v_{wk} = \frac{\Tilde{v}_{wk}}{\sum_l \Tilde{v}_{wl}}$. Note that we often overload $v_w$ in the equation below: in some cases, we treat it as a $D$-dimensional (number of categories) vector and in other cases we treat it as a $d$-dimensional (number of registers) vector. In the latter case, we would assign $v_{wm} = \frac{\Tilde{v}_{wk}}{\sum_l \Tilde{v}_{wl} \cdot d_k}$ for $m \in \{d_k, \dots, d_{k+1}\}$.

\subsection{\label{diffusion-si}Diffusion Equations}

Agents do not use the word until they have been exposed to it by a network neighbor at least once.

If agent $j$ was previously exposed to the word but is not exposed at iteration $t$, their attention to the new word, and their likelihood of adoption, fades \citep{Ellis2019}. Since attention tends to fade exponentially \citep{shalom2019fading}, we assume that agents retain fraction $r \in [0,1]$ of their attention to the word as modeled in Equation~\ref{decayeqsi}:
\begin{equation}\label{decayeqsi}
p_{j,w,t+1} = r \cdot p_{jwt}
\end{equation}

If $j$'s network neighbor $i \in N(j)$ uses the word at iteration $t$, $j$ updates their likelihood of using the word $p_{j,w,t+1}$. Per equation~\ref{modeleqsi}, the updated $p_{j,w,t+1}$ is proportionate to four things: (i) \textbf{Novelty}: a decaying function of the number of exposures $j$ has had to the word $\eta_{jwt}$; (ii) \textbf{Stickiness:} the ``stickiness'' of the word $S_w$; (iii) \textbf{Relevance:} the similarity between $j$'s identity and their understanding of the word's identity, $\delta_{jw}$; and (iv) \textbf{Relatability} and \textbf{Variety}: the fraction of their network neighbors to have adopted the word at iteration $t$, weighted by the similarity in their identity $\delta_{ij}$ and tie strength $w_{ij}$ \citep{Granovetter1973,valente1996social,watts2002simple}. 

\begin{equation}\label{modeleqsi}
p_{j,w,t+1} = \delta_{jw} S_w \eta_{jwt} \frac{\sum\limits_{i \in N(j) \cap adopt(t)} w_{ij} \delta_{ij}}{\sum\limits_{k \in N(j)} w_{kj} \delta_{kj}}
\end{equation}

We define the components of this equation. The first is a normalized cosine decaying function of $j$'s exposure. We denote the number of exposures $j$ has had to the word $w$ at time $t$ by $n_{jwt}$. Per Equation~\ref{eq:novelty}, $\eta_{jwt}$ starts off being equal to 1, when $j$ has never heard the word, and decays to 0 after $j$ has heard the word $\theta$ times. We describe how we picked $\theta$ and why we used a cosine decaying function in Section~\ref{param-est}.

\begin{equation}\label{eq:novelty}
\eta_{jwt} = \frac{1}{2} \cdot [\cos{(\frac{\min{(n_{jwt},\theta)}}{\theta}\pi)} + 1]
\end{equation}

The similarity between $j$'s identity and their understanding of the word's identity is one minus the difference in each component of identity, averaged using weights in $v_w$. Per Equation~\ref{eq:diffwordid}, $\delta_{jw}$ is 1 for the agent whose identity is most similar to the word in the categories of identity that are non-zero in $v_w$; since the distribution of similarity tends to be heavy-tailed (relatively few neighbors are very similar, many are not so similar), we log-scale each 

\begin{equation}\label{eq:diffwordid}
\delta_{jw} = v_w \cdot \frac{\log{(1 - |\Upsilon_{w} - \Upsilon_{j}|)}}{\max_{i \in V(G)}{\log{(1 - |\Upsilon_{w} - \Upsilon_{i}|)}}}
\end{equation}

The similarity between $j$'s identity and the identity of their neighbor $i$ is calculated in much the same way: 

\begin{equation}\label{eq:neighid}
\delta_{ij} = v_w \cdot \frac{\log{(1 - |\Upsilon_{i} - \Upsilon_{j}|)}}{\max_{k \in N(j)}{\log{(1 - |\Upsilon_{k} - \Upsilon_{j}|)}}}
\end{equation}

Finally, we derive empirical estimates of tie strength from dyadic communication frequency on Twitter. Specifically, the tie strength $w_{ij}$ is estimated based on the number of times $j$ mentioned $i$ in a tweet, relative to the maximum number of times $j$ mentioned any of its neighbors. Per Equation~\ref{eq:weight}, $w_{ij}$ near 0 means that $j$ weakly weighs input from $i$. 

\begin{equation} \label{eq:weight}
    w_{ij} = \frac{\textrm{log}(n_{ij})}{\textrm{max}_{k \in N(j)}\textrm{log}(n_{kj})}
\end{equation}

\section{\label{param-est}Parameter Estimation}

Five of our model's parameters---$S_w, Q, r, \eta_{jwt} \in [0,1]$ and $\theta \in \mathbb{Z}$---are not empirically measurable. For example, stickiness is an abstract concept rather than a measurable quantity. Since agents correspond to Twitter users we also cannot conduct any type of cognitive experiments at scale to assess, for instance, at what level of exposure users' attention starts to fade. Instead, we tune these parameters in a way that produces sensible outputs without overfitting. Specifically, each parameter is assigned such that the number of times the word was used in the model most closely matches the number of usages on Twitter; we do not maximize the study outcomes (e.g., Lee's L, likelihood of model pathways) in order to avoid overfitting the model.

The word's novelty$\eta_{jwt}$ is a cosine-decaying function. This function starts at 1 (0 exposures) and decays to 0 at $\theta$ exposures following the first quarter of a cosine curve with period $4\theta$. Other functions (e.g., exponential, quadratic, cubic) decay too quickly, not allowing the word to take off (i.e., the model terminated after 100 iterations with very few usages). Words did not take off because these functions all decay steeply before they decay gradually. The cosine function does the opposite (gradual then steep then gradual decay), which allowed the word to take off and produced the peak+decay pattern found in the empirical word time series (see Figure~\ref{word-ts}). A logistic function may have the same effect, although additional parameters are required in order to govern both the median and the speed of decay (2 parameters instead of 1 with cosine) --- which risks overfitting the model.

We tune global parameters associated with word identity and decision-making so they're the same across all words; specifically we choose the parameter values $Q \in \{0.7, 0.75, 0.8, 0.85,0.9,0.95\}$, $r \in \{0.2, 0.3, \dots, 0.9\}$, and $\theta \in 50, 100, 150, 200$ that minimize mean-squared error in number of uses across a random 20\% sample of words. We select a 20\% sample in order to minimize runtime and avoid overfitting. We find that the optimal values for the parameters are $Q=0.75$, $r=0.4$, and $\theta=100$. 

Since some words are inherently ``stickier'' than others (e.g., they experience higher adoption because they are related to topics of growing importance, used across a variety of semantic contexts, or have notable linguistic properties), the model's stickiness parameter is tuned for each word (i.e., the value is different for each word). We choose the value ($S_w \in \{0.10, 0.11, \dots, 1\}$) that most closely matches the word's number of uses. 

Importantly, the model's results are robust to small changes in most of these parameters. For instance, Lee's L declines only moderately in response to any changes in $\theta$ and $Q$ and small changes in $S_w$. By contrast, the model is highly sensitive to the choice of the rate of decay in attention, $r$. When $r>0.6$, the word never takes off (i.e., model terminated after 100 iterations with very few usages), and when $r<0.4$, the word diffuses far too widely, with the number of empirical uses far exceeding the number of simulated uses. 

\begin{table}
\centering
\caption{\label{parameters-table} The parameters and variables in our model. We indicate if the parameters are "tuned" to optimize the model's performance or provide a summary of how the values of the parameters are estimated from real data.}
\begin{tabular}{p{0.75in} p{0.75in} p{1in} p{3in}}
\textbf{Type of Variable} & \textbf{Parameter} & \textbf{Source, Range} & \textbf{Description} \\
\midrule
Agent $j$ & $\Upsilon_j$ & Twitter, ACS, Election, $[0,1]^d$ & 
The agent's sociodemographic identity, represented as their geolocation, estimated based on the geolocation the corresponding user has tweeted from; the percent of the agent's Census tract who fell into various categories of race, income, education, workforce participation, and languages spoken at home in the Census Bureau's 2018 American Community Survey; and the vote split of the agent's Congressional District in the 2018 House Election. \\ 
Word $w$ & $\Upsilon_w$ & Variable $[0,1]^d$ & 
Agents' shared mental representation of the sociodemographic identity signaled by the word $w$ at iteration $t$, determined based on early adopters (Section~\ref{model-si}). \\ 
Word $w$ & $v_w$ & Twitter $[0,1]^D$, $\sum_i v_{wi} = 1$ & The relative importance of similarity in each dimension of identity to agents' decisions to adopt the word $w$.\\
Agent $j$ & $\delta_{jw}$ & Variable $[0,1]$ & 
The difference in agent $j$'s identity and the identity signaled by the word, normalized to be in $[0,1]$. \\ 
Word $w$ & $S_w$ & Tuned $[0,1]$ & 
The stickiness of the word.\\ 
Global & $Q=0.75$ & Tuned $[0,1]$ & 
The threshold above which agents perceive a word to signal an identity.\\ 
Global & $r=0.4$ & Tuned $[0,1]$ & 
The fraction of the prior input that agents retain at each iteration, as their attention fades.\\ 
Global & $\theta=100$ & Tuned $\mathbb{Z}$ & 
The number of exposures an agent can have to the word before they stop adopting it.\\ 
Global & $\eta_{jwt}$ & Function $[0,1]$ & 
Agent $j$'s salience to the word at iteration $t$; this quantity decays with each subsequent exposure in a predictable way.\\ 
Dyad $(i,j)$ & $w_{ij}$ & Twitter $[0,1]$ & 
The weight of the edge from agent $i$ to agent $j$, estimated based on the number of times $j$ mentioned $i$, normalized to be in $[0,1]$. \\ 
Dyad $(i,j)$ & $\delta_{ij}$ & Twitter $[0,1]$ & 
The difference in agent $i$'s and agent $j$'s identities, normalized to be in $[0,1]$. \\ 
Agent $j$ & $p_{jwt}$ & Variable $[0,1]$ & 
The likelihood with which agent $j$'s  will use word $w$ at iteration $t$, updated at each model iteration (Section~\ref{model-si}). \\ 
\bottomrule
\end{tabular}
\end{table}
\FloatBarrier

\section{\label{new-words-si}Lexical Innovation}

In order to validate our model, we compare simulation outputs against 76 lexical innovations originating on Twitter after 2012; we seed the simulations using each word's initial adopters and compare its steady-state spatial distribution to the empirical one. In this section, we detail our methodology for finding these words and examine their properties.

\subsection{\label{si:identifying-new-words}Identifying New Words}

To identify lexical innovation on Twitter, we start with all 1,168,835 unique alphabetical tokens appearing in the crowdsourced slang catalog UrbanDictionary.com. We parse our Twitter sample to find occurrences of when one of the 4M users whose locations we know (i.e., one of the agents in our model) uses each of the UrbanDictionary.com words; since retweets may include unintentional uses of the word (e.g., a user was retweeting because of something else in the tweet and was not posting the tweet to signal that word's identity), our tally of occurrences excludes retweets---a high-precision filter to avoid diluting the identity signal. 

UrbanDictionary.com is a high-recall, low-precision source of lexical innovation---while many innovative words are added to the site via crowdsourcing, many of the words on the site are also not truly novel words, are infrequently used, or were innovated before 2013. To maintain high precision, we apply seven conservative filters to eliminate several of these lexical items:

\paragraph*{1- Dictionary Words.} Some words on UrbanDictionary.com also appear in, or share orthography with words that appear in, standard English dictionaries (e.g., goat). We eliminate 64,075 standard English terms that appeared in MeriamWebster.com before 2012. Since some words innovated in 2013-2020 may have subsequently been lexicalized in the dictionary (e.g., `amirite' which appeared in 2021,\footnote{\url{https://www.merriam-webster.com/words-at-play/new-words-in-the-dictionary}}) we do not eliminate words that appeared in MeriamWebster.com after 2012.

\paragraph*{2- Named Entities.} Named entities like products, characters, and people are often names assigned to entities that users on Twitter have to use if they want to refer to those entities; therefore, named entities are not truly lexical innovation. We eliminated 313,967 named entities and common phrases (e.g., iphone, lebron) cross-referenced from English-language WikiData, a crowd-sourced knowledge base, which includes many culturally significant or widely adopted named entities.

\paragraph*{3- Infrequent Words.} If new words are not used often enough, they may not have taken off or we may not get a clear enough signal about their adopters' spatial distribution. We eliminated 784,397 words that were not adopted at least 1,000 in our post-2012 sample (e.g., boofed). Words with fewer than 1,000 uses were removed from the analysis, because spatial distributions with such few observations were observed to be of poor quality.

\paragraph*{4- Words Commonly Used Before 2013.} In order to study the diffusion of lexical innovation, words in the study need to have been gained popularity after 2013. We eliminated 5,429 words used too often (i.e., over 1,000 times) before 2013 (e.g., awko). In order to keep consistency with filter 3 above, words should not be more than 1,000 times before 2013 to avoid having more uses of the word before 2013 than after.

\paragraph*{5- Words Innovated Before 2013.} Since the model is seeded with the word's first ten adopters, these initial adopters need to have used the word for the first time during or after 2013.  We eliminated 653 words whose first ten adopters (Section~\ref{init-adopters-si}) used the word before 2013. 

\paragraph*{6- Manual Filtration.} We manually inspect and filter the remaining words. One of the authors learned each word's meaning by conducting searches on several sites (UrbanDictionary.com, Wiktionary.com, KnowYourMeme.com, Fandom.com, Google.com). This author manually filtered the remaining 314 words, removing several named entities, common phrases, and typos or syntactic variants of dictionary words (e.g., affleck's, pikachus, beyhive, avacado, bathsalts, basicness). 

\paragraph*{7- Manual Combining.} Finally, a number of words were minor orthographic and syntactic variants of other words on the list. At times, small spelling or part-of-speech changes can be a normal part of the word's usage and incidental to the word's innovation \citep{grieve2019mapping} (e.g., a new verb will likely have present, present progressive, and past tenses all represented in the lexicon)---in which case, not combining the variants in the final list of words can artificially inflate the contribution those words have to the model's performance. And at other times, these small spelling or part-of-speech changes can, themselves, be further innovations designed to signal unique identities (e.g., the same word in a phonetically marked spelling can signal the linguistic identity associated with that phonology)---in which case, combining the variants in the final list of words can create a set of initial adopters and a spatial distribution that does not reflect the word's true origins and identity, and skewing the ``ground truth'' data in the model's evaluations. To summarize: always combining variants will invariably lead to false positives, while never combining variants will cause false negatives---both of which have negative consequences for downstream model evaluation tasks. In order to avoid both types of error, we combine only variants with very similar ($L>0.4$) spatial distributions (e.g., ``bonged'' and ``bonging,'' ``sksksk'' and ``sksksksk'') and do not combine others (e.g., ``schmood'' and ``shmood'', ``fleeked'' and ``fleeky''), treating similarity in spatial distribution as a high-precision indicator that the variants do, in fact, diffuse via the same mechanism (e.g., they signal the same identities or diffuse through the same parts of the network). 

These seven rounds of elimination left 76 unique lexical innovations coined on Twitter between 2013 and 2020 (Table~\ref{words-table}). These words range from terms describing cultural phenomena (e.g., meninist, ratioed, sksksk) to, often phonologically-motivated, spelling shifts (e.g., respeck, wottice, wypipo), part-of-speech changes (e.g., goated, ubering), shortenings (e.g., sjws, boffum), concatenations (e.g., cutecumber, situationship), and even new coinages (e.g., fleeky, yeeted). We publish the list of words and their spatial time series (see Datasets and Table~\ref{words-table}).

\subsection{\label{init-adopters-si}Initial Adopters}

A word's initial adopters are not necessarily the first users to adopt the word; e.g., in some cases, a word was used once and then not used again for several months or years, at which point it gained in popularity. In order to avoid such cases, initial adopters are defined as the first ten observed users of the word, after which the word was tweeted consistently until its peak adoption (`consistently' means we find at least one instance of the word in our sample each month). We set the threshold of 1+ uses per month, because users’ attention to words is likely to fade in 24-48 hours \citep{vilares2015sentiment}: simulating arrivals of tweets containing the word using a Poisson process, a 10\% sample of tweets tends to contain no instances of the word when arrival times are 24 to 48 hours (76\% - 86\% of the time), while it is likely to contain 1+ occurrences for much smaller arrival times (44\% at 12-hour and 67\% at 6-hour arrival time). 

Importantly, our model results are not sensitive to the first ten adopters that happen to be observed in our 10\% tweet sample; results do not significantly change if we perturb the users with whom we seed the model (Section~\ref{si-sensitivity-fb}).

\subsection{Word Identity}

\begin{figure}
    \includegraphics[width=\textwidth]{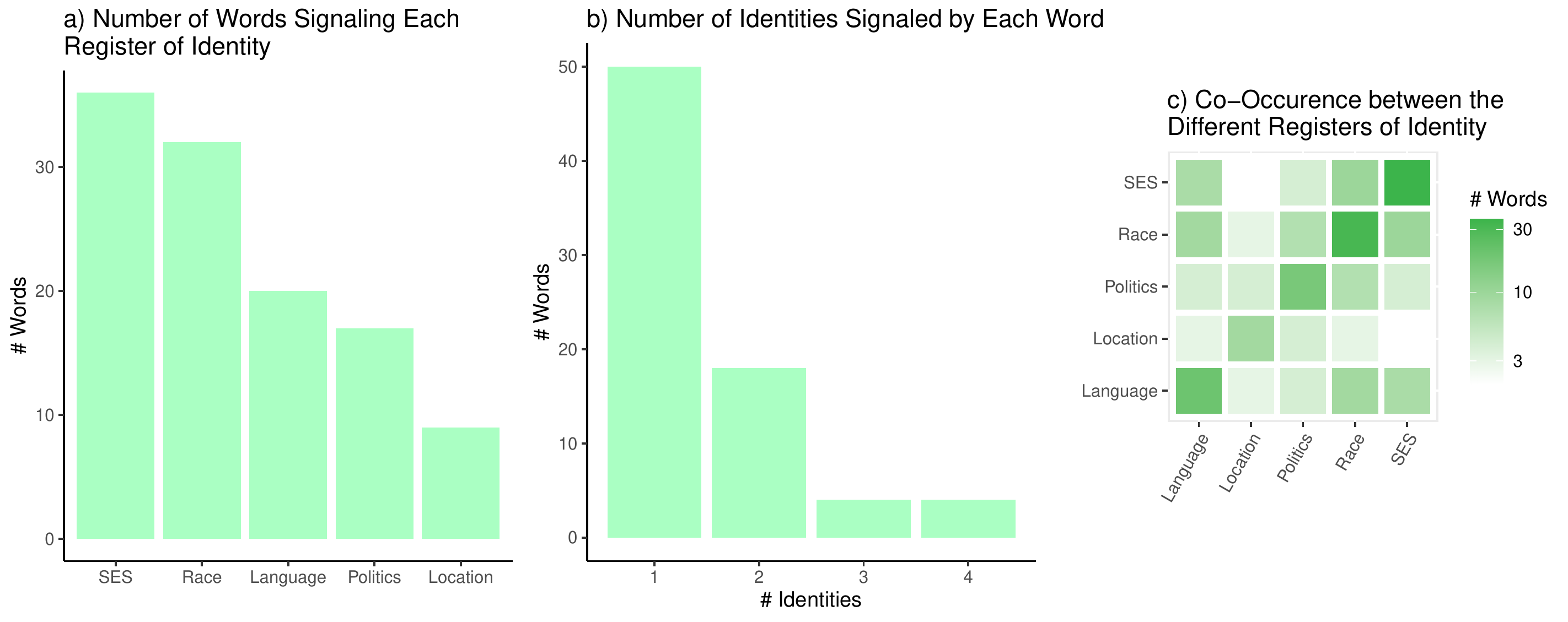}
    \caption{The 76 new words in our study signal a wide variety of identities: a) The number of words signaling each identity; b) The number of words signaling each pair of identities.}
    \label{words-summary}
\vspace{-12.50285pt}
\end{figure}

Applying the procedure described in Section~\ref{model-si}, we estimate which identities each word signals. We offer a sample of words with the identities signaled in Table~\ref{words-table}. In Figure~\ref{words-summary}, we describe how often the words tend to signal each category of identity (e.g., race, SES). Nearly two-thirds of the words signal just one category of identity (50 out of 76), while 18 signal two, 4 signal three, and another 4 signal four. For each category, we also show which other category it most often co-occurs with; unsurprisingly, we find that most categories co-occur with race most often.

\begin{table}\centering
\caption{A subset of the lexical innovations included in our study, including their definition and the identities they signal based on the initial adopters.}
\begin{tabular}{lrr}
\textbf{Word} & \textbf{UrbanDictionary.com's Definition} & \textbf{Word Identity} \\
\midrule
birbs & any bird that's “being funny, cute, or silly in some way." & SES \\
boffum & "both of them" & Race \\
caucasity & refers to the audacity of white people or things only white people are brave enough to do & SES, Language \\
challs & short for challenges & SES \\
chillay & combination of chillax and stay & Politics \\
cringiest & superlative form of cringy (characterized by causing feelings of embarrassment) & Race, SES, Language \\
cucked & a man who lets his wife or girlfriend treat him poorly (short version of cuckold) & SES \\
cutecumber & What you call a person when they're extremely adorable & Race \\
daddyish & characteristic of a father; paternal & Race \\
deerbra & fictional animal from a 2005 children's book; also featured in a 2013 Kevin Hart skit & Race \\
degular & the phrase "regular degular" means not worth remembering, not particularly noticeable & Politics \\
democrap & a pejorative name for Democrat, conveys distate for the U.S.A. Democratic party & Race, Language \\
dfwt & "Don't Fuck With That" is a phrase and a title of a 2015 song & Location, Race, Language, Politics \\
doggos & affectionate terms for dog used in the internet slang called DoggoLingo & SES \\
dumbassery & immature, foolish behavior; behavior typical of a dumbass & SES \\
earthers & the phrase "flat earthers" refers to someone who eschews science & SES \\
earthporn & photos of the luscious landscape/scenery of our mother nature. & Language \\
fineapple & "if you were a fruit you'd be a fineapple" is an old corny pickup line & Politics \\
fleeked & fixed to be cooler or better, fleeked out or fleeked up are common usages & Race \\
floof & ridiculously fluffy (often used to describe animals) & SES \\
gloing & the phrase "gloing up" mean growing up and becoming very attractive physically & Race \\
gmsfu & "get me so fucked up," used in reference to feelings about a situation & Location, Race, Language, Politics \\
goated & someone's a G.O.A.T, which means Greatest Of All Time & Location \\
gratata & machine gun sound he made in a rap video & Race, SES \\
grwm & "Get Ready With Me," a vlog filming one's daily routine. & Race \\
headassery & The act of doing stupid things & Race, SES \\
hewwo & saying "hello" in a cute way & Language \\
ikyfl & "I know you fuckin lying" & Location, Politics \\
iykyk & "if you know, you know" & Race, Politics \\
lewds & lewd photos & Race, Language \\
lituation & lit (cool) situation & Location, SES \\
meninist & members of the men's rights movement & SES, Language \\
nawfr & "No, For Real" used to clarify a statement that may seem untrue at first glance & Location, Race, SES, Politics \\
numbnuts & an incredibly stupid person & SES \\
periscoping & to post livestream videos on the Periscope platform & Race, SES, Language, Politics \\
pussification & the state in which a society becomes less and less tough & Race \\
qwhite & something that only a white person would do. & Race, SES, Language \\
ratioed & having significantly more replies than retweets or likes on Twitter, indicating public dislike & SES \\
respeck & alternative spelling of respect to add emphasis & SES \\
schmood & an exaggerated mood & Politics \\
shitbird & a person who regularly gets into trouble; an objectionable person & SES \\
shitstain & a stupid or contemptible person. & Language, Politics \\
shmood & a more exaggerated “mood” & Location, Language \\
situationship & a relationship in which the parties involved do not clearly define their relationship & Race \\
sjws & social justice warriors; used as an insult towards people of left-leaning opinions & SES \\
sksksk & VSCO Girl popular phrase, laughter & Race, SES \\
snowpocalypse & snow+apocalypse; An unusually severe blizzard or series of blizzards & SES, Politics \\
spoilery & involving or relating to spoilers (disclosures of a story's twists or ending) & Race, SES, Language \\
stummy & the combination of "Stomach" and "Tummy" & SES \\
thupid & stupid, to make fun of the person & Language \\
tiktoks & a video on TikTok & Politics \\
twatwaffle & an incompetent, contemptible person & SES, Language \\
ubered & means "took an uber ride [to get to a place]" & SES \\
udderly & an alternative form of "utterly," used when discussing cows or breastfeeding & SES \\
udigg & alternative spelling of "you dig?" meaning "Do you understand what I mean?" & SES \\
unstanning & stop STalking+fAN someone or something & Language \\
vibey & giving off good vibes & SES \\
wdym & short for "What do you mean?" & SES, Language \\
whomst & nonstandard form of who or whom, used humorously & SES \\
wokeness & awareness of issues concerning social and racial justice & Language \\
wottice & what is, written as British people pronounce it, used by Americans in mock formality & Race \\
wypipo & phonetic spelling of "white people" & SES \\
yeeting & making a violent motion of any variety or an exclamation of delight & Race \\
\bottomrule
\end{tabular}
\label{words-table}
\vspace{-16.60674pt}
\hspace{-16.55983pt}
\end{table}
\FloatBarrier

\FloatBarrier

\subsection{Empirical Regions}

\begin{figure}
    \includegraphics[width=\textwidth]{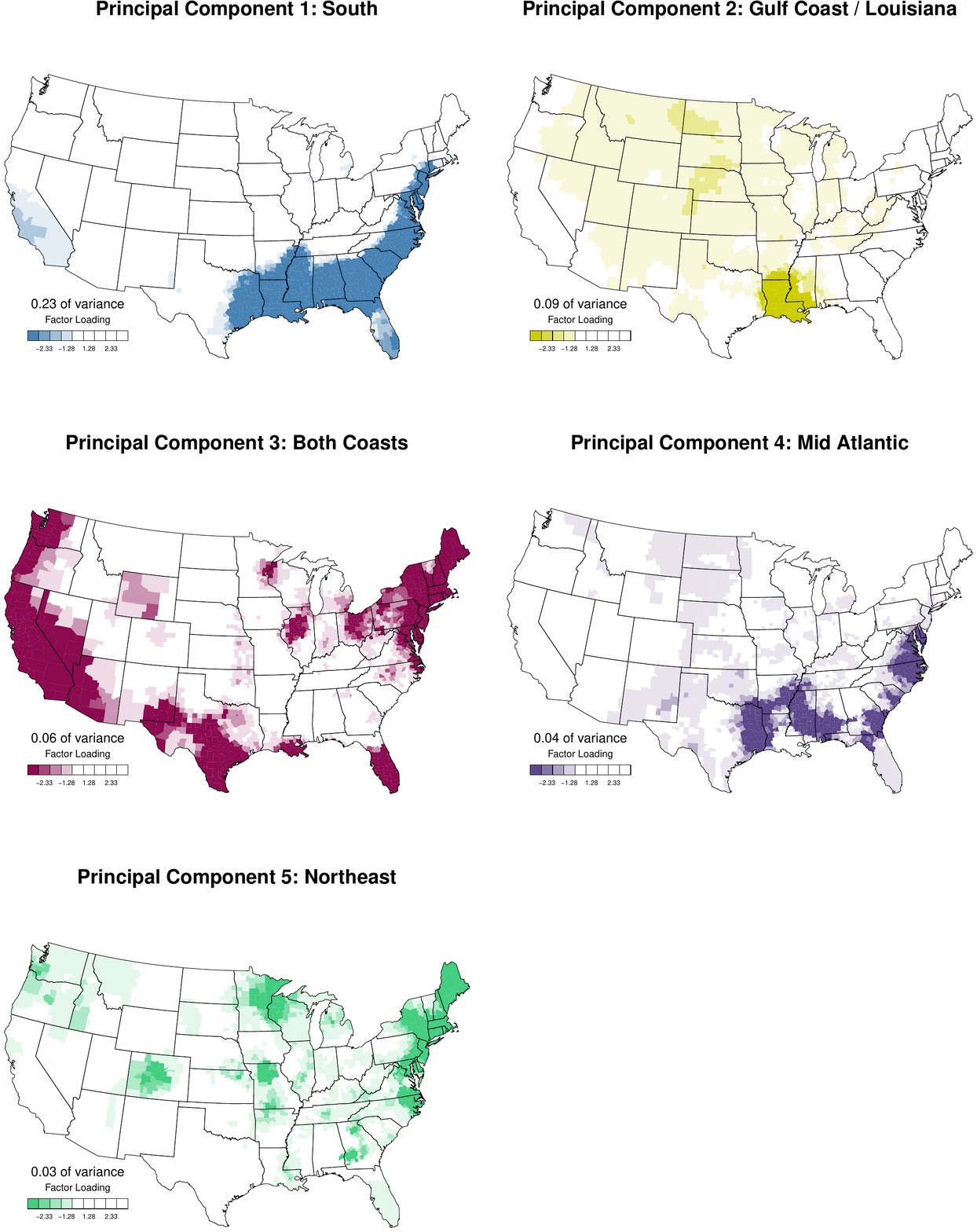}
    \caption{Top 5 dialect regions (principal components) corresponding to the 76 lexical innovations on Twitter. Our dialect regions match well with dialect regions found in other studies of linguistic variation (Figures~\ref{grieve} and ~\ref{labov}).}
    \label{top_pcs}
\vspace{-12.50285pt}
\end{figure}

Linguistic regions --- or the geographic areas in which linguistic variation tends to diffuse --- often carry cultural significance \citep{grieve2016regional}, including in their alignment with demographic distributions and historical events like major migrations. Linguistic regions are well-documented, including in the Phonological Atlas of North American English \citep{labov2008atlas} and the Dictionary of American Regional English (DARE) \citep{cassidy1985dictionary}. Following Grieve et al. 2017 \citep{grieve2017analyzing}, we quantify the key regions present in our data by computing the principal components of the 76 empirical distributions (Figure~\ref{top_pcs}). 

Our aggregate spatial distributions reflect key historical and cultural phenomena, many of which are known to play a role in linguistic diffusion. For instance, two among the top five principal components relate to major migration pathways: The first component (Figure~\ref{top_pcs}a) lines up with pathways of the early 1900s’ Great Migrations, when African American residents of the US South moved to the Northeast in pursuit of economic opportunity; these migrations sparked major African American linguistic and cultural shifts, and prior studies have shown that African American language terms on social media often diffuse through this region \citep{Jones2015}. Additionally, waves of immigrants from Scotland and Ireland in the 18th century settled in parts of Appalachia, corresponding roughly to the region in component four (Figure~\ref{top_pcs}d). This migration brought many Europeans seeking religious and political freedom, creating a distinctive culture that persists today \citep{woodard2011american}. Even decades afterward, linguistic diffusion is likely to happen along important historical migration pathways partly because of its lasting effects on cultural similarity and network topology \citep{kerswill2006migration}.

The other three components correspond to parts of the US with histories of language contact and or prevalent multilingualism, both known drivers of linguistic innovation \citep{vilares2015sentiment}. The second principal component (Figure~\ref{top_pcs}b) differentiates Louisiana’s dialect from the rest of the South. This region is home to speakers of Louisiana French, a creole with origins in the late 1600s, combining French spoken by European settlers, Algonquin spoken by Canadian migrants, and Mande spoken by their slaves from Senegambia \citep{valdman1997french}. The third component describes diffusion in the coasts, which are politically and racially distinctive from the middle of the country (Figure~\ref{top_pcs}c). The fifth component (Figure~\ref{top_pcs}e) delineates the U.S.A.’s Spanish-speaking population on the West Coast and in Florida, from the linguistically rich New England; notably, some of the words that diffuse in the west are Spanish-language borrowings (e.g., chunty, cayute) while those diffusing in the Northeast are phonologically-motivated orthographies corresponding to New England accents (e.g., cawfee).

\begin{figure}
    \centering
    \includegraphics[width=\textwidth]{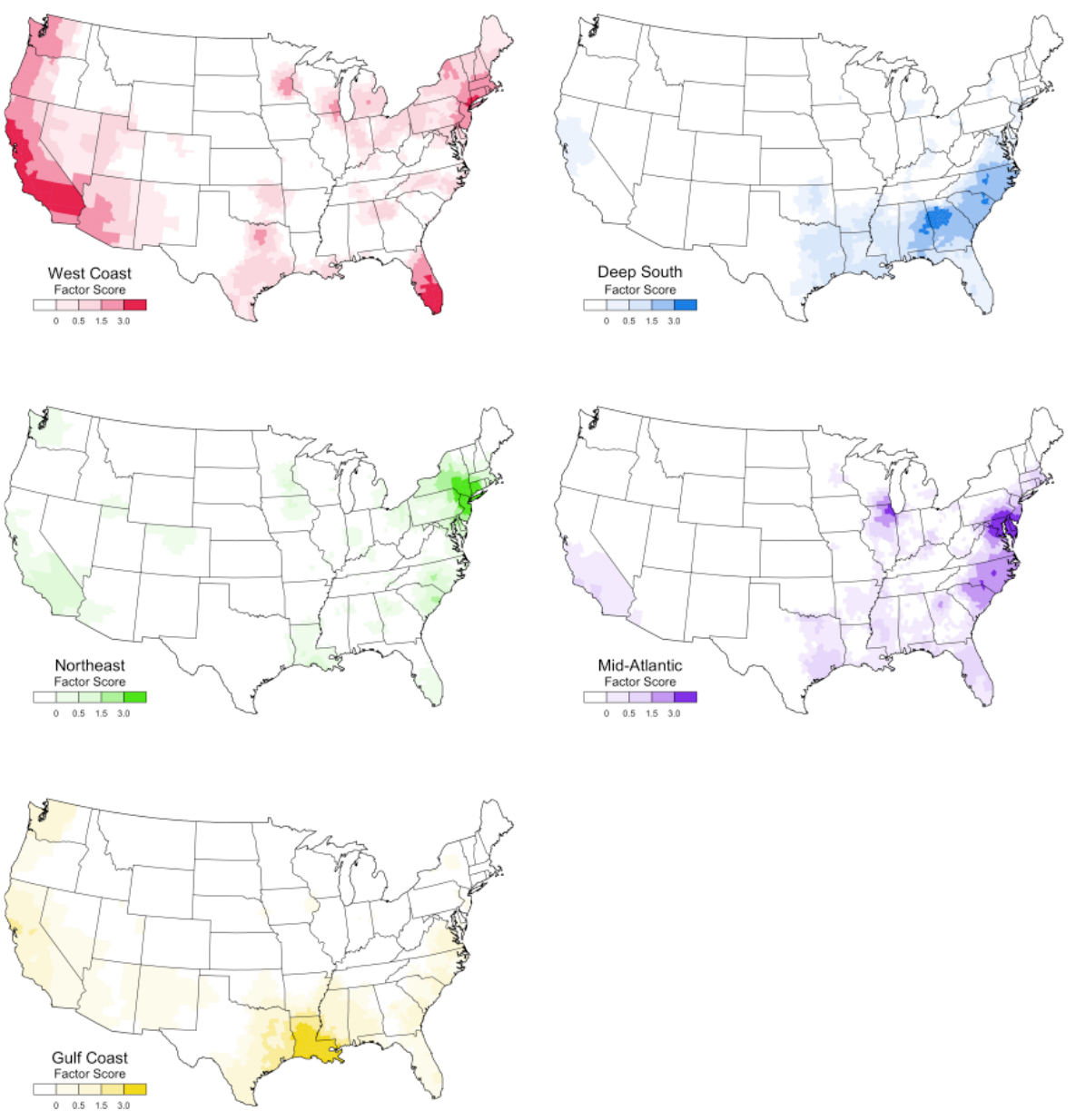}
    \caption{Dialect regions from Grieve et al. (2019)'s study of lexical innovation on Twitter \citep{grieve2019mapping} are highly similar to the regions identified in our data: a) West Coast matches Figure~\ref{top_pcs}c; b) Deep South matches Figure~\ref{top_pcs}a; c) Northeast matches Figure~\ref{top_pcs}e; d) Mid-Atlantic matches Figure~\ref{top_pcs}d; and e) Gulf Coast matches Figure~\ref{top_pcs}b.}
    \label{grieve}
\end{figure}

\begin{figure}
    \centering
    \includegraphics[width=\textwidth]{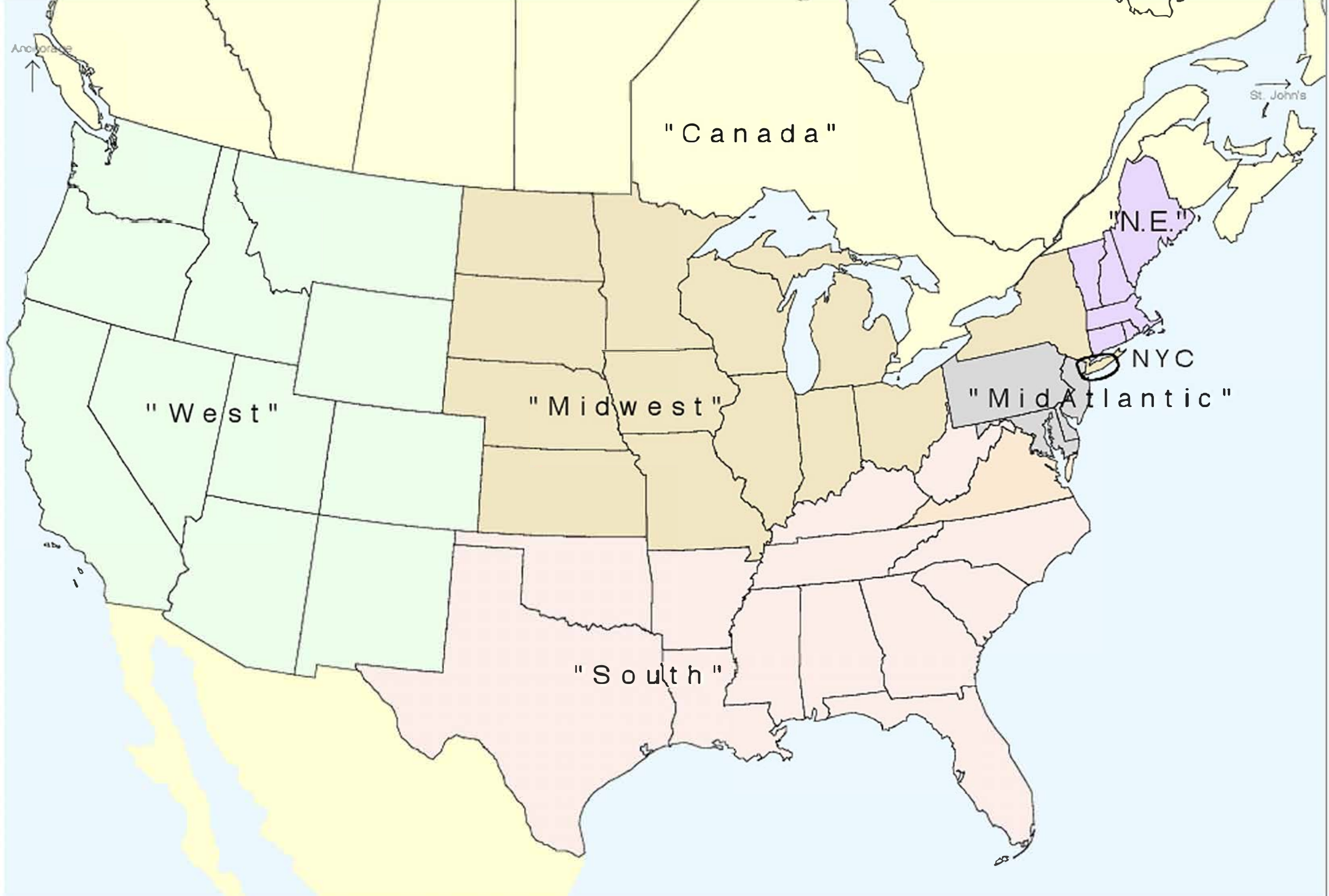}
    \caption{Dialect regions from Labov et al. (2008)'s study of phonological variation \citep{labov2008atlas} compare well with our regions. For instance, the South matches Figure~\ref{top_pcs}a, West matches Figure~\ref{top_pcs}c, Mid-Atlantic matches Figure~\ref{top_pcs}d}, and N.E. matches Figure~\ref{top_pcs}e.
    \label{labov}
\end{figure}

The spatial distributions in our 76-word sample also match linguistic regions identified in other studies \citep{eisenstein2012mapping,Jones2015,grieve2016regional}; our 76 words overlap with words used in other studies (e.g., boffum, fleeky, gmsfu), and have similar commonly appearing regions (Figure~\ref{top_pcs} vs. Figures~\ref{grieve} and~\ref{labov}). The alignment between our study and prior studies confirms the robustness and potential generalizability of our methods. Even using just 76 words (which is a relatively small set), we are still able to capture key geographic properties that appear in larger samples. That said, while many of the regions in Labov et al. (2008)'s Phonological Atlas are represented in our sample \citep{labov2008atlas}, some key linguistic regions are notably missing from ours and other studies of Twitter: 1) the Inland North, comprising parts of Western New England and the North Eastern Midwest; 2) the Midlands, spanning states along the Southern Midwest; and 3) the North Central, consisting of the upper Midwest (Figure~\ref{labov}).

\subsection{Empirical Time Series}

\begin{figure}
\centering
\includegraphics[width=\textwidth]{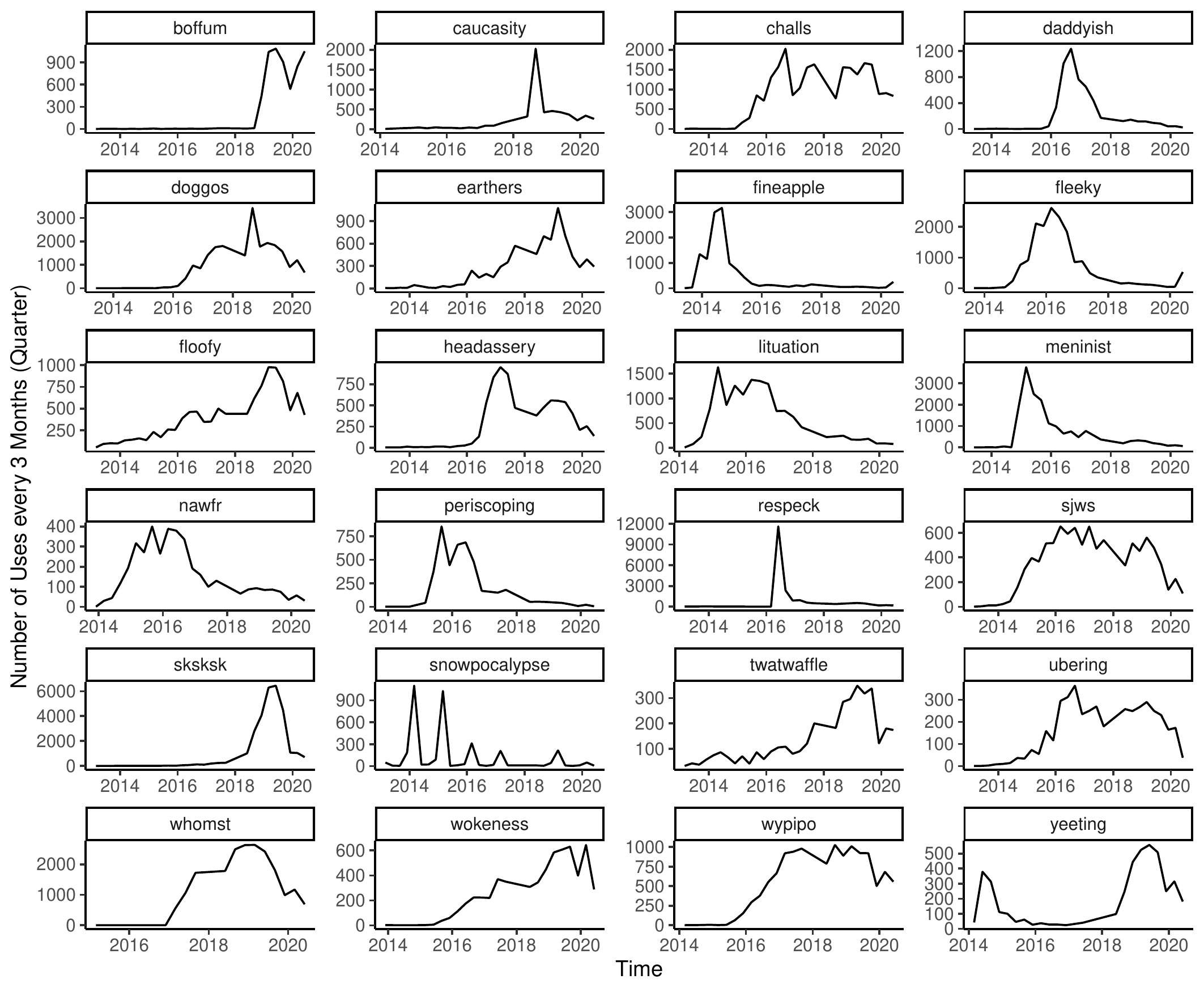}
\caption{Quarterly (i.e., 3-month) time series for new words in our study. These 24 words display common trends among the 76 words. Notably, most words are ``fads'' --- their adoption peaks and then decays to low usage. A few words are seasonal (e.g., snowpocalypse), and a handful have consistent usage after they are coined (e.g., ubered, challs, wypipo).}
\label{word-ts}
\end{figure}

We examine trends in the usage over time of the 76 words (Figure~\ref{word-ts}). Except for a few notable examples (e.g., ``ubering'' or ``wokeness''), most words do not appear to persist in the lexicon. Instead, their usage peaks and then decays, a pattern that is more associated with ``fads'' than with lasting cultural change.

\FloatBarrier

\FloatBarrier

\FloatBarrier

\FloatBarrier

\section{\label{eval-si}Evaluation}

\subsection{Data Preparation}

We create county-level spatial distributions of word adoption, using a procedure similar to Grieve et al. \citep{grieve2016regional,grieve2017analyzing,grieve2019mapping}. We (i) count the number of times a word was used in each county from the word's coinage to the end of our sample (empirical) or during the simulation, (ii) normalize the word's frequency of use to the number of agents (Twitter users) in the county and (iii) attenuate noise in our spatial distributions by smoothing the county-level maps using local Getis-Ord $G^*$ with a neighborhood size of 25 \citep{getis1992analysis}. The resulting spatial distribution 

We also create county-level spatial time series distributions of word adoptions---i.e., the number of times a word was used in each county at each time step, normalized to the number of users in the county and the number of uses at time $t$. The normalized adoption in county $i$, at time $t$, for word $w$ is modeled as $a_{i,t,w} = \frac{n_{i,t,w}}{n_{tw} n_i}$. Each time step $t$ represents a block of 10 iterations for simulated data (arbitrarily selected) and 3 months in the empirical data. Using this spatial time series, we are able to construct spatiotemporal pathways among pairs of counties $(i,j)$ -- connoting the strength of transmission from county $i$ to county $j$. We measure the strength of the pathway from county $i$ to county $j$ as the correlation between $i$'s adoption at time $t$ and $j$'s adoption at $t+1$. Since we are interested in the pathway strength between two counties, we concatenate the time series across all words in order to do this computation, correlating $[a_{i,1:T_1-1,1},a_{i,1:T_2-1,2},\dots,a_{i,1:T_W-1,W}]$ with $[a_{i,2:T_1,1},a_{i,2:T_2,2},\dots,a_{i,2:T_W,W}]$. Since $a_{i,t,w}$ is often 0, we use a version of Kendall's tau that accounts for zero-inflation $\hat{\tau}_{(i,j)}$. Following Pimentel et al. (2009)'s methodology \citep{pimentel2009kendall}, this metric is a weighted average of two quantities: 1) the strength of association in adoption (did $j$ frequently adopt the word at time $t+1$ iff county $i$ adopted at $t$?) and 2) the correlation in level of adoption, if both adopted (was $j$ level of adoption high at time $t+1$ iff county $i$'s was high at time $t$?).

\subsection{\label{lees-l-si}Comparing Spatial Distributions via Lee's L}

We use Lee's L \citep{lee2001developing} in order to assess how well our simulations approximate empirical spatial distributions. Lee's L is an extension of Pearson's $R$ correlation that controls for the effects of spatial autocorrelation using a Moran's I-like adjustment. If the Lee's L correlation between the empirical and simulated geographic distributions of a new word is $L>0.4$, the model's output is ``very similar'' to the empirical map; and if $L>0.13$, the model's output is ``broadly similar'' to the empirical map.

The thresholds for ``very'' and ``broadly'' similar distributions are based on results from Grieve et al.'s (2019) empirical evaluation of Lee's L index \cite{grieve2019mapping}, which compared the usage of 139 common words in spoken language and on Twitter. In a deeper evaluation of the metric, the authors concluded that spatial distributions where $L >= 0.4$ were ``very good matches,'' while spatial distributions with $L >= 0.13$ had ``broad alignment \dots [with] considerably more local variation.'' Grieve et al. (2019) used the fact that their median correlation was $L = 0.14$ and a handful of high Lee's L values to conclude that offline and online speech have similar enough spatial distributions that the former can be used to reason about the latter \cite{grieve2019mapping}. Other papers using Lee's L also tend to report strong associations when $L$ is in the $[0.15, 0.3]$ range \citep{heywood2008investigations,leepaper1,kim2018seoul,kim2018closer}. Some of these papers use a Mantel-like simulation test to reject the null hypothesis $L<=0$ \citep{lee2004generalized}. Since we are interested in the strength of association rather than significance, we opt not to use $p$ values: for instance, even very low values of $L$ that authors consider to be weak or non-correlations correspond to very small $p$ values. 

\begin{figure}
    \centering
    \includegraphics[width=6.25 in]{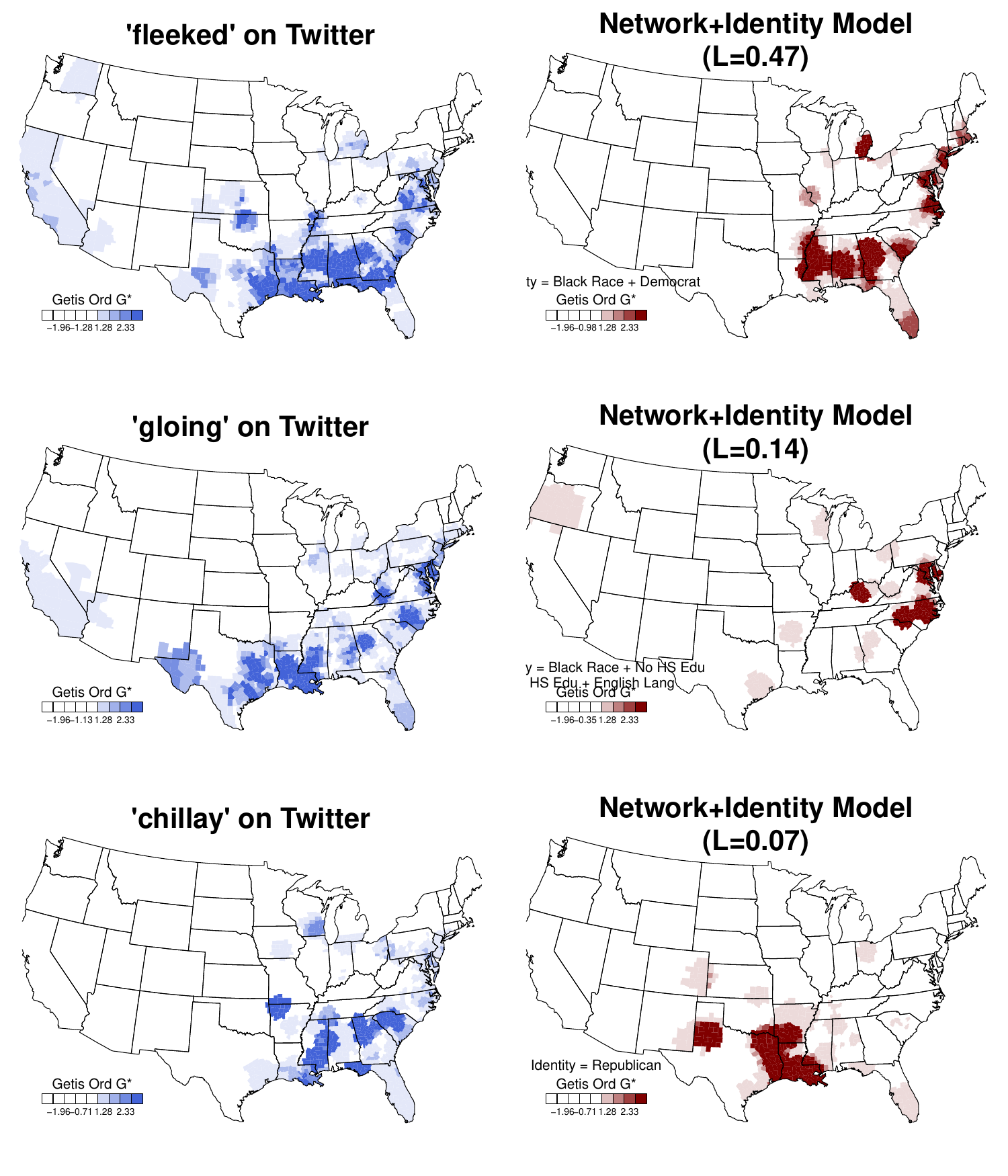}
    \caption{Visual intuition showing that Grieve et al. (2019)'s thresholds for ``very similar'' and ``broadly similar'' spatial distributions \citep{grieve2019mapping} apply when comparing empirical and simulated adoption in our model. The blue maps on the left are adoption on Twitter, the maroon maps on the right are one run of the Network+Identity model, and orange dots are the locations of the word's first ten adopters (Section~\ref{si:identifying-new-words}). In the examples pictured, a) both maps show dense adoption in the South and up the East Coast, and $L>0.4$ (``very similar'' maps per our thresholds); b) both maps have heavy adoption in the mid-Atlantic, but the map on the left also has adoption in the South, for which L>0.13 (``broadly similar'' per our thresholds); and c) the blue map shows adoption east of Mississippi in the South, while the red map has adoption in Louisiana and Texas, leading to a Lee's $L<0.13$ (``not similar'' per our thresholds).}
    \label{lees-l-eval}
\end{figure}

In our own visual review of our data, we find that Grieve et al. (2019)'s $L>0.4$ and $L>0.13$ thresholds frequently match our own intuition about how well simulated and empirical spatial distributions are matched. For instance, when $L>0.4$, the simulated and empirical adoptions often concentrate in very similar areas (e.g., both maps in Figure~\ref{lees-l-eval}a show dense adoption in the South and up the East Coast); when $L>0.13$, the simulated and empirical maps often have overlapping but still different areas of adoption (e.g., both maps in Figure~\ref{lees-l-eval}b have heavy adoption in the mid-Atlantic, but the map on the left also has adoption in the South); and when $L<0.13$, the two maps have minimal to no overlap (e.g., in Figure~\ref{lees-l-eval}c, the blue map shows adoption east of Mississippi in the South, while the red map has adoption in Louisiana and Texas).

\FloatBarrier

\subsection{\label{pathway-strength-si}Comparing Pathway Strengths}

In addition to assessing whether the models can reproduce empirical spatial distributions, we also evaluate whether the models can reproduce spread between pairs of counties or the spatiotemporal dynamics of diffusion. Let $(i,j)$ be the pathway from county $i$ to county $j$, and take any set of county-county pathways $S$. Let $M$ be any of the four models we developed: Network+Identity, Network-only, Identity-only, or Null. $M$'s ability to reproduce empirical pathways can be measured via $L_M(S)$, the likelihood of the empirical pathway strengths in $S$ given $M$'s simulated pathways strengths.\footnote{We use likelihoods instead of standard correlations, because of the sparsity of our empirical data: Since we estimate pathway strengths using a 10\% sample of tweets, diffusion between smaller counties (which consistently have fewer adoptions) often goes un- or under-detected.} In other words, $L_M(S)$

In order to calculate $L_M(S)$, we make four assumptions that are common in Bayesian modeling: 
\begin{enumerate}
\setlength{\itemsep}{0pt}
\setlength{\parskip}{0pt}
    \item $L_M(S)$ is normalized to the number of paths included in the sample, by taking the $|S|^{th}$ root of the probability; this ensures that the likelihood does not continue to shrink as $|S|$ gets large
    \item Each pathway in $S$ is selected randomly from the set of all pathways, and independently from the other pathways in $S$
    \item The probability any single empirical pathway $(i,j)$ is sampled into $S$ is proportional to its strength 
    $$P((i,j)) = \frac{\hat{\tau}_E^{(i,j)}}{\sum \hat{\tau}_E^{(k,l)}}$$
    \item The likelihood of any single empirical pathway $(i,j)$ given model $M$ is proportional to its strength 
    $$P((i,j) | M) = \frac{\hat{\tau}_M^{(i,j)}}{\sum \hat{\tau}_M^{(k,l)}}$$
\end{enumerate}

In the large $|S|$ limit, where the sampling variation becomes negligible, the likelihood of empirical pathways given model $M$'s pathways is:

\begin{equation} \label{eq:pathwaylikelihood}
\begin{split}
L_M(S) & = \lim_{|S| \longrightarrow \infty} \sqrt[|S|]{\prod_{(i,j) \in S} P((i,j) | M)} \\
 & = \lim_{|S| \longrightarrow \infty} \textrm{exp}(\frac{1}{|S|}\sum_{(i,j) \in S} \textrm{log} P((i,j) | M)) \\
 & = \lim_{|S| \longrightarrow \infty} \textrm{exp}(\sum_i \sum_j \frac{\hat{\tau}_E^{(i,j)}}{\sum \hat{\tau}_E^{(k,l)}} \textrm{log} P((i,j) | M)) \\
 & = \lim_{|S| \longrightarrow \infty} \textrm{exp}(\sum_i \sum_j \frac{\hat{\tau}_E^{(i,j)}}{\sum \hat{\tau}_E^{(k,l)}} \textrm{log} \frac{\hat{\tau}_M^{(i,j)}}{\sum \hat{\tau}_M^{(k,l)}}) \\
\end{split}
\end{equation}

Here, the first line follows from assumptions 1 and 2, the second is algebra, the third follows from assumption 3, and the fourth follows from assumption 4. 

\subsection{\label{urban-rural-si}Urban vs. Rural}

Studying urban/rural differences in pathways requires classifying geographic areas as urban vs. rural and then studying the flow between them. 

The first step is determining what unit of geography to use. According to Census Bureau, an \textit{urbanized area}\footnote{\url{https://www.census.gov/programs-surveys/geography/guidance/geo-areas/urban-rural.html}} is defined as a geographic unit (usually the equivalent of a city in terms of size) that has over 50,000 inhabitants and a population density of at least 1,000 people per square mile; Most urbanized areas have a dense urban core, and are surrounded by less-dense adjacent peripheries. Census Bureau considers any geographic area that is not an urbanized area to be ``rural.'' Unfortunately, Census Bureau supplies boundaries only for each urbanized area without dividing rural America into corresponding geographic areas, so urbanized areas cannot be used to study rural-rural or urban-rural diffusion. 

Since urbanized areas cannot be used to test hypotheses about urban/rural differences, another option could be to find a unit of geography that completely covers the U.S.A., where each unit is part of only one urbanized area. For instance, urbanized areas are subsets of contiguous Census blocks, so each Census block is either urban or rural; however, there are over 11 million Census blocks (i.e., there are 3-4 times as many Census blocks as agents in our model), so spatial time series over Census blocks would likely be too sparse to calculate pathway strengths. No larger units of geography can be classified as fully urban or rural using urbanized areas, so we would need to find another geographic area to classify as urban vs. rural. 

Counties are a natural choice for a geographic unit, as 1) counties completely cover the U.S.A. (e.g., some other units of geography, like core-based statistical areas, do not draw boundaries around all rural areas); and 2) the US Office for Management and Budget (OBM) publishes a method to classify each county as urban vs. rural.\footnote{\url{https://www.michigan.gov/documents/mdch/B-Urban_Rural_403741_7.pdf}} According to OBM's 2021 guidelines, each county is classified as ``urban'' if the urbanize areas contained within the county have at least 100,000 inhabitants.\footnote{\url{https://www.hhs.gov/guidance/document/defining-rural-population}}\footnote{\url{https://www.regulations.gov/document/OMB-2021-0001-0001}} Using these thresholds, 422 counties (about 13.10\% of the total) are considered urban, and the rest are rural. 

\section{\label{results-si}Results}

We include additional details of the results from the main paper.

\subsection{Sample Spatial Distribution}

\begin{figure}
    \centering
    \includegraphics[width=\textwidth]{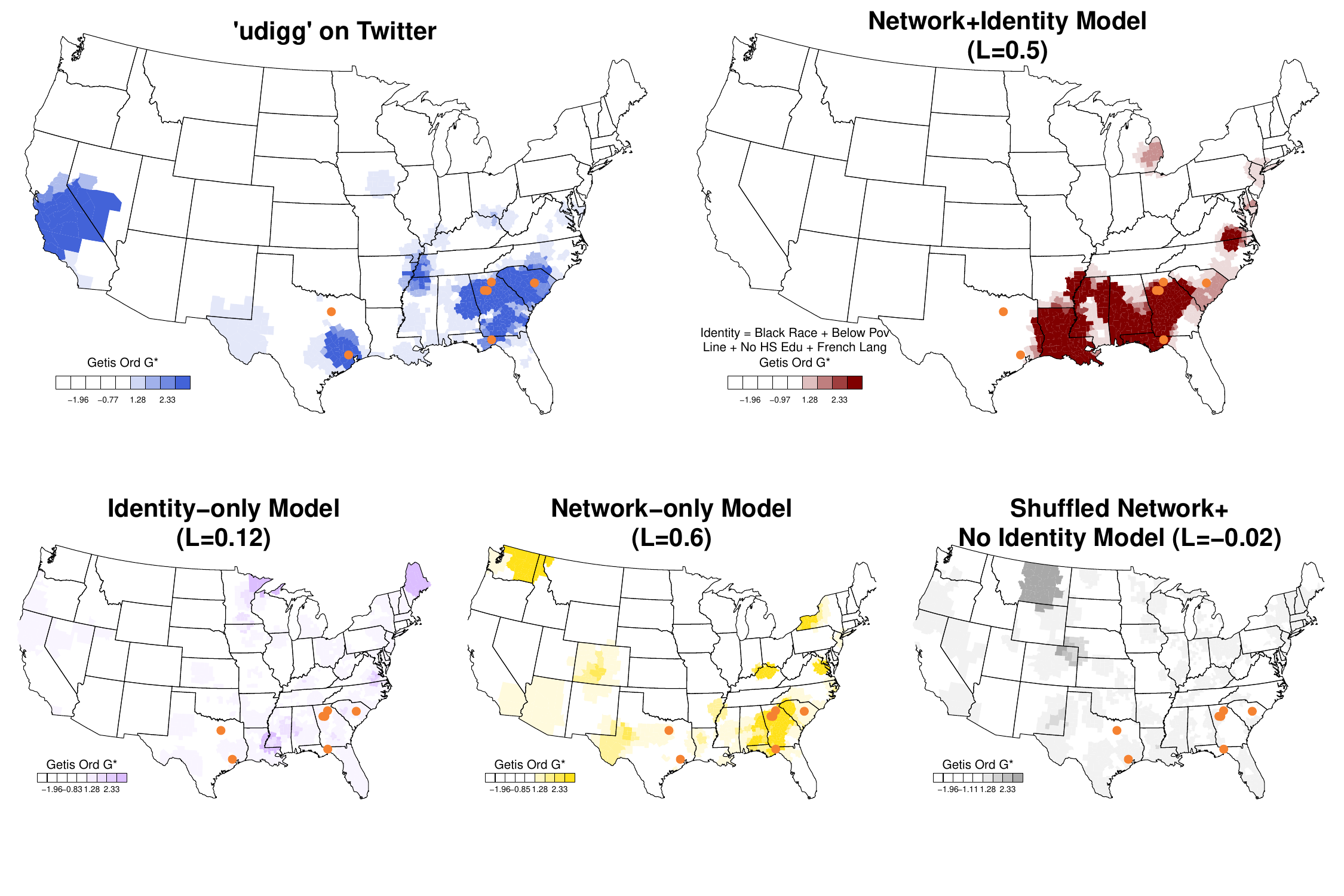}
    \caption{Comparing empirical and simulated spatial distributions for the word `udigg,' an alternative spelling of ``you dig?'' meaning ``Do you understand what I mean?'' Orange dots are the locations of the word's first ten adopters (Section~\ref{si:identifying-new-words}).}
    \label{udigg}
\end{figure}

\begin{figure}
    \centering
    \includegraphics[width=\textwidth]{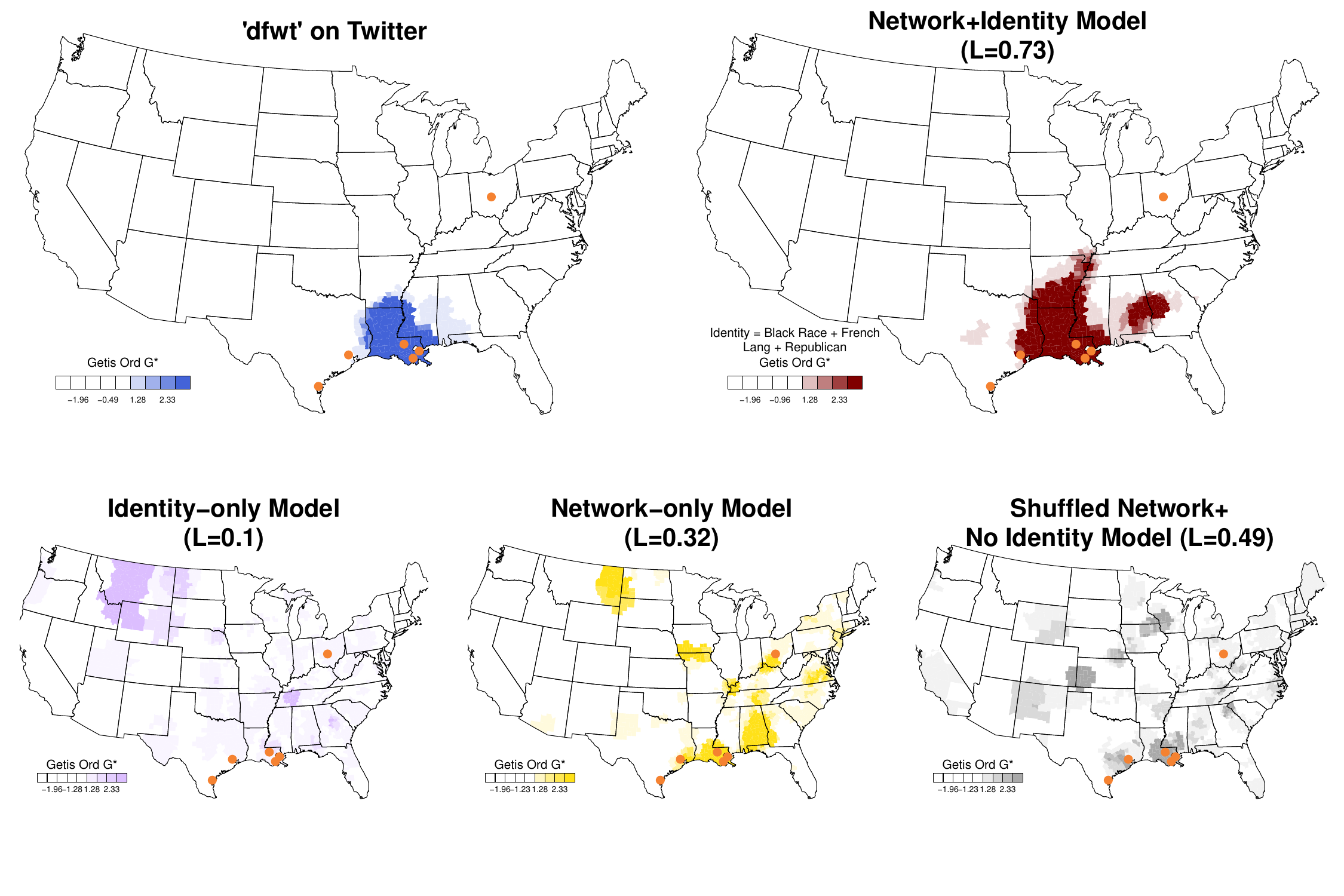}
    \caption{Comparing empirical and simulated spatial distributions for the word `dfwt,' abbreviated from ``Dont Fuck With That.'' Orange dots are the locations of the word's first ten adopters (Section~\ref{si:identifying-new-words}).}
    \label{dfwt}
\end{figure}

\begin{figure}
    \centering
    \includegraphics[width=\textwidth]{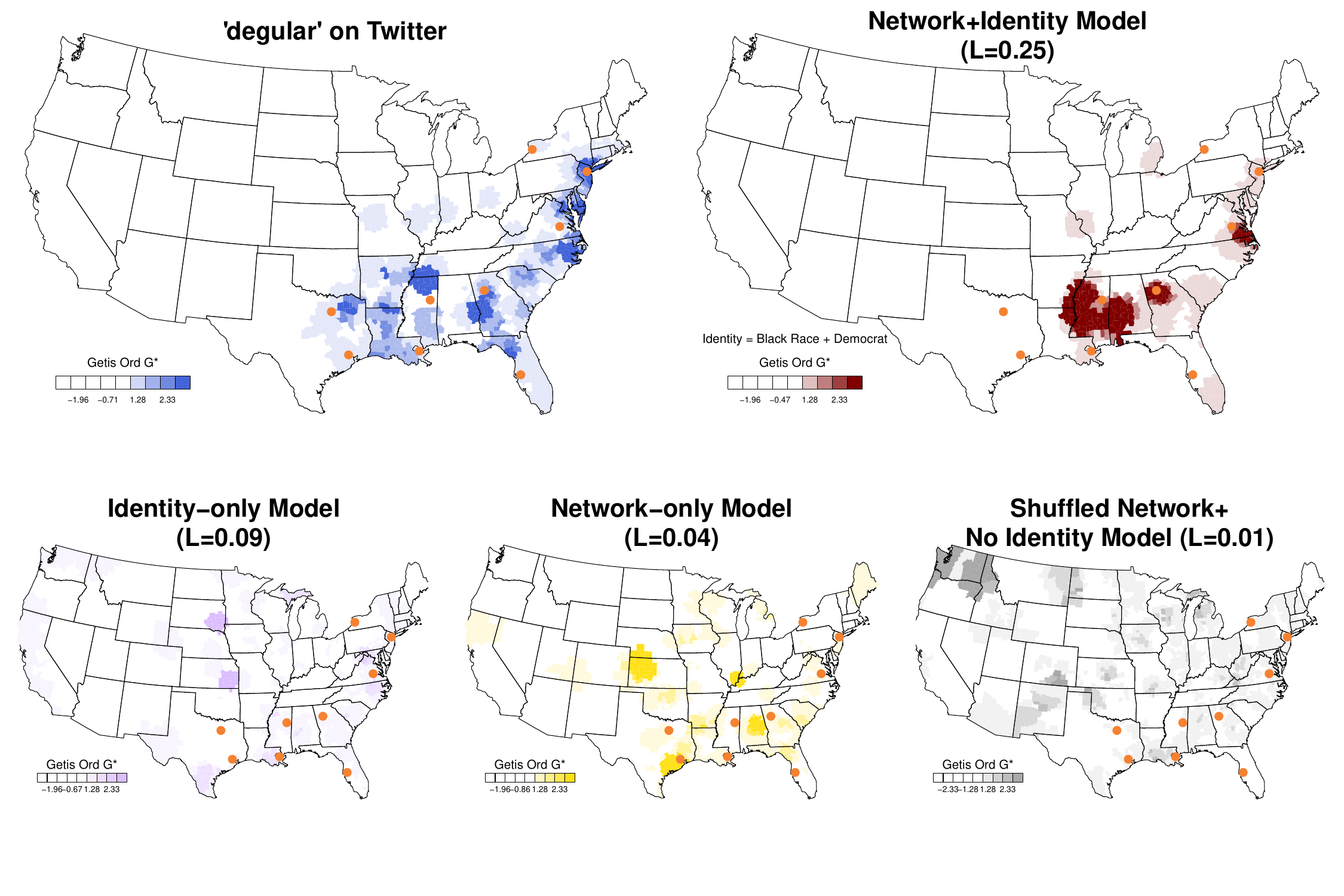}
    \caption{Comparing empirical and simulated spatial distributions for the word `degular,' from the phrase ``regular degular'' meaning ordinary or unremarkable. Orange dots are the locations of the word's first ten adopters (Section~\ref{si:identifying-new-words}).}
    \label{degular}
\end{figure}

\begin{figure}
    \centering
    \includegraphics[width=\textwidth]{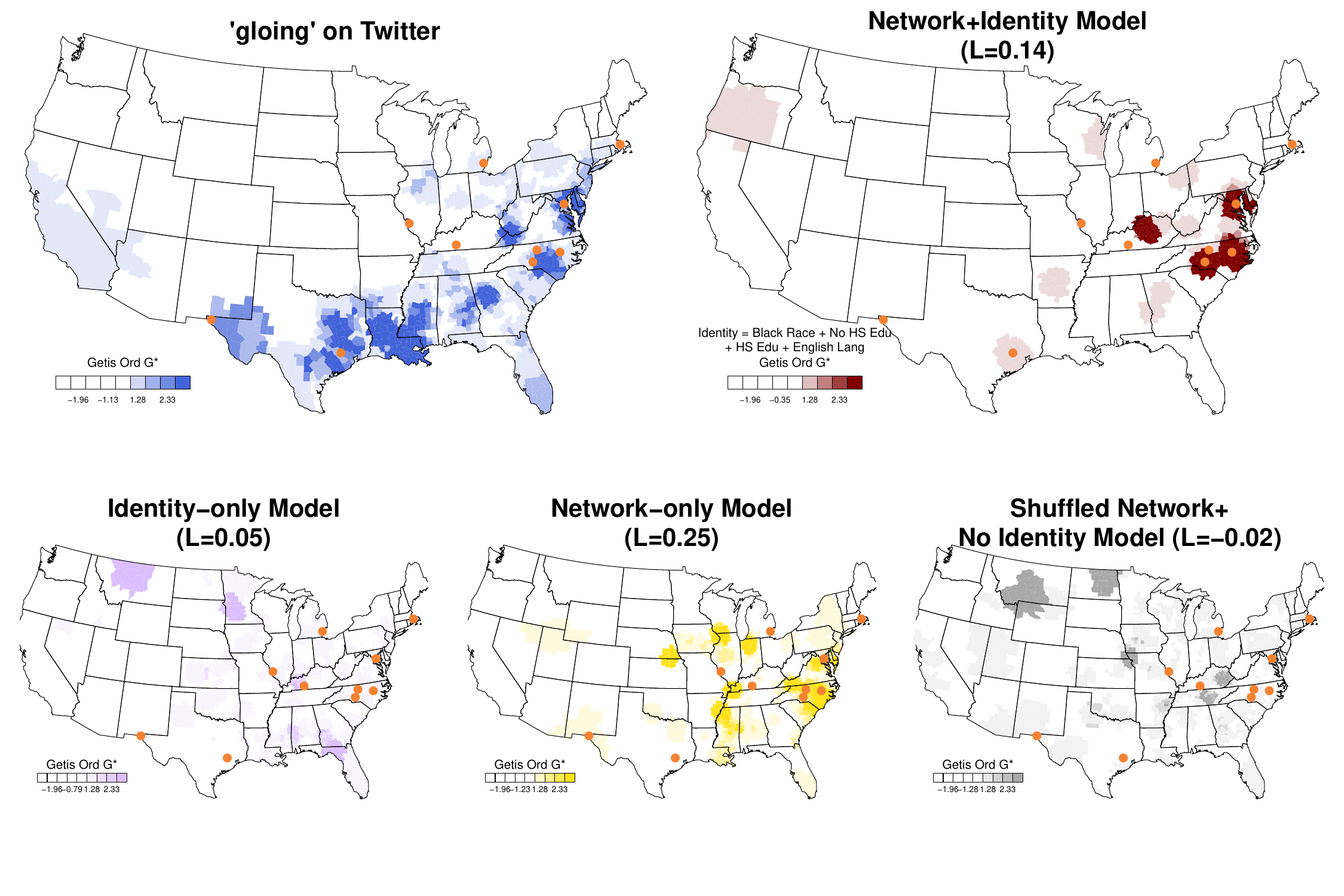}
    \caption{Comparing empirical and simulated spatial distributions for the word `gloing,' from the phrase ``gloing up'' mean growing up and becoming very attractive. Orange dots are the locations of the word's first ten adopters (Section~\ref{si:identifying-new-words}).}
    \label{gloing}
\end{figure}

\begin{figure}
    \centering
    \includegraphics[width=\textwidth]{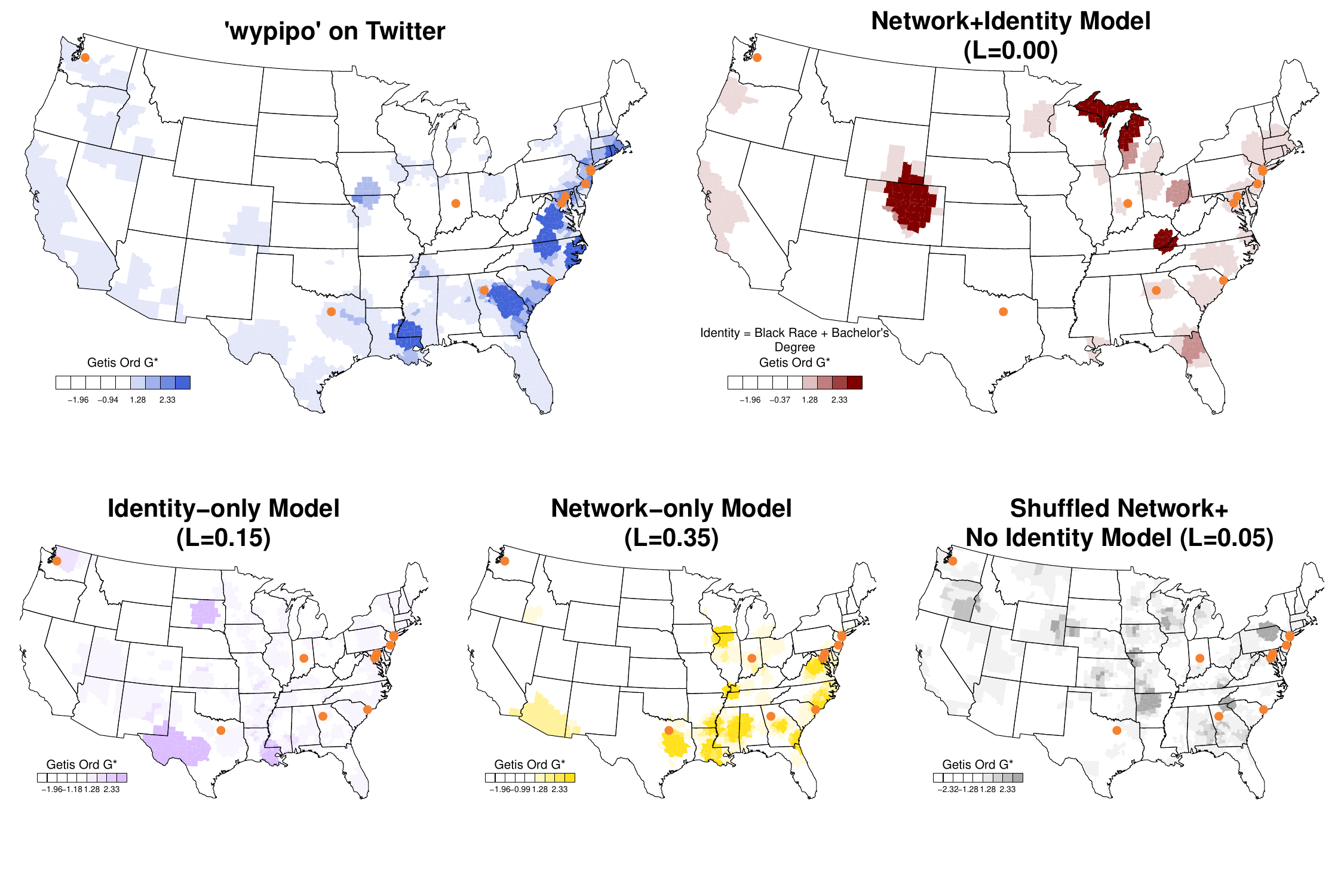}
    \caption{Comparing empirical and simulated spatial distributions for the word `wypipo,' meaning white people. Orange dots are the locations of the word's first ten adopters. (Section~\ref{si:identifying-new-words}).}
    \label{wypipo}
\end{figure}

We plot the smoothed spatial distributions for each of the 380 model runs and visually examined them to ensure the Lee's L results seemed consistent with our intuition about how well the empirical and simulated distributions matched each other. We display a few of these maps (including examples of good and poor matches) here; see Figures~\ref{udigg} -- \ref{wypipo}.

\subsection{Visualization of Pathways} 

We examine the strongest spatiotemporal pathways between pairs of counties in each of the three models, in order to visually assess whether they coincide with known cultural regions (Figure~\ref{pathways}). In order to avoid overcrowding our visual, we plot just a few of the top pathways (we find that the top 0.02\% provides the clearest visual); since large counties tend to account for many of the top pathways, we also exclude pathways going out of county $j$ that are not in the county's top 30 strongest pathways. Since the empirical pathways were sparsely sampled, we do not plot them here.

\begin{figure}
    \includegraphics[width=6in]{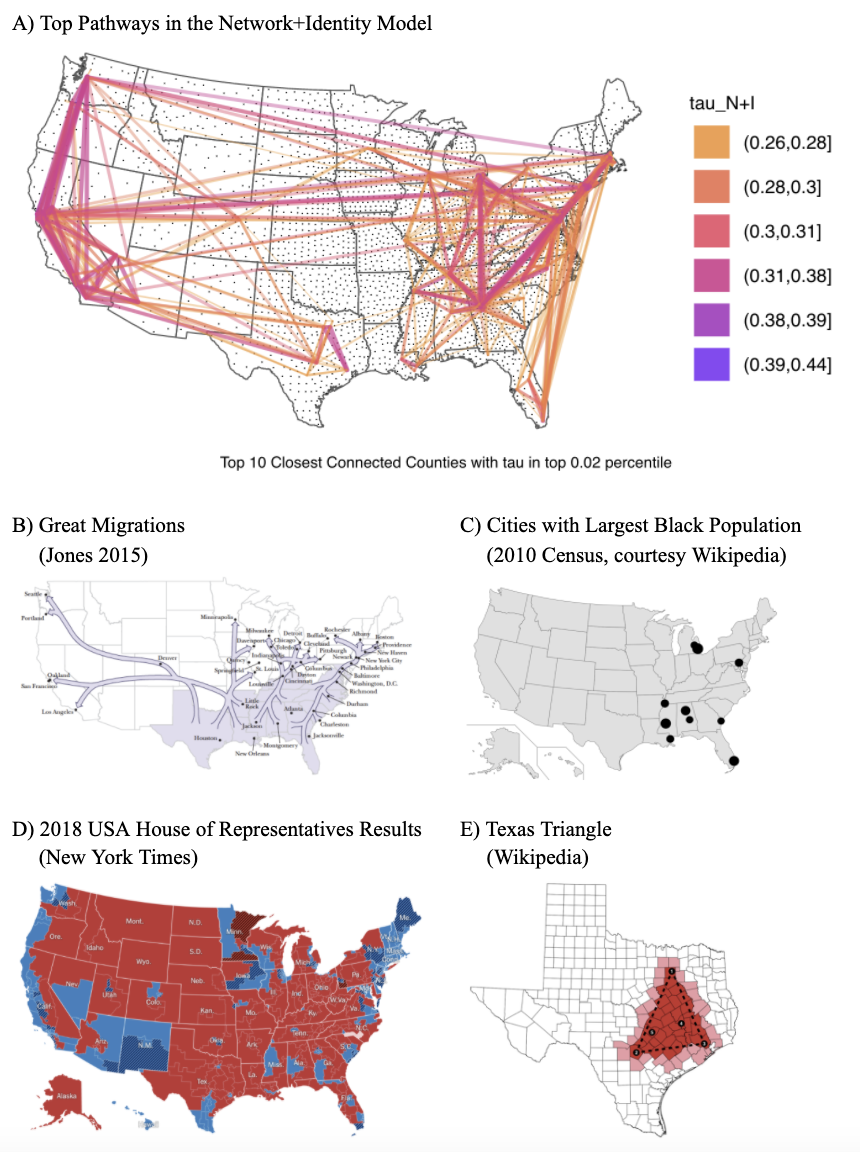}
    \caption{The strongest pathways in the Network+Identity model (part a) line up with some culturally significant regions: b) follow trajectories from the Great Migrations, pathways travel from Atlanta up the East Coast and to the Midwest, as well as from Texas to the West Coast; c) pathways also extend from Atlanta to several metropolitan areas with large Black populations; d) pathways travel between the Houston, Austin, Dallas, and San Antonio areas in Texas, a region known as the Texas Triangle; and e) pathways travel along each coast and between the East and West Coast, areas that are known for being politically liberal.}
    \label{pathways-cultural}
\vspace{-23.47865pt}
\end{figure}

\begin{figure}
    \centering
    \includegraphics[width=\textwidth]{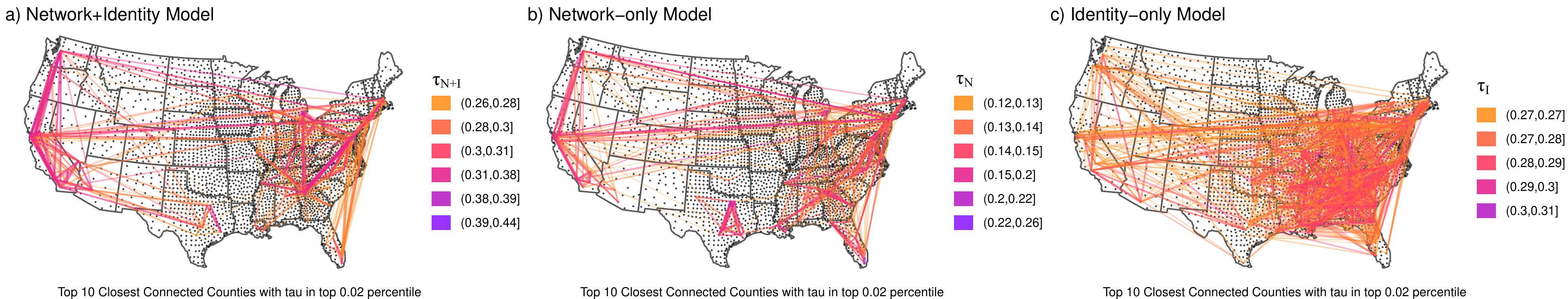}
    \caption{The strongest pathways across all three simulations. a) The Network+Identity model's pathways correspond to four major cultural regions (cf., Figure~\ref{pathways-cultural}); b) The Network-only model does not strong include pathways from Atlanta to metropolitan areas with large Black populations or from Texas to the West Coast; and c) The Identity-only model's pathways are poorly defined and do not include strong pathways along both coasts or in the Texas Triangle.}
    \label{pathways}
\end{figure}

We observe that pathways in the full Network+Identity model (Figure~\ref{pathways-cultural} A) coincide with some well-known geographic regions. Some pathways extend from the mid-Atlantic into the South, where African American Language is most adopted (Figure~\ref{pathways-cultural} B) \citep{Jones2015}; from Atlanta to other urban hubs in Eastern U.S.A., along pathways defined by the Great Migrations (Figure~\ref{pathways-cultural} B-C) \citep{Jones2015}; along and between both coasts, which are politically, linguistically, and racially distinctive from the middle of the country (Figure~\ref{pathways-cultural}D ) \citep{labov2012dialect,sylwester2015twitter}; and within the economically significant Dallas-Austin-Houston ``Texas triangle'' (Figure~\ref{pathways-cultural} E) \citep{cisneros2021texas}. 

Interestingly, the Network-only model appears to capture pathways of (i), (ii), and (iv) and the Identity-only model captures pathways of (iii) (Figure~\ref{pathways}). This suggests that the combined effect of network and identity increases the regions in which the model readily diffuses.

\FloatBarrier

\FloatBarrier

\FloatBarrier

\FloatBarrier

\FloatBarrier

\FloatBarrier

\FloatBarrier

\subsection{Model Error Analysis}

Characteristics of the new word are associated with the Network+Identity, Network-only, and Identity-only models' performance. A mixed-effects linear regression estimates the association between Lee's L for each trial and: (i) features of the model, like the word's stickiness, the identity signaled by the word, and the initial adopters' locations in the major U.S.A. divisions,\footnote{\url{https://www.census.gov/geographies/reference-maps/2010/geo/2010-census-regions-and-divisions-of-the-united-states.html}}; and (ii) how well the empirical distribution aligns with the top 5 dialect regions (i.e., the fraction of the empirical word's variance that is explained by each principal component, measured as the $R^2$ from regressing the PC loading against the word's spatial distribution). Since many trials correspond to identical initializations (the same initial adopters or simulation seed), random effects control for these potential confounds. There are over 35 variables in our error model, so we correct for multiple hypothesis testing using the Bonferroni correction.

All three models most closely approximate empirical distributions that align with the first principal component (corresponding to the U.S. South) or the fourth principal component (corresponding to the Mid-Atlantic, the region most similar to the first principal component) (Table~\ref{error-analysis}). Notably, the South is the most common region observed in not only our data but also other studies of lexical innovation (cf., \citep{grieve2016regional,grieve2017analyzing,cassidy1985dictionary}). Our error analysis suggests that both diffusion through the network and selection on the basis of shared identity provide mechanisms for words to diffuse to the South, and the fact that there are so many pathways to this region may partly explain why the South is so often a destination for innovation. The models also tend to better reproduce spatial distributions from words that are a little less sticky---i.e., words that diffused less broadly. 

\begin{figure}
    \includegraphics[width=\textwidth]{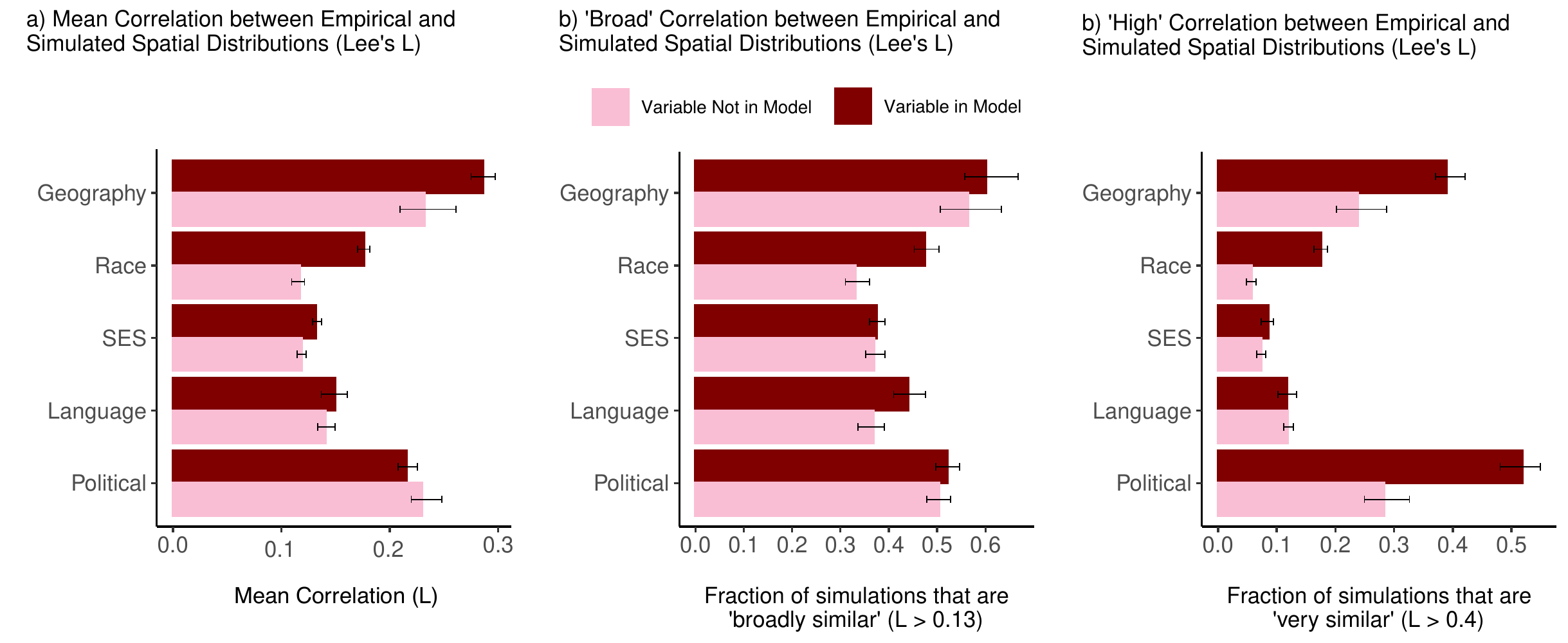} 
    \caption{The model performs best when four components of identity (geography, race, language, and politics) are included, while socioeconomic status does not seem to significantly affect the model's overall performance. The results are reported for just the words that signal that component of identity (e.g., the 20 words that signal race, with and without race in the model).}
    \label{ablation}
\vspace{-38.3148pt}
\end{figure}

Finally, we conduct ablation tests to assess the impact of removing each dimension of identity from the model. Specifically, we compare the Lee's L of the words signaling each component of identity to the Lee's L when that component of identity is removed from the model (e.g., the 20 words that signal race, with and without race in the model). See Figures~\ref{ablation}. The model performs best when Location, Race, Languages Spoken, and Political Affiliation are all included in the model -- the Lee's L is higher with these components than without. Interestingly, for SES, the performance of the model appears to be indistinguishable with or without that variable. The lack of significant impact could be because SES is so strongly correlated to Race and other dimensions of identity (cf., Figure~\ref{words-summary}). 

\begin{table}
    \centering
    \caption{Across all models, performance is associated with the word's stickiness and empirical geographic distribution, according to the mixed-effects regressions used for error analysis. For the listed models, each trial's Lee's L is the dependent variable while the trial characteristics are the independent variables. p-values are corrected for multiple hypothesis testing using the Bonferroni correction.}
    \begin{tabular}{@{} r *{2} l *{3}{S[table-format = -1.2]} @{}}
         & & \textbf{Network+
         Identity}& \textbf{Network-
         Only}& \textbf{Identity-
         Only}\\ 
         \midrule
        & Intercept & -0.06 & -0.11 & -0.15\\
        \midrule
        & Stickiness & -0.06\sym{***} & -0.02 & -0.04** \\
        \midrule
        & Location & 0.1 & -0.13 & 0.28 \\
        \midrule
        Race & White & -0.13 & 0.11 & -0.1 \\
        & Black & 0.01 & -0.02 & 0.01 \\
        & Hispanic & 0.79 & -0.17 & 1.31 \\
        & Native American & -0.96 & 0.18 & -0.95 \\
        & Native Hawaiian & 0.06 & 0.07 & 0.39 \\
        & Asian & 0.64 & 0.45 & 0.47 \\
        \midrule
        Languages Spoken & English & -0.01 & 0.16 & 0.11 \\
        & spanish & -1.07 & 0.42 & -2.23 \\
        & French & -0.12 & 0.19 & -0.08 \\
        & Chinese & -0.2 & -0.19 & -0.42 \\
        & Vietnamese & 0.42 & 0.42 & 0.56 \\
        & Tagalog & 0.1 & -0.04 & -0.01 \\
        \midrule
        Political Affiliation & Democrat & 0.13 & -0.04 & 0.33 \\
        & Republican & -0.64 & 0.46 & -0.23 \\
        & Other & -0.03 & 0.26 & -0.72 \\
        \midrule
        Empirical Dialect Region & PC1: South & 0.71\sym{***} & 0.76*** & 0.59*** \\
        & PC2: Louisiana & 0.07 & -0.09 & -0.01 \\
        & PC3: Coasts & 0.23 & -0.02 & 0.11 \\
        & PC4: Mid Atlantic & 0.36\sym{**} & 0.15 & 0.38*\\
        & PC5: Northeast & -0.13 & 0.05 & -0.05 \\
        \midrule
        Initial Adopters & EastNorthCentral & -0.07 & -0.04 & 0.13 \\
        & EastSouthCentral & -0.12 & 0.08 & 0.09 \\
        & MiddleAtlantic & -0.09 & -0.02 & 0.17 \\
        & Mountain & -0.06 & -0.15 & 0.01 \\
        & NewEngland & 0.03 & -0.11 & 0.21 \\
        & Pacific & -0.26 & -0.1 & 0.04 \\
        & SouthAtlantic & -0.23 & 0.04 & 0.07 \\
        & WestSouthCentral & 0.09 & -0.08 & 0.18 \\ 
         \bottomrule
    \end{tabular}
    
    \addtabletext{*p < 0.05}
    \addtabletext{**p < 0.01}
    \addtabletext{***p < 0.001} \\ 
    \label{error-analysis}
\end{table}
\FloatBarrier

\FloatBarrier

\subsection{\label{empirical-pathways-si}Empirical Pathways}

We present several empirical observations about the strength of pathways between counties on Twitter, which are critical to informing our hypotheses about the complementary mechanisms of network and identity in Section~\ref{complementary-roles}. To maintain a robust analysis, pathways with relatively few edges should be discarded. For 7,620,771 (89.87\%) pathways between pairs of counties $(i,j)$, no edges in the Twitter network start in county $i$ and end in county $j$; for an additional 720,597 (8.50\%) pathways, between 1 and 10 edges run from $i$ to $j$. When there are a small number of edges, the properties characterizing the pathway (e.g., distribution of edge weights and similarity, typically averaged over all of the edges) are very dependent on the properties of a few ties and have high variance. Therefore, to increase robustness, our analyses include just pathways $(i,j)$ where at least 10 edges run from $i$ to $j$. This leaves 138,376 pairs of counties, corresponding to 6,060 urban-urban, 41,420 urban-rural, and 90,896 rural-rural pathways.

\begin{figure}
    \includegraphics[width=\textwidth]{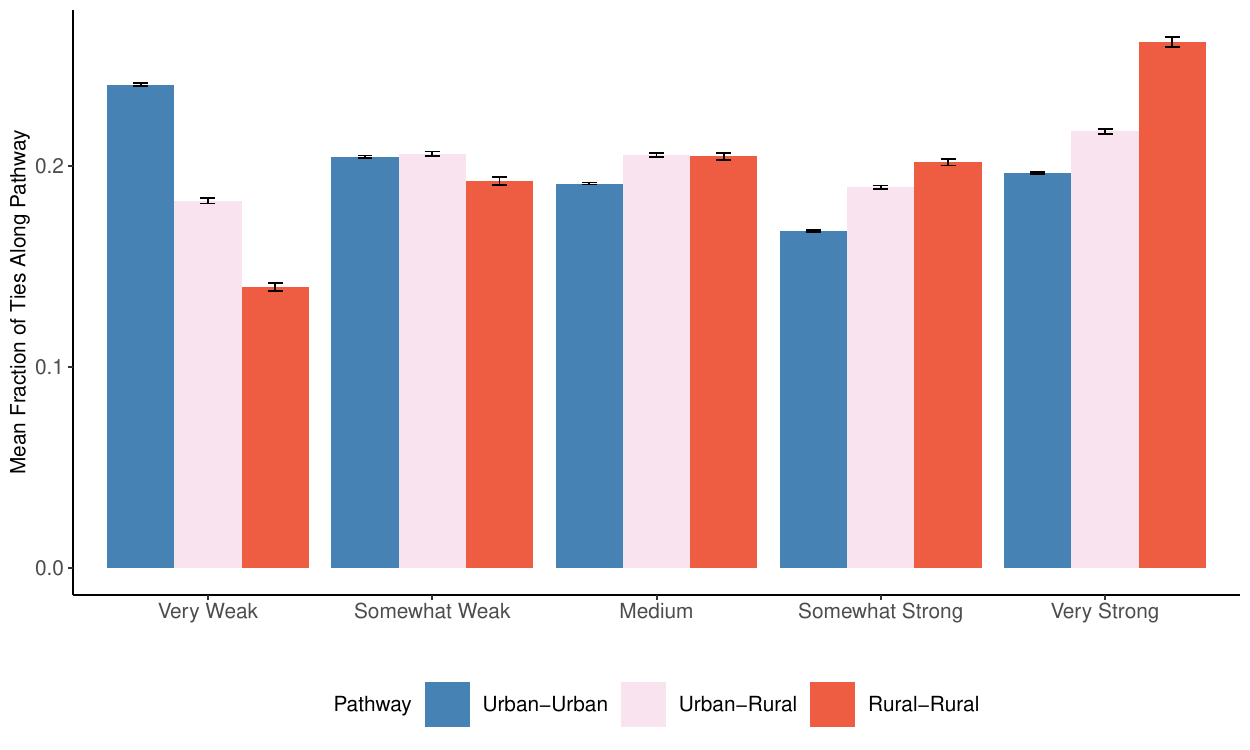} 
    \caption{Urban-urban pathways tend to have more weak ties, while rural-rural pathways tend to have more strong ties. Urban-rural pathways are in the middle. Error bars are 95\% bootstrap confidence intervals.}
    \label{fig-size}
\end{figure}

First, the strength of network ties (proportionate to the frequency of communication along the tie, as defined in Section~\ref{modeleqsi}, Equation~\ref{eq:weight}) is associated with the type of pathway. Figure~\ref{fig-size} plots how frequently each type of pathway is traversed by ties in each quintile of tie strength. Urban-urban pathways tend to have a higher fraction of weak ties (bottom two quintiles) and a lower fraction of strong ties (top two quintiles) running between them. By contrast, rural-rural pathways tend to have relatively more strong ties and fewer weak ties between them. Urban-rural pathways fall in between these two extremes, and tend to have a more even distribution of ties by strength. We define the strength of ties in the model equations.

\begin{figure}
    \includegraphics[width=\textwidth]{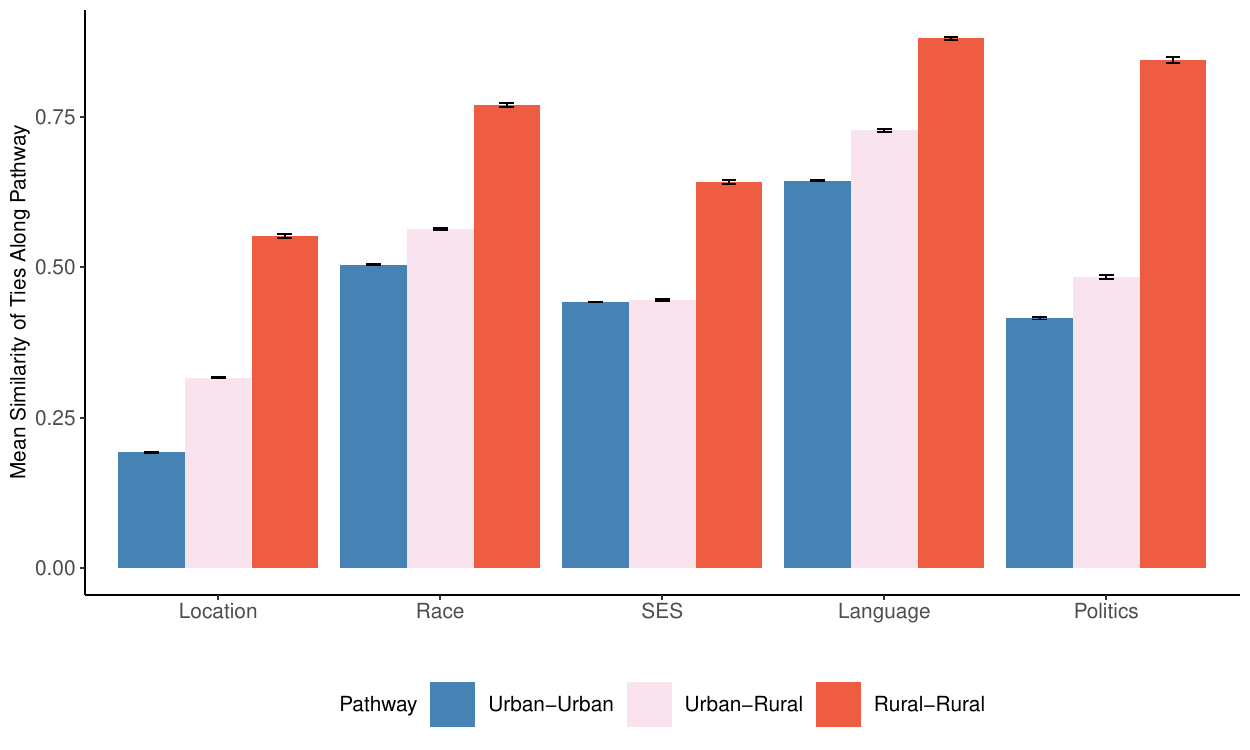}
    \caption{Urban-urban pathways tend to have more diverse ties (i.e., ties to individuals who are demographically different from them), while rural-rural pathways tend to have more similar ties. Urban-rural pathways are in the middle. Error bars are 95\% bootstrap confidence intervals.}
    \label{fig-sharedid}
\end{figure}

Second, urban-urban pathways tend to contain edges with less demographic similarity---that is, agents that are connected to residents of urban areas tend to be less similar to those residents (Figure~\ref{fig-sharedid}). By contrast, rural-rural pathways tend to contain edges with more demographic similarity. Urban-rural pathways fall in between these two extremes, though they tend to be closer to urban-urban pathways and have far less similarity than the rural-rural pathways do. The similarity in the identity of ties is the mean distance between users in the counties or mean similarity in race, SES, languages spoken, and politics using the same calculation for $\delta_{ij}$ (cf., Section~\ref{modeleqsi}, Equation~\ref{eq:neighid}).

\begin{figure}
    \centering
    \includegraphics[width=5.5in]{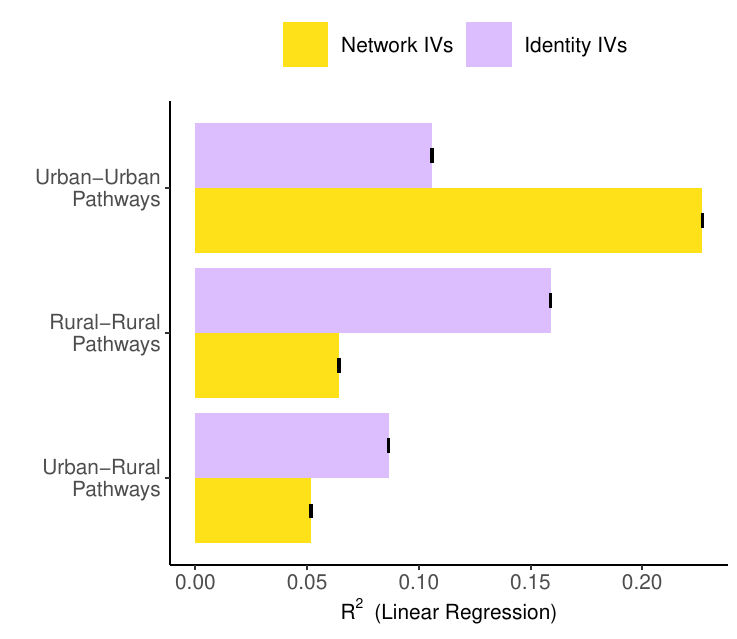}
    \caption{Using linear regression, network characteristics (number and strength of ties) explain a higher fraction of variance in urban-urban empirical pathway strengths and similarity in demographic identity explains a higher fraction of variance in urban-urban empirical pathway strengths.}
    \label{fig-netid}
\end{figure}

Third, network and identity tend to shape complementary sets of empirical pathways. To demonstrate the complementary roles of network and identity, we run four linear regressions explaining the strength of urban vs. rural empirical pathways using the two empirical characteristics about: 1) number of weak and strong ties running between the two counties (weak vs. strong are based on ties that fall into each quintile of pathway strength) or 2) similarity in identity. We find that, for urban-urban pathways, the network covariates explain a higher fraction of the variation in $\hat{\tau}_E$; for rural-rural pathways, shared identity explains a disproportionate fraction of the diversity in empirical pathway strength; urban-rural pathways fall in between (Figure~\ref{fig-netid} plots the $R^2$ for these six models, while we show the regression tables for each model in Table~\ref{reg-empnetwork}). 

Fourth, even though the network and identity play complementary roles in spatial diffusion (cf, Section~\ref{complementary-roles} and the prior paragraph), Network- and Identity-only pathways have similar pathway strengths. Indeed, the correlation between the strength of corresponding pathways in the Network- and Identity-only models is high (Pearson $R = 0.76$), suggesting that pairs of counties with a high propensity for transmission through the network are likely to also have robust diffusion on the basis of shared identity. This result is not surprising, since network characteristics, especially homophily, often correlate strongly to demographics \citep{mcpherson2001birds}. Notably, while Network- and Identity-only pathways through rural counties are often highly correlated (Pearson $R = 0.76$ for rural-rural pathways and $R = 0.77$ for urban-rural pathways), pathways among urban counties have weaker association ($R = 0.43$). In Section~\ref{complementary-roles}, we showed that urban-urban pathways become weaker when network and identity are both strong, while rural-rural pathways become stronger in this situation. In addition to the mechanism we proposed involving weak vs. strong ties, urban-urban pathways may become weaker when identity pathways are strong because Network- and Identity-only pathways reinforce each other less (i.e., when identity pathways become strong, network pathways are less likely to also be strong).

\begin{figure}
    \centering
    \includegraphics[width=\textwidth]{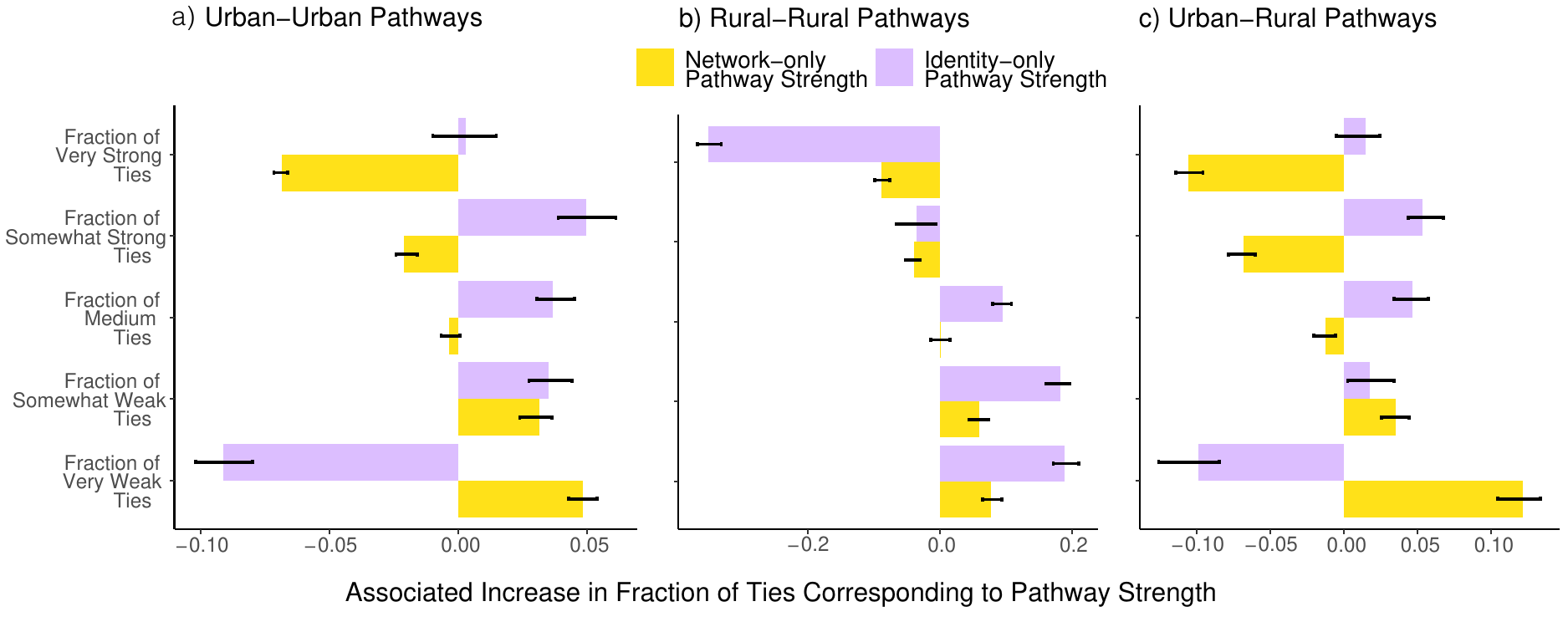}
    \caption{Network- and Identity-only pathways are poorly correlated in urban-urban pathways and strongly correlated in rural-rural pathways. a) Urban-urban pathways with a higher fraction of weak ties tend to have stronger network pathways and weaker identity pathways; conversely, pathways with a higher fraction of strong ties tend to have weaker network pathways and stronger identity pathways. b) On the other hand, rural-rural pathways with a higher fraction of weak ties tend to have stronger network \textit{and} identity pathways; conversely, pathways with a higher fraction of strong ties tend to have weaker network and identity pathways. c) Urban-rural follow the same pattern as urban-urban pathways.}
    \label{fig-weight-mechanism-si}
\end{figure}

Fifth, and relatedly, although ties with higher edge weight tend to connect users with higher demographic similarity (Table~\ref{reg-weight-id}), the network and identity pathways tend to fall out of sync in urban areas. We run ten regressions on pathways between counties, where the dependent variables are the fraction very weak (lowest quintile of edge weight), somewhat weak, medium strength, somewhat strong, and very strong edges connecting the two counties; independent variables are the strength of the network-only pathway ($\hat{\tau_N}$) and the strength of the identity-only pathway ($\hat{\tau_N}$) (Figure~\ref{fig-weight-mechanism-si}). In the urban-urban and urban-rural cases, pathways with more weak ties tend to have strong network and weak identity; conversely, pathways with more strong ties tend to have strong identity and weak network. In other words, urban pathways tend to be strong in one but not both mechanisms, and weak ties are associated with the typical weak-tie diffusion (strong diffusion through the network, between dissimilar nodes) \citep{Granovetter1973}. By contrast, the rural-rural pathways have stronger alignment between network and identity: pathways with more weak ties tend to be strong network \textit{and} identity pathways, while those with more strong ties tend to have weaker network and identity transmission.

\begin{figure}
    \includegraphics[width=\textwidth]{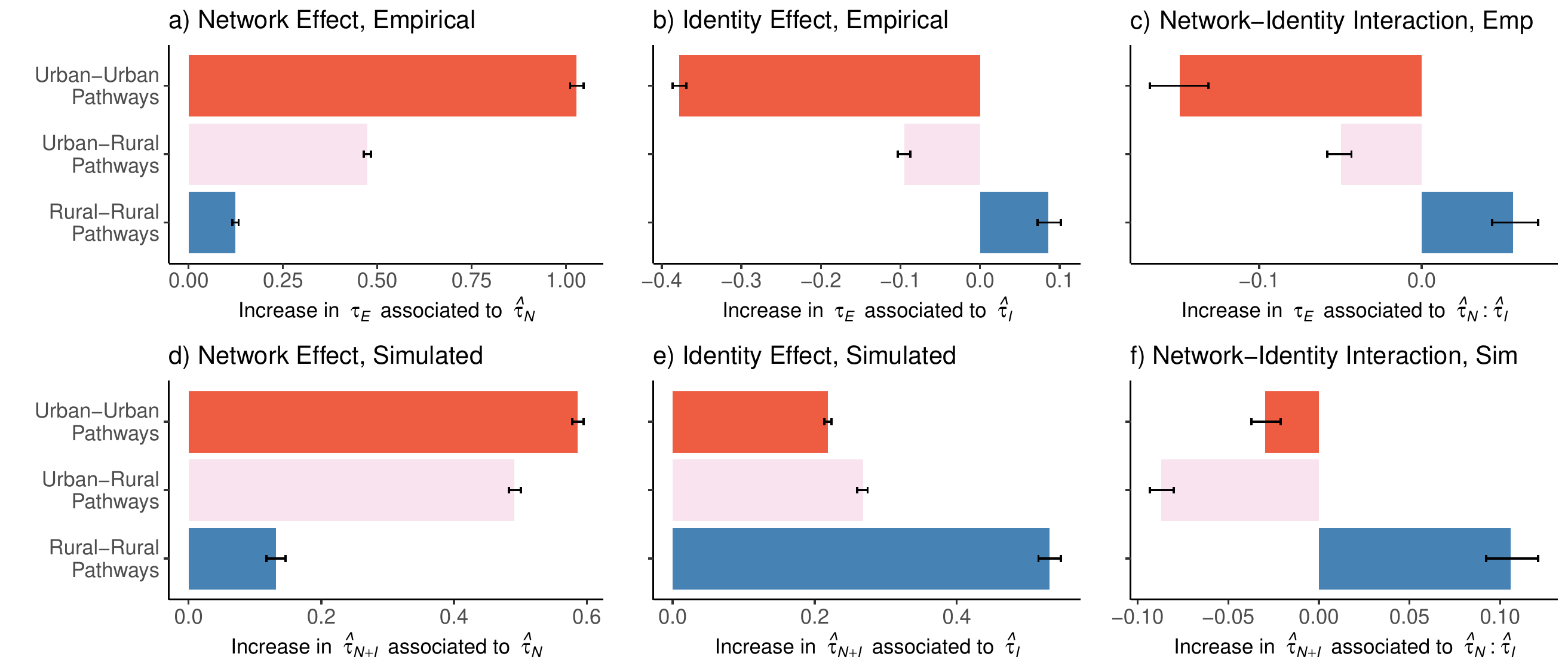} 
    \caption{Estimated slopes from a linear regression, predicting empirical pathway strength ($\tau_E$) from the strength of the pathways in the Network- and Identity-only models ($\hat{\tau_{N}}$,$\hat{\tau_{I}}$), interacted with the type of pathway (urban vs. rural county). We find that a) the strength of the Network-only model's pathways have the largest effect on the strength of the urban-urban empirical pathways than the Identity-only model; b) conversely, identity has the largest effect on the strength of rural-rural pathways; and c) urban-urban strong network pathways are dampened by strong identity pathways---and conversely, rural-rural strong identity pathways are amplified by strong network pathways. The same trends hold when we use the strength of pathways in the Network+Identity model ($\hat{\tau_{N+I}}$) as the dependent variable (d-f). Error bars are 95\% bootstrap confidence intervals.}
    \label{fig-emerge-si}
\end{figure}
\FloatBarrier

\subsection{\label{urban-rural-emerge}Urban vs. Rural Differences are Explained by Network Topology and Demographic Distributions}

We show that the complementary roles of network and identity are emergent properties of the spatial distribution of network and identity characteristics. Using linear regression, we compare the network-identity interactions required to reproduce empirical pathway strengths (Figure~\ref{fig-emerge-si}a-c), to those required to explain pathway strengths in the full Network+Identity model (Figure~\ref{fig-emerge-si}d-f).\footnote{Just as with empirical pathways, the variance in the full model's pathways can be almost entirely explained by the interaction between network, identity, and urban vs. rural pathways ($R^2=0.85$).} Full model pathways emerge from the interaction of network, identity, urban/rural classification in almost the same way as empirical pathways.\footnote{A key difference between the empirical and full model pathways: Identity pathway strength is negatively correlated to empirical pathway strength for urban-urban pathways, but positively correlated to full model pathway strength. Although both fit our theoretical expectations that identity matters less than network for urban-urban pathways (the slope is smaller for urban-urban than rural-rural pathways), the negative association also matches the hypothesis that identity inhibits weak tie diffusion. The failure of the full model to capture the negative association suggests that the distribution of network ties and demographics may not fully reproduce the diversity of exposure in urban areas.} Since our model equations do not explicitly specify differences between urban and rural agents, the differences we observe must be the result of the elements in our formulation---namely, the core modeling assumptions (e.g., fading of attention, diffusion through a network, performance of identity) and the underlying data (e.g., joint spatial distributions of network ties and demographics). 

Notably, the emergence of urban/rural differences in our model contradicts two factors often cited as the drivers of urban/rural dynamics: (i) that behavioral variation drives different ways of choosing which words to adopt  \citep{glenn1977rural,gimpel2020urban}; and (ii) that properties like population size and the number of incoming and outgoing ties drive different levels of exposure \citep{fischer1978urban,labov2003pursuing}. Our findings indicate that these two factors do not fully explain urban/rural differences. First, behavioral variation is not necessary to explain urban/rural differences, because our model, which reproduced these differences, has all agents using the same rules to decide whether to adopt the word. Second, in spite of having the same population, degree distribution, and core modeling assumptions, the Null (Shuffled Network+No Identity) model never accurately predicts pathway strength across all three types of pathways (Figure~\ref{fig-pathways})---suggesting that an explanation for urban/rural differences must include shared identity and network homophily. Based on our findings, we propose a new reason for the urban/rural patterns in the adoption of innovation: namely, the spatial distribution of ties and demographic characteristics. For instance, urban areas may tend to lead in language change, because they get early exposure via their numerous, diverse weak ties. Rural areas might adopt words from demographically similar or weakly tied urban areas, and, if it matches predominant local identities, transmit the word to other rural counties via strong ties.

\FloatBarrier

\FloatBarrier

\FloatBarrier

\begin{table}
\centering
\caption{Standardized coefficients from a regression showing the association between edge weight (dependent variable) and similarity in identity (independent variables). Stronger ties tend to share more demographic similiarities than weaker ones.}
\begin{tabular}{lrr}
  & Coefficient & \\
\midrule
Intercept & 0.00 & \\
Geography Similarity & 0.055 & *** \\
Race Similarity & 0.047 & *** \\
SES Similarity & 0.0013 & *** \\
Language Similarity & 0.0044 & *** \\
Politics Similarity & 0.0044 & *** \\
\bottomrule
\label{reg-weight-id}
\end{tabular}

\addtabletext{*** $p < 10^{-15}$}
\addtabletext{** $p < 0.01$}
\addtabletext{* $p < 0.05$}

\end{table}
\FloatBarrier

\FloatBarrier

\begin{table}
\centering
\caption{Displaying standardized coefficients from six linear regressions, predicting empirical pathway strength $\tau_E$. Each regression either uses network characteristics (the distribution of the number of ties by quintile of tie strength) or similarity in identity ($\delta_{ij}$ for each component of identity, as defined in the model equations). And each regression is run on the subset of pathways that run between urban counties, between rural counties, or from urban to rural counties.}
\begin{tabular}{lrrrrrr}
 & Urban-Urban Coefficient & & Rural-Rural Coefficient & & Urban-Rural Coefficient & \\
\midrule
Intercept & -1.45 & *** & -0.19  & *** & -0.51 &  \\
Log \# Weakest Ties (Quintile 1) & 0.43 & *** &  0.22  & *** & 0.34 & *** \\
Log \# Weak Ties (Quintile  2) & 0.15 & *** & 0.07  & *** & 0.10  & *** \\
Log \# Medium Ties (Quintile 3) & 0.09 & *** & 0.03  & * & 0.00  & \\
Log \# Strong Ties (Quintile 4) & -0.06 & *** & -0.02  & * & -0.06 & *** \\
Log \# Strongest Ties (Quintile 5) & -0.07 & *** & -0.17  & *** & -0.09 & *** \\
\midrule
$R^2$ & 0.48 & & 0.06  & & 0.10 & \\
\midrule
\midrule
Intercept & 1.16 & *** & 1.18  & *** & 0.48 & ***\\
Closeness in Geography &  0.59  & *** & -0.89 & *** & -1.77 & ***\\
Similarity in Race &  -0.98 & *** & -1.07  & *** & -0.66 & ***\\
Similarity in SES & -0.49  & *** &  0.06  & *** & -0.26 & ***\\
Similarity in Language & -1.03  & *** &  0.28  & *** & 0.43 & ***\\
Similarity in Politics &  0.13 & *** & -0.45  & *** &  0.16 & ***\\
\midrule
$R^2$ & 0.09 & & 0.13 & & 0.06 & \\
\bottomrule
\label{reg-empnetwork}
\end{tabular}

\addtabletext{*** $p < 10^{-15}$}
\addtabletext{** $p < 0.01$}
\addtabletext{* $p < 0.05$}

\end{table}
\FloatBarrier

\begin{table}
\centering

\caption{Displaying standardized coefficients from a linear regression, predicting empirical or simulated pathway strength from the strength of the corresponding pathways in the Network- and Identity-only models ($\hat{\tau_{N}}$,$\hat{\tau_{I}}$), interacted with the type of pathway (urban-urban vs. rural-rural vs. urban-rural; urban-rural is reference level).}
\begin{tabular}{lrrrr}
 & Dependent Variable: Empirical Pathway $\tau_E$ & & Dependent Variable: Simulated Pathway $\hat{\tau_{N+I}}$ & \\
\midrule
Intercept & -0.11 & & 0.03 & *** \\
rural & -0.52 & *** & -0.51 & *** \\
urban & 0.85 & *** & 0.39 & *** \\
$\hat{\tau_{N}}$ & 0.58 & *** & 0.53 & *** \\
$\hat{\tau_{N}}$ : rural &  0.39 & *** & -0.34 & *** \\
$\hat{\tau_{N}}$ : urban & -0.44 & *** & -0.02 & *** \\
$\hat{\tau_{I}}$ & -0.08 & *** & 0.33 & *** \\
$\hat{\tau_{I}}$ : rural & 0.24 & *** & 0.25 & *** \\
$\hat{\tau_{I}}$ : urban & -0.50 & *** & -0.16 & *** \\
$\hat{\tau_{N}}:\hat{\tau_{I}}$ & -0.06 & *** & -0.08 & *** \\
$\hat{\tau_{N}}:\hat{\tau_{I}}$ : rural & 0.17 & *** &  0.19 & *** \\
$\hat{\tau_{N}}:\hat{\tau_{I}}$ : urban & -0.06 & *** & 0.08 & *** \\
\midrule
$R^2$ & 0.71 & & 0.85 & \\
\bottomrule
\label{tab-confluence}
\end{tabular}

\addtabletext{*** $p < 10^{-15}$}

\end{table}
\FloatBarrier

\section{\label{limitations-si}Model Limitations}

We describe the limitations of our model and how we hedge against the types of methodological and ethical challenges large-scale computational studies tend to have \citep{olteanu2019social}.

\subsection{Data from Social Media}

A limitation of our study is its underlying social media dataset. In working with a 10\% Twitter sample, we are likely to have an incomplete and inaccurate account of many key parameters: agents, network edges, word adopters, etc. Additional concerns include not only the biased nature of the Twitter data---both the user base and speech patterns of users are unrepresentative of offline activity \citep{hargittai2015bigger,hargittai2020potential}---but also the lack of research systematically assessing under what conditions one can make inferences about offline phenomena from users' online behavior in general \citep{shaw2015big}. 

Although unaccounted gaps in the data may affect results, social networking sites are well-suited to studying the diffusion of linguistic innovation: The majority of present-day lexical innovation happens online and on a fast timescale \citep{Maity2016}, speakers leave a written trace of the words they use, and sites track features like user location and connections between users. Moreover, in the context of our study, computer-mediated communication generates relevant empirical datasets \citep{androutsopoulos2006introduction}, underscoring the validity and generalizability of our parameter estimation strategy. Prior research indicates that Twitter mirrors behaviors of interest and key assumptions underlying our model: for instance, online linguistic regions are similar to offline ones \citep{Grieve2018}, including in how a word diffuses over space \citep{eisenstein2012mapping,Eisenstein2014}; linguistic variation is associated with sociodemographic identities like place \citep{eisenstein2010latent,huang2016understanding}, race \citep{Jones2015,blodgett2016demographic,bokanyi2016race}, SES \citep{abitbol2018socioeconomic,abitbol2018location}, multilingualism \citep{vilares2015sentiment,stewart2018si}, and political ideology \citep{sylwester2015twitter}; networks on Twitter tend to be denser between locations with greater offline traffic flow \citep{takhteyev2012geography}; and mechanisms like enregisterment \citep{squires2010enregistering}, linguistic homophily \citep{bryden2013word}, and attention shifting \citep{shalom2019fading} are also found online.

\subsection{Construction of the Network}

There are many limitations to using the mutual-mention network, often consequences of the conservative decisions we took in order to maximize data integrity: First, the network we constructed does not capture all exposures a user may have to the new word. Users on Twitter see many tweets beyond those posted by their neighbors in the mutual-mention network, including users they follow and their followees, users whose posts they have engaged with even if the engagement is not mutual, promoted content, public tweets they find using the search feature, and, of course, words they hear used outside Twitter. Additionally, we are using a 10\% sample of tweets to construct the network, so some nodes and edges in the ``actual'' mutual-mention network are likely missing from the network we used for our simulation. Moreover, a Twitter user's network evolves over time, often co-evolving with linguistic use \citep{kovacs2020language}, while, for simplicity, we chose for our network to remain static. Finally, while some studies have shown similarities between online and offline networks \citep{adamic2003friends,takhteyev2012geography}, it is not clear that our results will generalize to linguistic innovation outside of Twitter. 

\subsection{Size of Validation Data}

In evaluating our model against the 76 new words on Twitter, potential methodological concerns are that (i) such a small sample may exclude important linguistic regions, and (ii) including only users with GPS-tagged tweets introduces bias both in which users are included in our study and in the linguistic attributes of their tweets \citep{malik2015population,pavalanathan2015confounds}. However, we verify the spatial distributions in our sample match those from other studies \citep{eisenstein2012mapping,Jones2015,grieve2016regional}; they contain overlapping words (e.g., boffum, fleeky, gmsfu), and have similar commonly appearing regions (Figure~\ref{top_pcs} vs. Figures~\ref{grieve} and~\ref{labov}). Since some of our words have only 1K uses in our dataset, we also attenuate noise in our spatial distributions by analyzing county-level maps, smoothed using local Getis-Ord G \citep{getis1992analysis}.

\subsection{Conceptualization of Identity}

We model the identity of each agent $\Upsilon_j$ using sociodemographic characteristics. Linguistic variables often index niche and dynamic identities, such as membership in a community of practice, interest areas, and intersectionality \citep{eckert2008variation}. Using time-invariant demographic markers as the sole proxy for identity is a limitation of our approach, reflecting assumptions about language variation that are simpler than conventions presently adopted in the sociolinguistics literature \citep{Eckert2012}. For instance, sociolinguists often study how language is used to signal non-demographic identity such as membership to a high school clique \citep{eckert2000language} or an online community \citep{ilbury2020sassy}. Nonetheless, we model identity using only sociodemographic characteristics, since identities do form on the basis of attributes like place, race, and socioeconomic status; additionally, non-demographic identities indexed by language style have not been shown to affect the geographic localization of linguistic variables in the U.S.A.

\subsection{Model Abstractions}
To maintain parsimony, we also did not incorporate several known factors about linguistic diffusion into our model. For instance, we did not consider the impact of phenomena like interdependencies in the diffusion of related lexical items \citep{friedkin2016network}, the co-evolution of networks and language adoption \citep{kovacs2020language}, structural diversity in the network \citep{ugander2012structural}, non-demographic identities \citep{eckert2008variation}, and audience accommodation \citep{giles1973accent,pavalanathan2015audience,kovacs2020language} on an agent's decision to adopt a new word; nor do we look at how adoption of cultural innovation itself shapes identity and networks \citep{lizardo2006cultural}. Although these assumptions oversimplify the mechanisms underlying diffusion, we chose to include them in order to isolate the effects of network homophily and performance of identity (e.g., some factors like structural diversity may confound the effects of homophily while audience accommodation may confound the effects of identity). Future work could explore the effects of one or all of these additional phenomena in modeling the diffusion of innovation.

\subsection{Ethical Considerations}
Many Twitter users do not know that the text and GPS coordinates of their statuses could be used by researchers, due to a lack of transparency in Twitter's terms of service \cite{fiesler2018participant,fiesler2020robots}. We chose to use data like the mention network and spatial time series for each word, fully aware of questions around the adequacy of consent on Twitter, because surveys have found that the users of social networking sites tend to be comfortable with researchers analyzing aggregates of language in public tweets \citep{fiesler2018participant}. However, we wholeheartedly support efforts to develop clearer standards for the use of online data in research \citep{vitak2016belmont,vitak2017ethics}, so users can be aware of, and have agency over, potential applications of their data.

\section{\label{sensitivity}Sensitivity Analyses}

We describe the procedures used to test for the robustness of our model to changes in some of the assumptions. We show that these changes do not meaningfully alter the Lee's $L$ correlation between the full Network+Identity model's spatial distribution and the empirical results. 

\subsection{\label{si-sensitivity-fb}Sensitivity to Network Topology}

A robustness analysis shows that our results generalize to other network topologies. Specifically, we hypothesize that our high-level results will not change when using a network constructed based on the county-to-county friend network provided in Facebook's Social Connectedness Index (SCI) \citep{bailey2018social} instead of the Twitter mention network. Facebook's Social Connectedness Index provides a measure for $SC_{ij}$, the \textit{social connectedness} between each pair of counties $(i,j)$:

\begin{equation}
\label{eq:sci}
SC_{i,j} = \frac{FBConnections_{i,j}}{FBUsers_i \cdot FBUsers_j}
\end{equation}

Since Facebook does not publish a user-to-user network, the Facebook user-user network used for sensitivity analysis is a synthetic graph constructed using three key assumptions:
\begin{itemize}
    \item The edge distribution between each pair of counties is proportionate to $FBConnections_{i,j}$ from Equation~\ref{eq:sci}.
    \item Nodes are just the same agents from the Twitter network (the same number of agents and each agent has the same geolocation), since SCI does not specify the number of users in each county. Therefore $FBUsers_i = TwitterUsers_i := N_i$.
    \item The synthetic Facebook network has the same number of edges as the Twitter network, in order to control for network size.
\end{itemize}

Using these three assumptions, the number of edges between each pair of counties is

\begin{equation}
\label{eq:fb-connections}
FBConnections_{i,j} = SC_{i,j} \cdot N_i \cdot N_j \cdot \frac{\sum N_i}{\sum_{k,l} SC_{k,l} \cdot N_k \cdot N_l} 
\end{equation}

Equation~\ref{eq:fb-connections} simulates the impact of a different network topology while controlling for agent distribution and network size. While differences in either of these characteristics could also alter patterns of diffusion, Facebook does not reveal these pieces of information for us to use. Additionally, there is a very strong association between agent distribution in the Twitter network and population size, both at the county-level and the Census tract-level (Pearson's $R > 0.98$), so we assume that Facebook and Twitter likely have similar user distributions since they both probably approximate the population distribution. 

In order to test the impact of network topology, we run all Network+Identity model trials as described in Section~\ref{experiments}, using five randomly generated Facebook user-user networks instead of the Twitter mutual-mention graph. In each synthetic Facebook user-user network, a pair of counties $(i,j)$ has $FBConnections_{i,j}$ ties running between them. Those ties are generated by randomly selecting, with replacement, $FBConnections_{i,j}$ agents in each county and drawing an edge between them. No multi-edges or self-edges were created in this process. The performance of the full model does not significantly change between the Twitter and Facebook networks (Figure~\ref{fig-sensitivity}). 

\subsection{\label{si-sensitivity-init}Sensitivity to Initial Adopters}

Since our data includes a 1\% to 10\% sample of tweets, there is a high likelihood that we will not actually capture the word's first ten adopters (Section!\ref{init-adopters-si}). However, as we will show, this shortcoming likely does not alter our results because the Network+Identity model's performance is not significantly different when run with different seeds. 

We run the Network+Identity model with ten different initial adopters. To account for the variability in the tweet sample, these initial adopters are a subset of the unique agents responsible for the first 50 uses of the word, since at least one of these 50 uses appears in a 10\% sample with probability 0.995. If there are fewer than 10 distinct users responsible for the first 50 users, the second through tenth unique adopters of the word are added to the list of users being sampled from. We draw five random samples of 10 unique users from this set, and run the full Network+Identity model seeded with each of these groups.

\begin{figure}
    \includegraphics[width=5.5in]{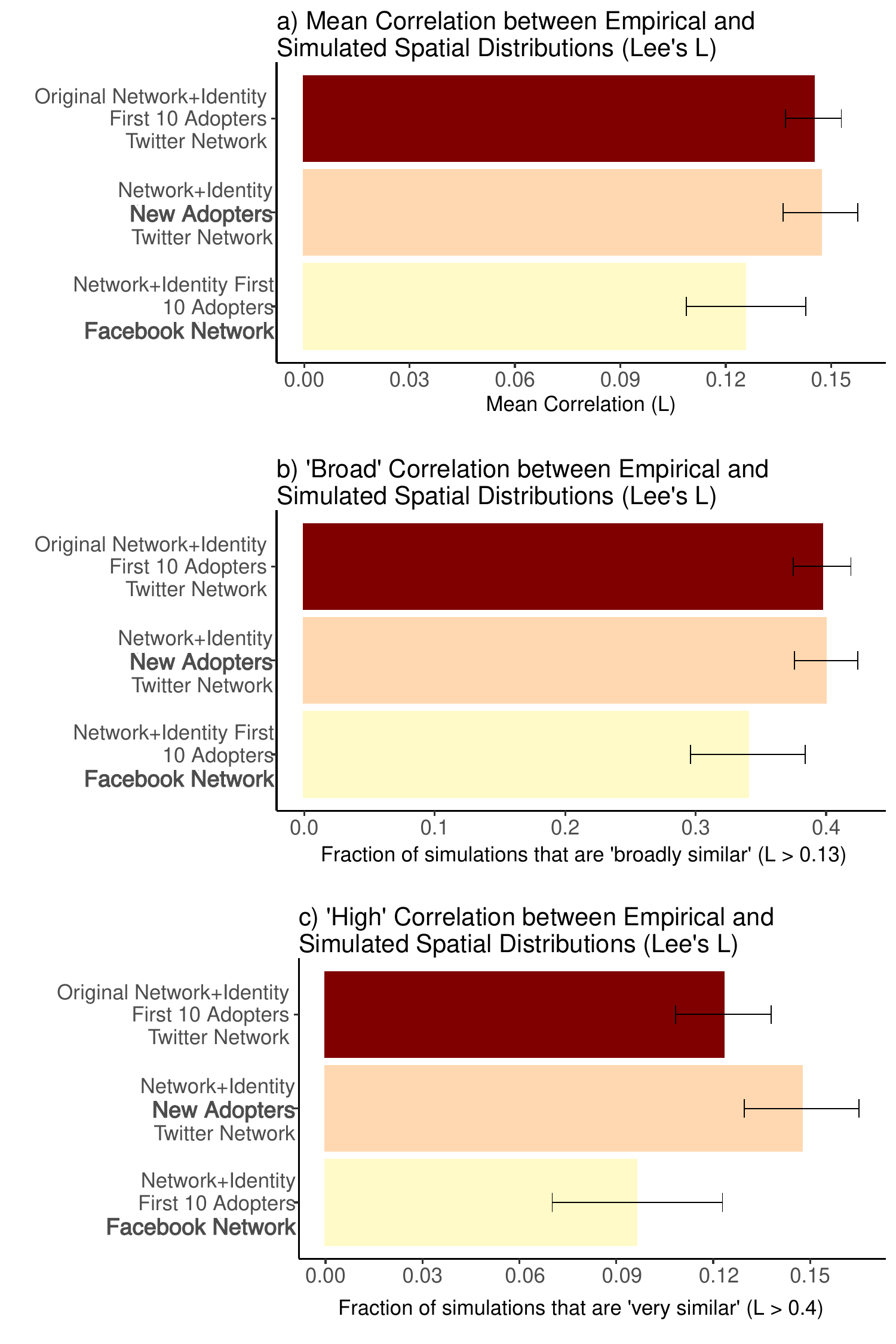} 
    \caption{The full Network+Identity model's performance is not sensitive to selection of initial adopters and network topology. The original Network+Identity model (Twitter network, first 10 observed adopters, Section~\ref{model-si}) does not reproduce empirical spatial distributions significantly better or worse with initial adopters randomly selected from the first 50 users of the word (Section~\ref{init-adopters-si}) and the Facebook network (Section~\ref{network-si}).}
    \label{fig-sensitivity}
\end{figure}

Compared to the Network+Identity model seeded with the first 10 observed adopters of the word, the performance does not significantly change when using these randomly sampled adopters (Figure~\ref{fig-sensitivity}).

\subsection{Sensitivity to Thresholds}

\begin{figure}
    \includegraphics[width=5.5in]{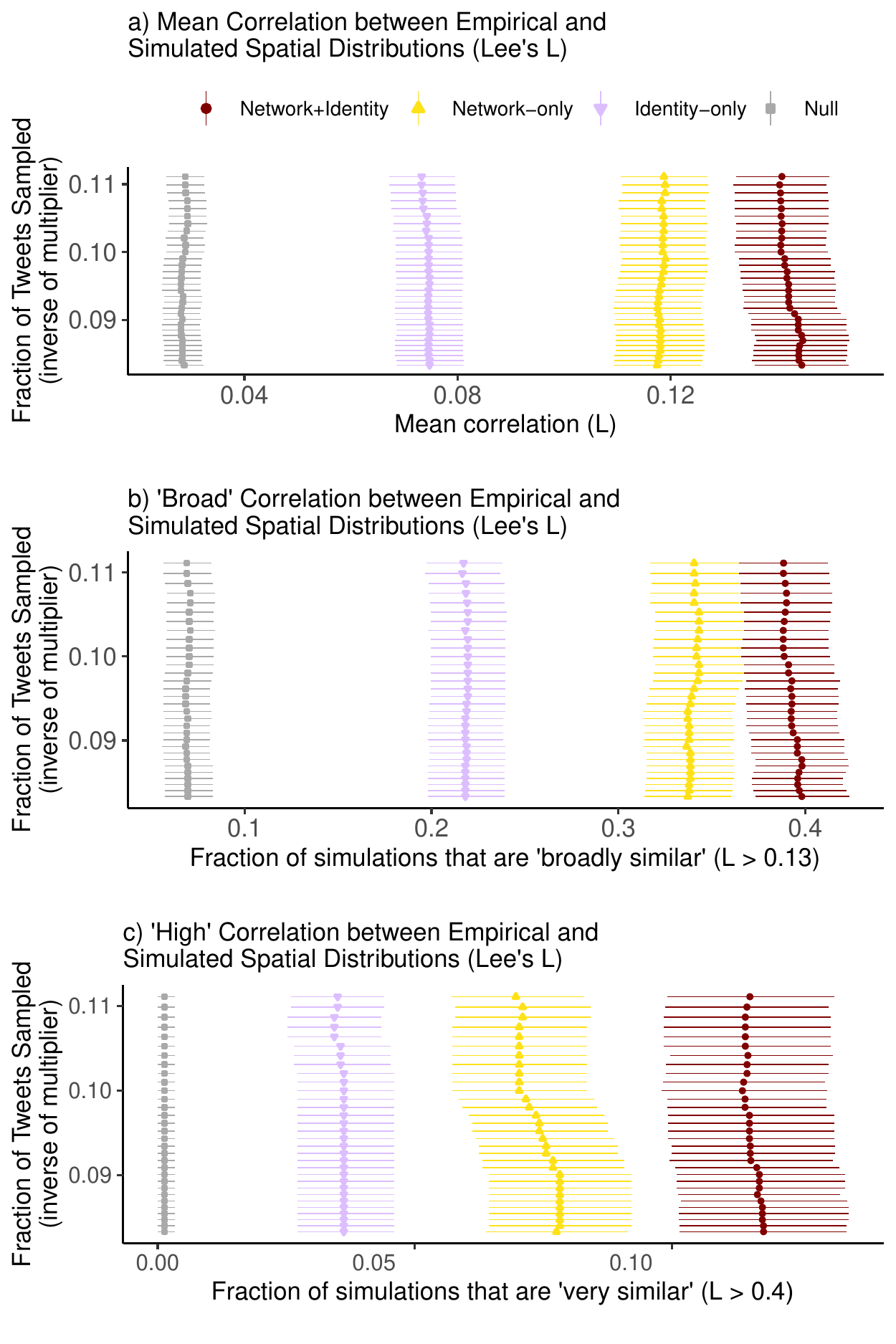} 
    \caption{The comparison between the four models' performance (in particular, the fact that the Network+Identity model performed best) is not sensitive to the choice of multiplier (i.e., how many more times was a word was used on Twitter than in our Decahose sample.)}
    \label{fig-scaling}
\end{figure}

The optimal parameters for each word correspond to trials that most closely matched the empirical level of usage (Section~\ref{param-est}). Specifically, the word needed to be used roughly 10x as often in the model compared to the usage in our Twitter sample. Using a multiplier of 10 is logical when using a 10\% sample of tweets from the Decahose. However, as detailed in Section~\ref{twitter-si}, the Twitter archives from which we pulled our sample sometimes logged less than 10\% of all tweets, so the ``true'' multiplier may be slightly higher than 10. Accordingly, we test for sensitivity to our main results (full model's Lee's L is higher than others) under different values of the multiplier, finding that the results are robust with factors as small as 8 and as large as 14 (Figure~\ref{fig-scaling}).

We stop the model once the growth in adoption slows to under 1\% increase over 10 iterations. For convenience, the iteration at which this condition is met will be denoted as $t_{STOP}$ Sensitivity analysis on a random 10\% sample of words (we select a 10\% sample in order to minimize runtime and avoid overfitting) suggests that spatial distributions of adoption do not meaningfully change once the growth of the model has slowed to this point. Specifically, the spatial distribution at $t_{STOP}$ and at $t_{STOP}+100$ are very highly correlated with Lee's $L>0.8$ in all of our runs.

We also have to set a threshold in order to determine which pathways to plot on maps in Figure~\ref{pathways}. Although plotting the top 0.02\% of ties overall and top 30 ties per county provides the clearest visual, we find that our conclusions remain the same if we have fewer (up to 0.01\%, 10 ties) or more (up to 0.05\%, 50 ties) pathways on the map.

\section{Datasets}

\subsubsection*{twitter-new-words-list.txt} The list of 76 innovative words we used in this study.

\subsubsection*{twitter-new-words-county-ts.txt}The empirical spatial time series for each of the 76 new words in our studies.

\FloatBarrier

\FloatBarrier

\end{document}